\documentclass[lettersize,journal]{IEEEtran}
\usepackage{algorithm}
\usepackage{stfloats}
\usepackage{url}
\usepackage{verbatim}
\usepackage{cite}
\usepackage{amsmath,amssymb,amsfonts}
\usepackage{algorithmic}
\usepackage{gensymb}
\usepackage{amsthm}
\newtheorem*{remark}{Remark}  

\usepackage{booktabs}
\usepackage{array}
\usepackage{adjustbox}
\usepackage{tabularray}
\usepackage{diagbox}
\usepackage{colortbl}
\usepackage{bbding}
\usepackage{wasysym}

\usepackage{graphicx,subfigure}
\usepackage{textcomp}
\usepackage{xcolor}
\usepackage{flushend}

\begin{document}
	
	\title{Frontiers of Generative AI for Network Optimization: Theories, Limits, and Visions}
	
	\author{Bo~Yang,~\IEEEmembership{Senior Member,~IEEE,} Ruihuai~Liang,~\IEEEmembership{Student Member,~IEEE,}  Weixin Li, Han Wang, \\Xuelin Cao,~\IEEEmembership{Senior Member,~IEEE,} Zhiwen Yu,~\IEEEmembership{Senior Member,~IEEE,}  Samson Lasaulce,~\IEEEmembership{Member,~IEEE,}  \\ M\'erouane Debbah,~\IEEEmembership{Fellow,~IEEE}, Mohamed-Slim Alouini,~\IEEEmembership{Fellow,~IEEE}, \\ H. Vincent Poor,~\IEEEmembership{Life Fellow,~IEEE},  and Chau~Yuen,~\IEEEmembership{Fellow,~IEEE}  
		\thanks{

			B. Yang, R. Liang, W. Li, and H. Wang are with the School of Computer Science, Northwestern Polytechnical University, Xi'an, Shaanxi, 710129, China (email: yang$\_$bo@nwpu.edu.cn, liangruihuai$\_$npu, li$\_$weixin, wang$\_$han@mail.nwpu.edu.cn). 
			
			Z. Yu is with the School of Computer Science, Northwestern Polytechnical University, Xi'an, Shaanxi, 710129, China, and Harbin Engineering University, Harbin, Heilongjiang, 150001, China (email: zhiwenyu@nwpu.edu.cn).
			
			X. Cao is with the School of Cyber Engineering, Xidian University, Xi'an, Shaanxi, 710071, China (email: caoxuelin@xidian.edu.cn). 
			
			S. Lasaulce is with Khalifa University, UAE, and CRAN, Nancy, France (email: samson.lasaulce@univ-lorraine.fr).
			
			M. Debbah is with the Center for 6G Technology, Khalifa University of Science and Technology, P O Box 127788, Abu Dhabi, United Arab Emirates (email: merouane.debbah@ku.ac.ae). 
			
			M.-S. Alouini is with the Computer, Electrical and Mathematical Sciences and Engineering Division, King Abdullah University of Science and Technology, Thuwal 23955, Saudi Arabia (email: slim.alouini@kaust.edu.sa).
			
			H. V. Poor is with the Department of Electrical and Computer Engineering, Princeton University, Princeton, NJ 08544, USA (email: poor@princeton.edu).
			
			C. Yuen is with the School of Electrical and Electronics Engineering, Nanyang Technological University, Singapore (email: chau.yuen@ntu.edu.sg). 
			
			(\textit{Corresponding author:} Chau Yuen)
	}}
	
	\markboth{Journal of \LaTeX\ Class Files,~Vol.~XX, No.~XX, May~2025}%
	{Shell \MakeLowercase{\textit{et al.}}: A Sample Article Using IEEEtran.cls for IEEE Journals}
	

	\maketitle
	
	\begin{abstract}
		While interest in the application of generative AI (GenAI) in network optimization has surged in recent years, its rapid progress has often overshadowed critical limitations intrinsic to generative models that remain insufficiently examined in existing literature. This survey provides a comprehensive review and critical analysis of GenAI's role in network optimization. We focus on the two dominant paradigms of generative models, including generative diffusion models (GDMs) and large pre-trained models (LPTMs), and organize our discussion around a categorization we introduce, dividing network optimization problems into two primary formulations: one-shot optimization and Markov decision process (MDP). We first trace key developments, including foundational contributions from the broader AI community, and systematically categorize current efforts in networking. We also briefly review frontier applications of GDMs and LPTMs in other related networking tasks, providing additional context for this survey. {Building on this, we present theoretical generalization bounds for GDMs in both one-shot and MDP settings. The established bounds also enable an equivalent analysis of LPTMs.} Most importantly, we reflect on the overestimated perception of GenAI’s general capabilities and caution against the all-in-one illusion it may convey. We highlight critical limitations, including difficulties in satisfying hard constraints, limited concept understanding, and the inherently probabilistic nature of outputs. We also propose key directions for future research, such as bridging the gap between generation and optimization. Although these two tasks are increasingly integrated in practical applications, they differ fundamentally in both objectives and underlying mechanisms, necessitating a deeper understanding of their theoretical connections. Ultimately, this survey aims to provide researchers and practitioners with a structured overview and a deeper insight into the strengths, limitations, and potential of GenAI in network optimization.
	\end{abstract}
	
	
	\section{Introduction}
	\IEEEPARstart{S}{ince} the proposal of the 6G roadmap \cite{roadmap6G}, the vision for future networks has evolved beyond traditional connectivity to encompass native support for intelligence, computing, and sensing. These new demands place unprecedented emphasis on advanced, flexible, and autonomous network optimization capabilities \cite{shi2023optimizationIn6G,AI2024MultipleAccessSurvey}. Current architectures such as 5G radio access networks (RANs) have already transitioned to virtualized forms like virtualized RANs (vRANs), where standardized network slicing enables differentiated service delivery. The open RAN (O-RAN) architecture takes this further by emphasizing openness, modularity, and intelligence, introducing the RAN intelligent controller (RIC) as the key layer for network optimization. The RIC supports both near-real-time (on the millisecond scale) and non-real-time (on the second scale and above) control through advanced algorithmic strategies \cite{oRAN2024Evolution6G}. However, despite this progress, existing systems still suffer from limitations such as rigid interfaces and interoperability challenges between heterogeneous software and hardware components, which constrain optimization efficiency and scalability. The next-generation architecture of 6G, therefore, calls for more intelligent, software-defined, and fully programmable approaches to network optimization.
	
	{\textit{Network optimization}} refers to a set of mechanisms, algorithms, and models designed to monitor, manage, and enhance the performance of communication networks (whether wireless or wired) \cite{mec2017mao}. The targeted performance metrics may vary across scenarios and typically include latency, coverage, throughput, reliability, and energy efficiency. While some network improvements involve high-level modifications to protocols or architectural mechanisms \cite{NetProvision2022NSDI,Taking5GRAN}, much of the optimization effort in practice focuses on parameter control and decision-making within existing operational frameworks, especially in the context of near-real-time network management (e.g., in the RIC). In this survey, we concentrate on how Generative AI (GenAI) contributes to this type of practical network optimization, where network control tasks are typically formulated as mathematical optimization problems. These formulations involve: (i) optimization variables, representing the parameters to be adjusted; (ii) input parameters, capturing system states or environmental information; (iii) an objective function, describing the performance goal; and (iv) constraints, encoding operational or physical limits. Depending on the temporal dependency structure of decisions, these problems can be modeled either as \textit{one-shot optimization}, where each decision is made independently, or as a \textit{Markov decision process (MDP)}, where decisions are made sequentially and influenced by state transitions.
 	
	Current network optimization tasks frequently address scenarios such as computation offloading \cite{yb-tmc}, throughput maximization in reconfigurable intelligent surface (RIS)-assisted networks \cite{MADRL2024xURLLCris}, RAN slicing for vehicular networks \cite{DyRANSlicing2021}, energy-efficient cloud services \cite{TanGo2023infocom}, traffic load balancing \cite{Transmitting2023Mobicom}, and edge-assisted satellite–terrestrial systems \cite{Edge2023LEO}. These approaches predominantly fall into two categories: data-driven methods \cite{yb-tmc,MADRL2024xURLLCris,DyRANSlicing2021} and numerical optimization algorithms \cite{TanGo2023infocom,Transmitting2023Mobicom,Edge2023LEO}. However, with the advent of 6G and its anticipated demands including access network slicing for diverse services, large-scale swarm coordination, and intelligent interpretation of emerging service requirements, traditional methods face growing limitations. Traditional data-driven models often struggle with generalization to unseen scenarios, while numerical techniques face scalability challenges as problem complexity increases.
	
	Against this backdrop, the rapid advancement of GenAI offers promising opportunities for addressing complex network optimization problems in a more scalable and generalizable manner. GenAI, in contrast to discriminative AI, focuses on learning the underlying global probability distribution of data, enabling the generation of new samples through distribution-based sampling (e.g., in image synthesis). Discriminative AI, by comparison, typically aims to learn from labeled data to distinguish between input samples, as in classification or regression tasks. Because GenAI models the full data distribution, they inherently possess discriminative capabilities and often exhibit strong cross-domain generalization, even when their outputs are probabilistic. It also inherits the scalability of deep learning for high-dimensional data and has become increasingly adaptable as large-scale models have emerged. Earlier generative models, such as variational auto-encoders (VAEs) \cite{VAE2020NIPS}, generative adversarial networks (GANs) \cite{GAN2024NIPS}, and normalizing flow (NF) \cite{NormalizingFlow2021}, primarily evolved from density estimation approaches. More recent frameworks, such as the generative flow network (GFlowNets) \cite{GFlowNets2021Bengio}, have been developed to handle discrete combinatorial optimization tasks in AI for science (AI4S). However, these models still face limitations in architecture and applicability. Despite ongoing use in areas such as channel estimation and noise modeling \cite{VAE2022channelEst,GAN2022IPheader,NF2019NoiseModeling,gflownet2023selection}, they have produced few results in complex network optimization scenarios.
	
	Today, the two most prominent classes of GenAI are {\textit{generative diffusion models (GDMs)}} and {\textit{large language models (LLMs)}}. GDMs, originally introduced for image generation with denoising diffusion probabilistic models (DDPMs) \cite{DDPM} as the representative example, generate samples by iteratively denoising from a known noise distribution. LLMs \cite{GPT3}, known for their scalability, have shown remarkable capabilities in natural language processing and reasoning. While LLMs have shown promising results in communication and networking applications, their scope is largely limited to language-centric tasks and fails to fully encompass the diverse modalities (such as channel state information, graph structures, and signal traces) common in network optimization. To better capture this broader range of generative modeling efforts, we adopt the term \textit{LPTMs} \cite{LMM-PM2024} in this survey. The main scope difference between LPTMs and LLMs is that LPTMs cover a wider range of communication modes. Accordingly, we treat GDMs and LPTMs as two distinct paradigms throughout the rest of this paper. GDMs refer to models trained using forward-noising and reverse-denoising mechanisms, while LPTMs denote large-scale pre-trained (mainly autoregressive) models trained on various types of data beyond natural language, including, but not limited to, multimodal and domain-specific representations relevant to networking. Both paradigms support diverse neural architectures and are capable of addressing a wide range of optimization tasks across modalities. {Tables \ref{tab_GDM_LPTM_comp} and \ref{tab_GDM_LPTM_advan} provide a more intuitive comparison of GDMs, LPTMs, and traditional methods, clearly illustrating the distinctive characteristics of GDMs and LPTMs, as well as their respective advantages over traditional approaches.}
	
	Researchers have begun leveraging these powerful generative paradigms to tackle optimization tasks across various layers of the network stack. These models are used not only for modeling and decomposing optimization problems, but also for directly generating high-quality solutions and enabling stochastic decision-making. In the past two years, GenAI-driven approaches—especially those based on GDMs and LPTMs—have rapidly evolved in both breadth and depth, leading to promising practical deployments and theoretical insights in network optimization.
	
	\begin{table*}[t]
		\centering
		\caption{Key abbreviations.}
		\begin{tblr}{
				colspec={X[0.2,l] X[0.52,l] X[0.2,l] X[0.52,l]},
				cells = {c},
				hlines, vlines,
				rowsep = {2pt},
				row{1} = {0ex},
			}
			\textbf{Abbreviation} & \textbf{Full term} & \textbf{Abbreviation} & \textbf{Full term} \\
			AI & Artificial Intelligence & ISAC & Integrated Sensing And Communication \\
			AI4S & AI for Science & LLM & Large Language Model \\
			AIGC & AI Generated Content & LoRA & Low-Rank Adaptation \\
			AoI & Age of Information & LPTM & Large Pre-Trained Model \\
			BBO & Black-Box Optimization & MDP & Markov Decision Process \\
			BLO & Bi-Level Optimization & MEC & Multi-access Edge Computing \\
			CFG & Classifier-Free Guidance & MLP & Multi-Layer Perceptron \\
			CSI & Channel State Information & NF & Normalizing Flow \\
			DDIM & Denoising Diffusion Implicit Model & OOD & Out-Of-Distribution \\
			DDPG & Deep Deterministic Policy Gradient & OPRO & Optimization by PROmpting \\
			DDPM & Denoising Diffusion Probabilistic Model & O-RAN & Open RAN \\
			DiT & Diffusion Transformer & QoS & Quality of Service \\
			DRL & Deep Reinforcement Learning & RAN & Radio Access Network \\
			GAN & Generative Adversarial Network & RIC & RAN Intelligent Controller \\
			GAT & Graph Transformer & RIS & Reconfigurable Intelligent Surface \\
			GCN & Graph Convolutional Network & RLVR & RL with Verifiable Rewards \\
			GDM & Generative Diffusion Model & RSMA & Rate-Splitting Multiple Access \\
			GenAI & Generative AI & RSU & RoadSide Unit \\
			GFlowNet & Generative Flow Network & UAV & Unmanned Aerial Vehicle \\
			IoT & Internet of Things & UE & User Equipment \\
			IP & Internet Protocol & VAE & Variational Auto-Encoder \\
			IQL & Implicit Q-Learning & VR & Virtual Reality \\
            IRS & Intelligent Reflecting Surface & vRAN & Virtualized RAN \\
		\end{tblr}
		\label{tab_key_abbreviations}
		\vspace{-0.3cm}
	\end{table*}
	
	\begin{table*}[t]
		\centering
		\caption{{Comparative summary of GDMs and LPTMs in our taxonomy.}}
		\begin{tblr}{
				colspec={X[0.1,l] X[0.5,l] X[0.45,l]},
				cells = {c},
				hlines, vlines,
				rowsep = {2pt},
				row{1} = {0ex},
			}
			\textbf{Dimension} & \textbf{GDMs} & \textbf{LPTMs} \\
			\textbf{Training Objective} & Denoising score matching & Next-token prediction \\
			\textbf{Generative Mechanism} & Iterative denoising via stochastic differential processes & Direct autoregressive decoding \\
			\textbf{Data Modality} & Continuous or discrete signals (solutions, images, trajectories, graphs, etc.) & Discrete tokens (language, code, multi-modal pairs) \\
			\textbf{Model Scale} & Medium or small-scale (mostly $<$10B parameters) & Large-scale ($>$10B parameters) \\
			\textbf{Typical Applications} & Solution synthesis, trajectory planning, channel-signal generation & Text-based reasoning, multi-modal understanding and decision-making \\
		\end{tblr}
		\label{tab_GDM_LPTM_comp}
	\end{table*}

    \begin{table*}[t]
	\centering
		\caption{{Comparison among GDM-based, LPTM-based and traditional (mathematical and conventional learning-based) network optimization.}}
		\begin{tblr}{
				colspec={X[0.2,l] X[0.15,l] X[0.15,l] X[0.4,l]},
				cells = {c},
				hlines, vlines,
				rowsep = {2pt},
				row{1} = {0ex},
			}
			\textbf{Dimension} & \textbf{GDMs} & \textbf{LPTMs} & \textbf{Traditional (mathematical and conventional learning-based)} \\
			\textbf{Building Cost} & Medium & High & Medium  \\
			\textbf{Inference Latency} & Medium & High & Low (low-dim) / High (high-dim)  \\
			\textbf{Interpretability} & Low & Medium & High  \\
			\textbf{Constraint Awareness} & Medium & Medium & High \\
			\textbf{Scalability} & Medium & High & Low  \\
			\textbf{Generalization} & Medium & High & Low  \\
		\end{tblr}
		\label{tab_GDM_LPTM_advan}
		\vspace{-0.38cm}
    \end{table*}
	
	\begin{table*}[t]
		\centering
		\caption{Comparison with relevant surveys or magazines in terms of coverage of scopes.}
		\begin{tblr}{
				colspec={X[0.18,c] X[0.32,c] X[0.25,c] X[0.2,c] X[0.2,c] X[0.21,c]},
				hlines, vlines,
			}
			\textbf{Reference} 
			& \textbf{Field} 
			& \textbf{Network Optimization Covered} 
			& \textbf{Consideration of GDM} 
			& \textbf{Consideration of LPTM} 
			& \textbf{Theoretical Bounds Analysis} \\
			\cite{survey2_pushing6G} & Telecom Base Model & \Circle  & \Circle  & \CIRCLE  & \Circle  \\
			\cite{survey3_shi6G} & Networking & \CIRCLE  & \Circle  & \Circle  & \Circle  \\
			\cite{survey5_iotj} & Telecom Base Model & \CIRCLE  & \CIRCLE  & \CIRCLE  & \Circle  \\
			\cite{survey6_LLMTelecom} & Telecom Base Model & \CIRCLE  & \Circle  & \CIRCLE  & \Circle  \\
			\cite{survey7_GenerativeOpimization} & AI & \Circle  & \CIRCLE  & \CIRCLE  & \Circle  \\
			\cite{survey12_FromLLM2AutoAIAgent} & AI & \CIRCLE  & \Circle  & \CIRCLE  & \Circle  \\
			\cite{survey13_KnowledgeDrivenDL} & Networking & \CIRCLE  & \Circle  & \CIRCLE  & \Circle  \\
			\cite{survey15_GAIforOptXG} & Networking & \CIRCLE  & \CIRCLE  & \CIRCLE  & \Circle  \\
			\cite{magazine3_LLMMultiAgent6G} & Agentic Network & \Circle  & \CIRCLE  & \CIRCLE  & \Circle  \\
			\cite{magazine4_LLMAgent6GPerception} & Agentic Network & \Circle  & \Circle  & \CIRCLE  & \Circle  \\
			\cite{survey10_InternetOfAgent} & Agentic Network & \CIRCLE  & \CIRCLE  & \CIRCLE  & \Circle  \\
			\cite{survey11_AGINative} & Telecom Base Model & \Circle   & \Circle  & \Circle  & \Circle  \\
			\cite{magazine1_AgentStandard} & Agentic Network & \CIRCLE  &\Circle   & \Circle  & \Circle  \\
			\cite{magazine2_MobileAIAgent6G} & Agentic Network & \CIRCLE  & \CIRCLE  & \CIRCLE  & \Circle  \\
			\cite{survey1_foundation} & Telecom Base Model & \Circle  & \Circle  & \CIRCLE  & \Circle  \\
			\cite{survey4_du} & Networking & \CIRCLE  & \CIRCLE  & \Circle  & \Circle  \\
			\cite{survey8_jiang} & Telecom Base Model & \CIRCLE  & \CIRCLE  & \CIRCLE  & \Circle \\
			\cite{survey9_wirelessLargeAI} & Telecom Base Model & \CIRCLE  & \CIRCLE  & \CIRCLE  & \Circle \\
			\cite{survey14_FromLAM2Agentic} & Agentic Network & \CIRCLE  & \CIRCLE  & \CIRCLE  & \Circle \\
			Ours & Networking & \CIRCLE  & \CIRCLE  & \CIRCLE  & \CIRCLE  \\
		\end{tblr}
		\label{tab_existing_surveys}
		\vspace{-0.38cm}
	\end{table*}
	
	\subsection{Motivation}
	Although many recent studies have achieved impressive results by applying GenAI to network optimization tasks, there is still a lack of comprehensive surveys that systematically cover this emerging direction, as well as a shortage of critical reflection on its underlying assumptions and limitations. 
	
    {Although many existing surveys and journal articles have discussed the intersection of GenAI and network optimization \cite{survey2_pushing6G,survey3_shi6G,survey5_iotj,survey6_LLMTelecom,survey7_GenerativeOpimization,survey12_FromLLM2AutoAIAgent,survey13_KnowledgeDrivenDL,survey15_GAIforOptXG,magazine3_LLMMultiAgent6G,magazine4_LLMAgent6GPerception}, this specific topic has not been their primary focus of research.} Some works place greater emphasis on designing system-level frameworks for applying GenAI \cite{survey11_AGINative,magazine1_AgentStandard,magazine2_MobileAIAgent6G}, rather than examining the construction and underlying mechanisms of GenAI itself. Others tend to treat GenAI (particularly LPTMs) as a general-purpose solution across a wide range of networking tasks \cite{survey10_InternetOfAgent,survey11_AGINative,magazine4_LLMAgent6GPerception}. While a number of works provide valuable high-level overviews that reflect on limitations and outline future directions \cite{survey1_foundation,survey4_du,survey8_jiang,survey9_wirelessLargeAI,survey14_FromLAM2Agentic}, they often fall short of offering in-depth analyses at the theoretical level. 
	
	Emerging work from both the networking community \cite{liang2025pad,WirelessMathBench2025XinLi} and the artificial intelligence (AI) community \cite{GSM-Symbolic,DRL2025ReasoningLLM,DoPhD2025LLM,ReasoningBeyondLimits2025Debbah} has begun to critically examine the fundamental capabilities and limitations of GenAI when applied to complex reasoning and optimization tasks. Notably, LPTMs often struggle with rule learning, symbolic reasoning, and mathematical accuracy \cite{GSM-Symbolic,DRL2025ReasoningLLM,DoPhD2025LLM,ReasoningBeyondLimits2025Debbah,WirelessMathBench2025XinLi}. Transformer-based models face well-documented theoretical bottlenecks in compositionality \cite{FaithAndFate2023,Representational2023TFM}, and GDMs tend to perform poorly in tasks that demand high precision and constraint satisfaction due to their probabilistic generative mechanisms \cite{onMemorizationOfDiff,OnGeneralization2023Diffusion}. These findings suggest that existing GenAI architectures are inherently limited in addressing many of the advanced demands of network optimization, such as strict adherence to formal rules, interpretability, and robust generalization under domain constraints. Specifically, the availability of background and precisely labeled data for network optimization is far more limited than the abundance of text, video, and image data in the broader AI domain, which may fundamentally constrain the capabilities of GenAI. Moreover, network optimization problems often involve complex domain knowledge across logic, physics, and mathematics related to communication systems, making it challenging for off-the-shelf GenAI models and neural architectures to be directly applicable. Network optimization also imposes stringent requirements on time and space efficiency, necessitating lightweight and distributed model designs that are compatible with communication network infrastructures.
	
	Therefore, a holistic overview of the trends, challenges, and limitations of GenAI for network optimization is urgently needed. In addition, we critically assess the current state of GenAI in network optimization and offer perspectives for future directions.
	
	\begin{figure*}[t]
		\centering
		\centerline{\includegraphics[width=6.0in]{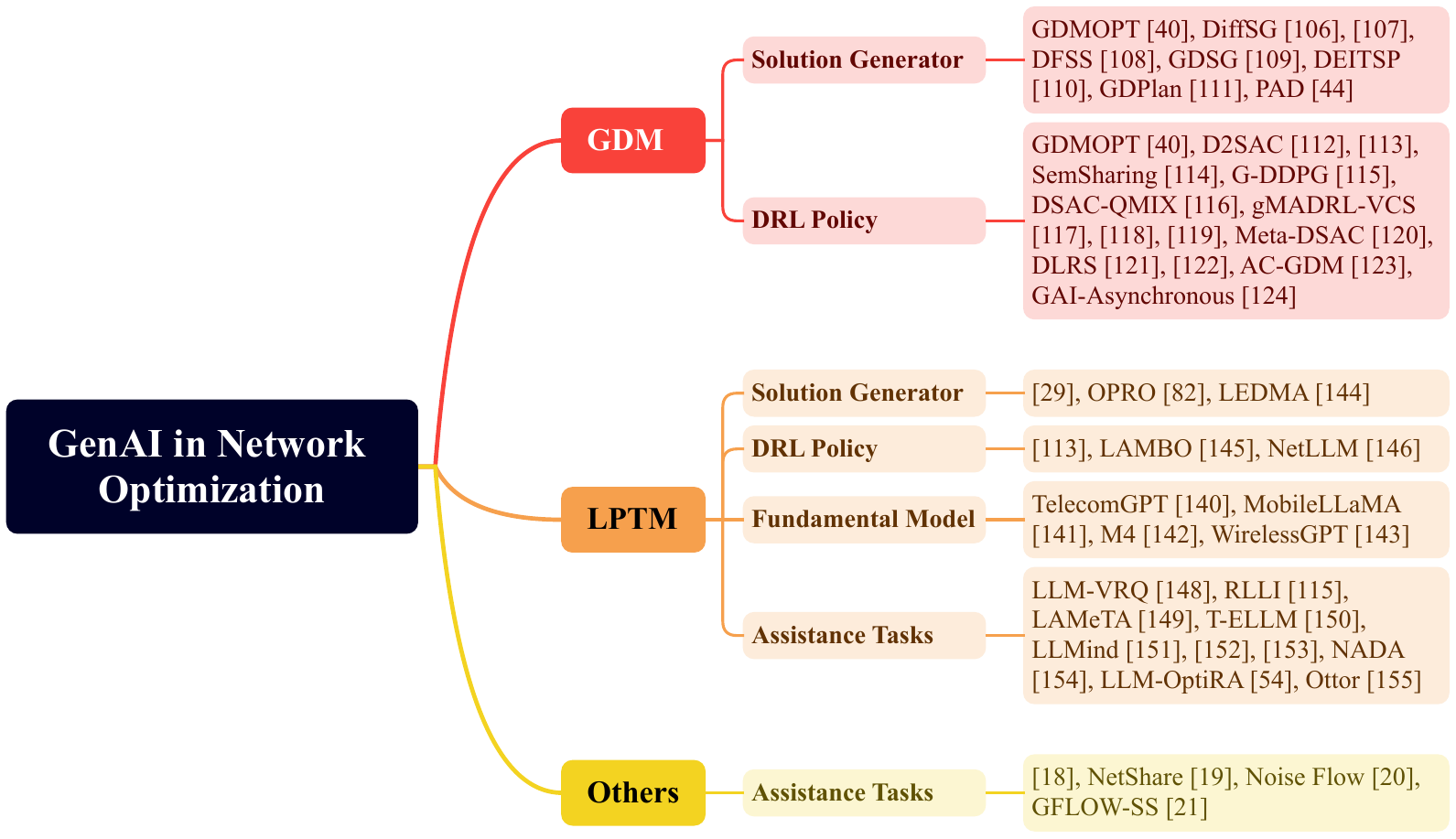}}
		\caption{GenAI in network optimization : taxonomy map.}
		\label{fig_taxonomy_mind_map}
        \vspace{-0.3cm}
	\end{figure*}
	
	\subsection{Contribution}
	In this paper, we focus on the role of GenAI in network optimization. We provide a brief classification of GenAI's work in network optimization in Fig. \ref{fig_taxonomy_mind_map}. Starting from a comprehensive survey of existing works on GenAI for network optimization, we provide a theoretical analysis of its strengths and limitations. As shown in Table \ref{tab_existing_surveys}, unlike existing relevant surveys and magazine articles, this paper offers a more in-depth and systematic contribution in terms of coverage and theoretical insight. In particular, we theoretically examine the gap between the practical development of GenAI optimizers and the desired characteristics of network optimization. The specific contributions of this paper are as follows.
	\begin{enumerate}
		\item We provide a comprehensive review of recent advancements in applying two major categories of GenAI, GDMs and LPTMs, to network optimization problems. Drawing on representative works from the AI community, we summarize key challenges and outline emerging trends in this research direction.
		\item Focusing on GDMs as a representative case,we analyze two primary paradigms {(One-shot optimization and MDP)} for their application to network optimization tasks and establish theoretical generalization boundaries for each. This analysis identifies direct factors that influence model performance, thereby offering practical guidance for future implementations. In parallel, we discuss both the strengths and limitations of GenAI in network optimization.
		\item Inspired by recent advances across multiple disciplines, we reflect critically and argue that both GDMs and LPTMs still face limitations in their application to network optimization, particularly in terms of output uncertainty and complex task reasoning. With the empirical studies summarized, we also discuss future directions, including the development of fundamental theories, rule-guided generation, and domain-specific benchmarks.
	\end{enumerate}
	
	\subsection{Paper Architecture}
\begin{figure*}[t]
		\centering
		\centerline{\includegraphics[width=5.9in]{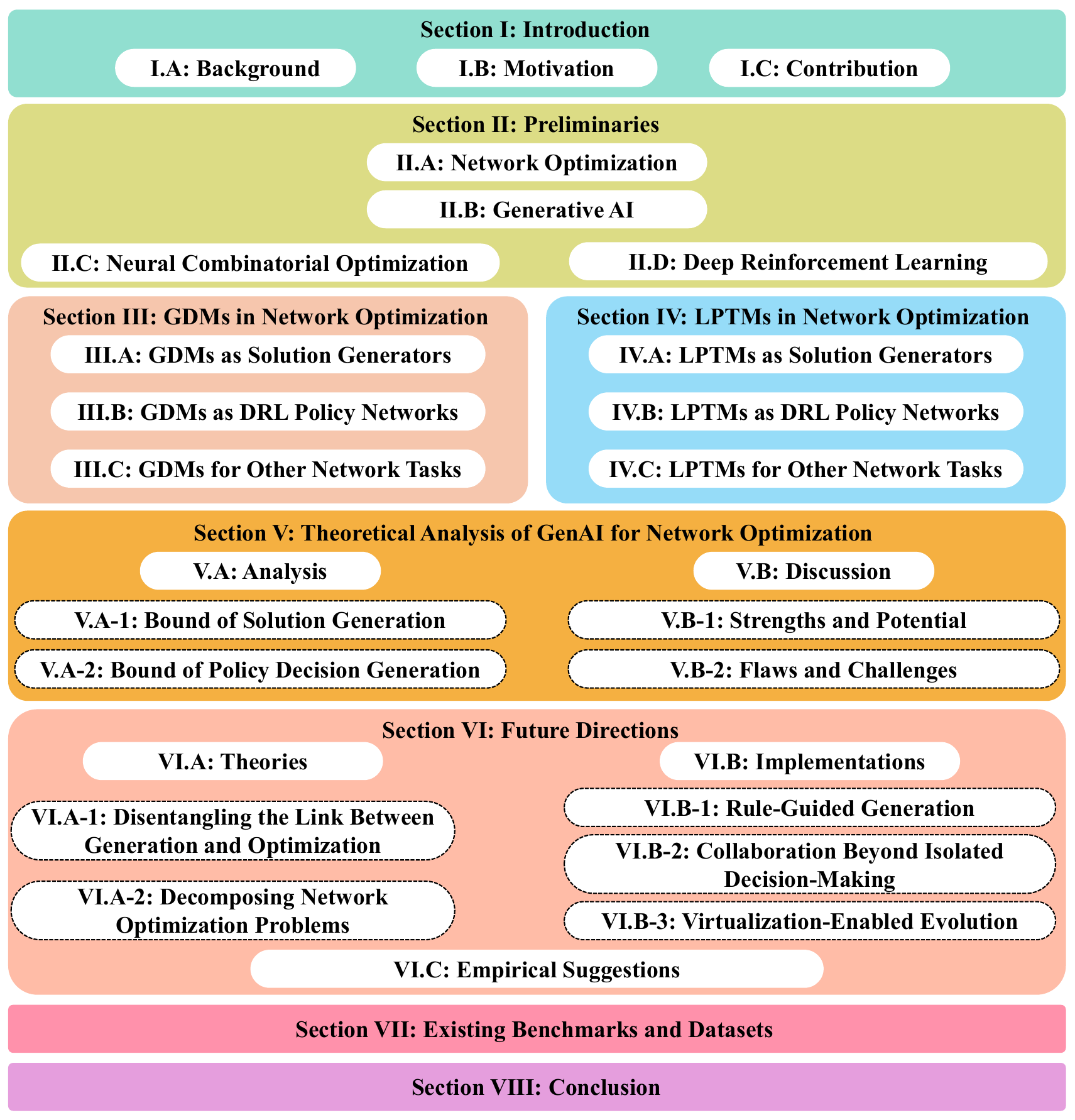}}
		\setlength{\abovecaptionskip}{-0.05cm}
		\caption{{Roadmap of this survey.}}
		\label{fig_section_framework}
		\vspace{-0.38cm}
	\end{figure*}

	The remainder of this paper is organized as illustrated in Fig. \ref{fig_section_framework}. In Section \ref{sec_preliminaries}, we provide essential background on network optimization problems, GenAI, and two major categories of GenAI technologies that are highly relevant to network optimization. Sections \ref{sec_GDM} and \ref{sec_LPTM} review and summarize representative works on applying GDMs and LPTMs, respectively, to network optimization tasks. Going further, Section \ref{sec_theory} presents theoretical boundaries for two application paradigms of GDMs in network optimization, along with an extended discussion on the architectural and theoretical limitations of GenAI. Sections \ref{sec_future} and \ref{sec_bench} outline future research directions and existing reference benchmarks, respectively. Finally, Section \ref{sec_conclusion} concludes this paper. The key abbreviations are summarized in Table \ref{tab_key_abbreviations}.
	
	\section{Preliminaries}\label{sec_preliminaries}
	
	\subsection{Network Optimization}
	Network optimization in communication systems is caused by the pursuit of optimal allocation and control of limited physical resources and optimal estimation of the environment \cite{mec2017mao}. It has high-dimensional, dynamic, and uncertain characteristics, posing challenges such as high computational complexity, strict real-time demands, difficulty in finding a global optimal solution, and limited generalization ability \cite{shi2023optimizationIn6G}. Common types include signal recovery, resource allocation, and semantic optimization problems. These problems often involve various mathematical properties, such as non-convex optimization {\cite{LLMOptiRA2025}}, mixed-integer programming \cite{yb-tmc}, bi-level optimization (BLO) \cite{BLO2023Introduction}, and stochastic optimization {\cite{StochasticOptimization}}.
	
	Based on the influence intensity of adjacent output solutions on the current environment parameters, network optimization problems can be divided into two categories according to the problem's mathematical modeling approach. First, one-shot optimization problems typically use supervised learning methods in deep learning or specified numerical algorithms. In this paradigm, the problem usually has a clear optimal solution, but consecutive decisions are weakly correlated. In contrast, MDP problems, which involve stochastic decision-making, are more suited to unsupervised deep reinforcement learning (DRL) methods. These methods enable neural networks to learn optimal policies through interaction with the environment, thereby maximizing long-term rewards. It is noteworthy that MDP problems do not admit a single deterministic optimal policy due to probabilistic state transitions. Instead, the objective is to find a policy that maximizes the expected return. Thus, MDP problems emphasize the strong correlation between consecutive decisions. 
	
\begin{figure}[t]
		\centering
		\centerline{\includegraphics[width=3.6in]{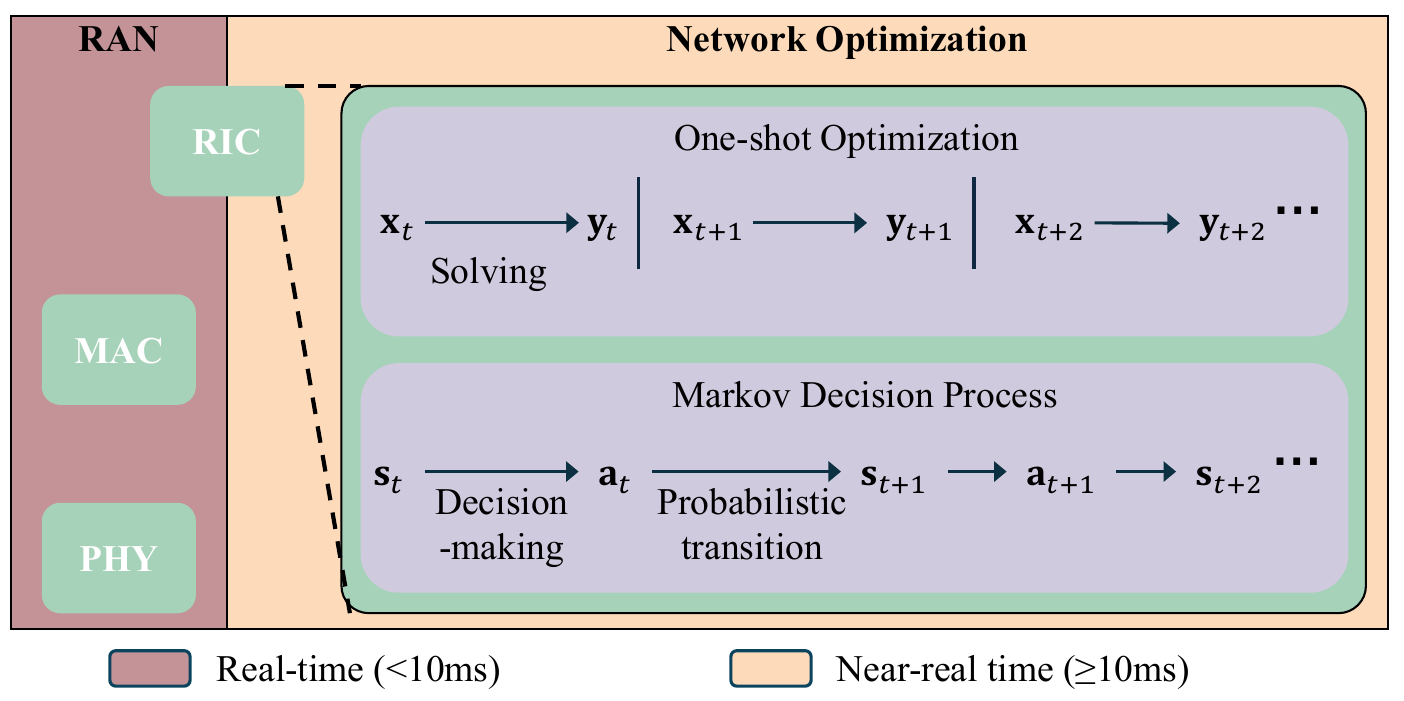}}
		\caption{{The two modeling paradigms for network optimization and the location in network systems.}}
		\label{fig_modeling_paradigms}
		\vspace{-0.38cm}
	\end{figure}

	As shown in Fig. \ref{fig_modeling_paradigms} and Table \ref{tab_SG_DRL_comparison}, both approaches must meet near-real-time inference demands but differ significantly in training and solution characteristics. One-shot optimization is more effective when a clear optimal solution exists, whereas DRL policies are better suited for sequential decision-making in dynamic environments.
	
	\begin{table*}[thb]
		\centering
		\caption{Comparison of the two main roles of GenAI in network optimization.}
		\begin{tblr}{
				colspec={X[0.2,l] X[0.52,c] X[0.3,c] X[0.42,c] X[0.43,c] X[0.35,c]},
				cells = {c},
				hlines, vlines,
				rowsep = {2pt},
				row{1} = {0ex},
			}
			\textbf{Role of GenAI} & \textbf{Problem Modeling Form} & \textbf{Typical Training Paradigm} & \textbf{Solution Characteristics} & \textbf{Correlation between two consecutive decisions} & {\textbf{Inference Frequency}} \\
			Solution generator & Optimization problem to minimize/maximize the objective function under given constraints & Supervised & Optimal solution exists & Weak & Near real-time \\
			DRL policy & Markov decision process to maximize the expected reward & Unsupervised & No known optimal solution, only expected high-quality actions exist & Strong & Near real-time \\
		\end{tblr}
		\label{tab_SG_DRL_comparison}
		\vspace{-0.38cm}
	\end{table*}

	\subsection{Generative AI}
	GenAI aims to learn the probabilistic distribution of data, i.e., to model the joint probability {\(P(\textbf{x},\textbf{y})\)} to be able to synthesize new samples in the original distribution. In contrast, discriminative AI only focuses on conditional probability {\(P(\textbf{y}|\textbf{x})\)}, which is used to complete discriminative tasks such as classification and regression. Generative models are originally designed for multimodal generation tasks, such as images, text, and audio. And generative models can approximate conditional probability distributions when guided by conditions or prompts.
	
	\subsubsection{GDMs} GDMs have emerged as one of the most prominent generative frameworks in recent years. Initially designed for image synthesis, GDMs have since been extended to a wide array of domains, including the generation of solutions for complex optimization problems. The widely acknowledged starting point of this research line is the DDPM \cite{DDPM}, which progressively adds Gaussian noise to data samples and trains a U-Net model \cite{Unet2015} to reverse this process and reconstruct clean outputs. Building on this foundation, Classifier-Guidance \cite{ClassifierGuidance} and classifier-free guidance (CFG) \cite{ClassifierFree} introduced two of the most influential mechanisms for conditional generation. A substantial body of work has since been dedicated to improving the efficiency of GDMs, including denoising diffusion implicit model (DDIM) \cite{DDIM2022song} and DPM-Solver \cite{DPM_solver}, along with subsequent developments such as \cite{GaussianSolvers2023NIPS,ConsistencyModel2023,ParallelGDM2023NIPS}. Breakthroughs in high-resolution image synthesis were achieved with Stable Diffusion \cite{StableLatentDiffusion}, which introduced the latent diffusion paradigm. Meanwhile, ControlNet \cite{ControlNet} significantly enhanced the controllability and quality of conditional image generation. Beyond continuous data domains, GDMs have also been extended to support discrete data generation \cite{DiscreteDiffusion}. Additionally, the backbone neural architectures have evolved to include advanced alternatives such as graph transformer (GAT) \cite{GraphTransformer2019} and diffusion transformer (DiT) \cite{DiT_Xie}, reflecting growing architectural diversity and task specialization.
	
	\subsubsection{LPTMs} LLMs are the most important part of LPTMs, represented by GPT \cite{GPT3}, LLaMA \cite{LLaMA}, Gemini \cite{Gemini}, DeepSeek \cite{DeepSeekV3}, etc. They typically adopt an autoregressive approach to model the conditional probability {$P(\mathbf{x_t}|\mathbf{x_{<t}})$} at each token position in a sequence, where {$\mathbf{x}_t$ }represents the $t$th token in the output sequence. This allows them to generate a wide range of outputs, including natural language text, programming code, structured instructions, and more. 
	
	Some of the latest foundation models have evolved from traditional LLMs into what are now called large reasoning models, such as DeepSeek-R1 \cite{DeepSeek-R1}. These models are typically fine-tuned on top of LLMs using reinforcement learning to promote step-by-step reasoning or tree-search-like problem-solving behaviors, often with a focus on enabling reflection and self-correction. DeepSeek-R1, for example, is built using group relative policy optimization (GRPO) training guided by feedback from DeepSeek-V3 \cite{DeepSeekV3}’s judgment feedback. Closely related are mathematical LLMs, such as rStar-Math \cite{rStarMath2025}, DeepSeek-Prover-V2 \cite{DeepSeekProverV2}, MINIMO \cite{FormalMath2024Intrinsic}, and LLMOPT \cite{LLMOPT2025}, which specialize in mathematical reasoning tasks. These models often adapt general reasoning methods to math problems, commonly incorporating inference-time scaling (akin to extending thinking time in humans) and leveraging tree search to increase logical depth.
	
	In parallel, some models aim to fundamentally rethink the architecture and training strategies of LLMs, particularly through diffusion-based language models and world models. LLaDA \cite{LLaDA2025} and MMaDA \cite{MMaDA} use diffusion processes instead of autoregressive generation, enabling controllable output lengths, denoising-style masked generation, and multi-modal input-output support. World models, on the other hand, aim to learn latent representations of physical environments and predict object dynamics within them. Notably, the latest V-JEPA 2 \cite{V-JEPA-2} is pre-trained on large-scale unlabeled video data and then fine-tuned on a small set of robotic control videos, enabling V-JEPA 2 AC to support zero-shot robotic control.
	
	These recent advancements in reasoning models, diffusion language models, and world models hold promising implications for network optimization tasks. However, how to effectively adapt and tailor them to networking scenarios remains an open challenge. In this stage, LPTMs have been increasingly applied across diverse scenarios, including mathematical modeling \cite{ModelingAgent2025ModelingBench}, analysis of optimization problems \cite{LLMOptiRA2025}, and even approximate problem-solving \cite{LLMasOptimizer2024}. 
	
	\subsubsection{Others}
	{GenAI is not only applied to the above two types of generative models, but also has applications in other fields.} Beyond images and text, GenAI has continued to advance across various multi-modal domains, including image-text contrastive learning (CLIP \cite{CLIP}), visual segmentation (SAM \cite{SegmentAnything}), tabular data (CARTE \cite{CARTE}), graph-structured data (GraphEdit \cite{GraphEdit}), and multi-modal composition (CoDi \cite{any2any}). {These methods have also provided many new ideas for network optimization, especially the issue of multi-modal composition in the network, which has received much attention.}
	
	\subsubsection{Summary} 
\begin{figure}[t]
		\centering
		\includegraphics[width=1.015\linewidth]{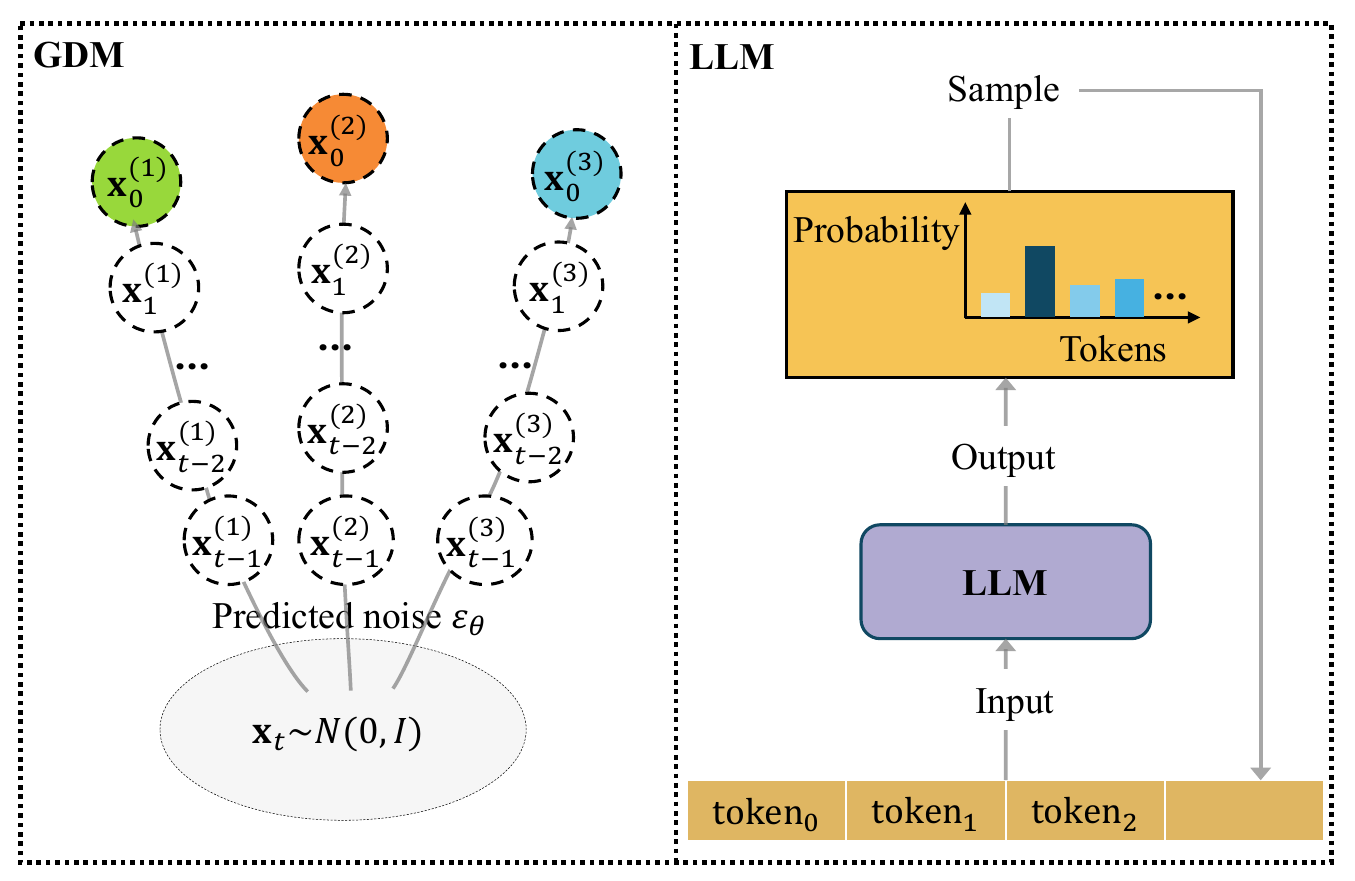}
		\caption{{Comparison between GDMs and LLMs in terms of generation mechanism. The sources of randomness in GDMs and LLMs differ: in GDMs, randomness arises from the initial noise sample and the stochastic nature of the denoising trajectory, whereas in LLMs, it stems from sampling predicted tokens. Moreover, LLMs are autoregressive, as each predicted token is fed back into the model for subsequent predictions, while most GDMs are non-autoregressive.}}
		\label{fig_GDM_vs_LLM}
        \vspace{-0.38cm}
	\end{figure}

Overall, GDMs and LPTMs constitute the two dominant paradigms in the current landscape of Generative AI. For network optimization tasks, their key distinction lies in their generative mechanisms: LLMs are autoregressive, while GDMs are not. Specifically, the stochasticity in LLM outputs arises from sampling over the predicted probability distribution of the next token. In contrast, GDMs introduce diversity through the initial noise in the solution and the stochastic nature of the denoising trajectory, as illustrated in Fig. \ref{fig_GDM_vs_LLM}.
	
	\subsection{Neural Combinatorial Optimization}
	Combinatorial optimization refers to the class of problems in which the goal is to find the optimal configuration or sequence from a discrete, often combinatorially large, solution space, under given constraints. Classical examples include the traveling salesman problem (TSP), maximum independent set (MIS), and vehicle routing, all of which are NP-hard and difficult to scale with traditional solvers. Combinatorial optimization can be viewed as a class of problems encompassed by network optimization.
	
	Neural combinatorial optimization aims to tackle these problems using neural networks to learn heuristics or direct solution mappings. GDMs have recently emerged as a powerful and popular approach for solving challenging combinatorial optimization problems \cite{sun2023difusco,accelerating2023afterDIFUSCO,li2023t2t,unsupervised2024CO,boosting2025energyGuided}. These studies primarily focus on large-scale instances of classical combinatorial optimization problems, often involving hundreds of thousands of nodes, and aim to develop pre-trained models on canonical problem types that can generalize effectively to problem variants. Unlike reinforcement learning-based solvers \cite{DIMES2022}, GDMs can directly model the solution distribution and generate near-optimal candidates in a {single inference}, making them well-suited for one-shot optimization scenarios common in network settings. This growing body of work has offered valuable inspiration for how GenAI, especially GDMs, can be applied to generate high-quality solutions in network optimization contexts.
	
	\subsection{Deep Reinforcement Learning}
	DRL combines deep learning's perception capabilities with reinforcement learning's decision-making capabilities. It involves learning optimal policies through interactions between an agent and its environment to maximize cumulative rewards. By utilizing deep neural networks as function approximators, DRL is capable of addressing complex decision-making problems \cite{Diffusion_robot_Survey,GDM_DRL_survey}.
	
	GDMs are one of the earliest GenAI techniques applied in DRL, showcasing significant potential in robot control applications \cite{isConditional2023,IDQL2023,DiffusionQL2023,he2023MTDIFF,AdaptDiffuser2023,EDP2023NIPS,DiffusionPolicy2024OOD,safeOfflineRL2024,DiffuserLite2024NIPS,taskAgnosticDiffusionPlanner2025}. Its unique capabilities in modeling complex data distributions have since enabled gradual expansion to broader DRL tasks. Recent advancements highlight GDM's effectiveness in trajectory generation, providing smooth, diverse motion paths for agents \cite{DiffusionQL2023}. In multi-task learning scenarios, GDMs facilitate knowledge transfer across different tasks, improving overall learning efficiency \cite{taskAgnosticDiffusionPlanner2025}. Moreover, through data augmentation, GDMs enhance the quality and diversity of training datasets, which is crucial for improving the generalization ability of DRL agents \cite{he2023MTDIFF}. The application of GDMs in DRL is being widely migrated to the MDP framework of network optimization modeling \cite{survey4_du}. Ongoing innovations in the AI community are driving performance improvements in MDP-based network optimization.
	
	\section{GDMs in Network Optimization}\label{sec_GDM}
	Most applications of GDMs in network optimization are inspired by advances in the AI community. These applications primarily fall into two categories: using GDMs as solution generators for combinatorial optimization problems \cite{sun2023difusco,accelerating2023afterDIFUSCO,li2023t2t,unsupervised2024CO,boosting2025energyGuided}, and using GDMs as policy networks in DRL for robotic control \cite{isConditional2023,IDQL2023,DiffusionQL2023,he2023MTDIFF,AdaptDiffuser2023,EDP2023NIPS,DiffusionPolicy2024OOD,safeOfflineRL2024,DiffuserLite2024NIPS,taskAgnosticDiffusionPlanner2025}, as shown in Table \ref{tab_GDM_SG_DRL_AI}.
	
	When used as solution generators for combinatorial optimization, GDMs are typically discrete diffusion models targeting large-scale problems (often involving over 100,000 nodes). In the AI domain, GDMs have achieved near-optimal performance on classical combinatorial optimization tasks such as TSP and MIS. Recent efforts have focused on accelerating solution generation \cite{accelerating2023afterDIFUSCO} and enabling cross-problem generalization by leveraging shared knowledge across problem instances \cite{boosting2025energyGuided}. In the DRL setting, GDMs are commonly applied to generate action-state trajectories for robotic control. In robotic control, GDMs are mostly used in offline reinforcement learning, where they learn to predict the target trajectory from the history of trajectories observed during interaction with the environment. Most existing approaches execute only the first action of the generated trajectory, leading to inefficient exploration and significant time overhead. Current research is therefore exploring ways to refine generated trajectories through planning to improve action-generation efficiency \cite{EDP2023NIPS,DiffuserLite2024NIPS}. General-purpose DRL using GDMs has also been explored as a promising research direction \cite{he2023MTDIFF,AdaptDiffuser2023,taskAgnosticDiffusionPlanner2025}.
	
	\subsection{GDMs as Solution Generators}
	Network optimization problems are inherently more diverse and complex than classical combinatorial optimization. They encompass not only discrete-variable problems but also non-convex optimization, mixed-integer programming, and related areas. Compared to traditional combinatorial tasks, network optimization problems often involve more intricate objective functions and constraints, as they must account for various factors such as connectivity, latency, and total available resources.
	
	GDMOPT \cite{survey4_du} is one of the earliest works to apply GDMs as solution generators in the network domain, conducting a case study on a classical convex optimization problem. DiffSG \cite{liang2024diffsg} demonstrates the effectiveness and cross-domain generalization of GDMs on a variety of more complex network optimization tasks. \cite{liang2024iotj} further provides a theoretical explanation of why GenAI can outperform discriminative models in solution generation, offering insights into GDMs' underlying mechanisms and analyzing the impact of key hyperparameters in network optimization settings, with DiffSG as a baseline. DFSS \cite{SG_2024DFSS} aims to address the security of channel state information (CSI)-based sensing in integrated sensing and communication (ISAC) scenarios. It tackles unauthorized sensing or spoofing attacks by adversaries trying to exploit CSI signals. DFSS first uses a discrete graph diffusion model to generate connection topologies for ISAC device deployment, minimizing operational costs while preserving sensing performance. Then, it applies a continuous diffusion model to generate safeguarding CSI values, embedding them into pilot signals to significantly reduce the sensing accuracy of unauthorized devices. 
	
	GDSG \cite{liang2024gdsg} considers the common challenge in network optimization where optimal solutions are difficult to obtain. It extends GDM training to support robust learning from sub-optimal solution samples only. Trained solely on efficiently constructed sub-optimal datasets, GDSG successfully optimizes multi-server multi-user computation offloading problems in multi-access edge computing (MEC) scenarios, involving discrete diffusion for offloading decisions and continuous diffusion for resource allocation. DEITSP \cite{DEITSP2025KDD} focuses on reducing the computational complexity of autoregressive GDM solvers. DEITSP employs a one-step diffusion mechanism that enables inference from any timestep toward the target solution. It also introduces a novel modification to the DDIM \cite{DDIM2022song} process: each solution candidate undergoes a one-step prediction, followed by renoising and reintegration into the denoising process, to enhance exploration. Additionally, it streamlines the GAT backbone to reduce inference time while maintaining solution quality, achieving faster results compared to DIFUSCO \cite{sun2023difusco} and T2T \cite{li2023t2t}. GDPlan \cite{GDPlan2025TON} addresses network planning by incorporating an external energy score-based model to guide graph diffusion for Internet Protocol (IP) link capacity configuration. High-quality solutions are further refined through external solvers in GDPlan. 
	
	The most recent work, PAD \cite{liang2025pad}, leverages large pre-trained mathematical LLMs to encode problem-specific structures in network optimization. By enabling GDMs to recognize and understand mathematical features of different problems, PAD enhances cross-problem generalization performance, demonstrating the potential of combining mathematical information with solution generation.
	
	Overall, the use of GDMs as solution generators in network optimization benefits from their generative nature and strong generalization capabilities. Key advantages include superior solution quality under comparable conditions compared to discriminative models, robustness to imperfect training data (including purely sub-optimal solutions), and promising potential for cross-problem generalization. So far, GDMs have demonstrated effectiveness across many common network optimization problems, and foundational theoretical work has begun to emerge. Future directions will likely focus on developing more comprehensive theoretical frameworks and more efficient engineering implementations.
	
	\begin{table*}[t]
		\centering
		\caption{Representative works on GDMs as solution generators and DRL policies for the AI community.}
		\begin{tblr}{
				colspec={X[0.14,l] X[0.22,c] X[0.15,c] X[0.63,c]},
				cells = {c},
				hlines, vlines,
				rowsep = {2pt},
				row{1} = {0ex},
				cell{2}{1} = {r=5}{}, 
				cell{7}{1} = {r=10}{}, 
			}
			\textbf{Type} & \textbf{Algorithm} & \textbf{Neural Backbone} & \textbf{Contributions} \\
			Solution generator & DIFUSCO \cite{sun2023difusco} & GCN & Early implementation of graph diffusion models to solve large-scale TSP and MIS problems. \\
			& Distilled DIFUSCO \cite{accelerating2023afterDIFUSCO} & GCN & Accelerated the denoising process of DIFUSCO using progressive distillation. \\
			& T2T \cite{li2023t2t} & GCN & Incorporated the gradient direction of automatically differentiated objective functions for TSP and MIS into each denoising step, leading to consistently better solution quality than DIFUSCO. \\
			& DiffUCO \cite{unsupervised2024CO} & GCN & Employed relaxed objective functions and randomly sampled input-cost pairs to enable unsupervised training. \\
			& Energy-Guided Sampling \cite{boosting2025energyGuided} & GCN & Proposed a general energy-guided sampling framework that enables a pre-trained GDM for the TSP to generate solutions across TSP variants. \\
			DRL policy & Decision Diffuser\cite{isConditional2023} & U-Net & Pioneered the use of conditional generative models to directly solve sequential decision-making tasks by modeling the policy as a return-conditioned diffusion model. \\
			& IDQL \cite{IDQL2023} & MLP & Demonstrated that the commonly used simple Gaussian policy struggles to perform well in learning the Q-function solely from offline datasets under implicit Q-learning (IQL) settings. At the same time, GDMs can effectively take on this role with their powerful multimodal distribution-learning capability. \\
			& Diffusion-QL \cite{DiffusionQL2023} & MLP & Implemented a conditional diffusion model as a policy network to directly maximize the estimated action-value function. \\
			& MTDIFF \cite{he2023MTDIFF} & Transformer & Leveraged GDMs for both trajectory planning and data synthesis in multi-task DRL. \\
			& AdaptDiffuser \cite{AdaptDiffuser2023} & MLP & Adopted a pre-trained GDM policy to generate trajectories for fine-tuning, forming an iterative loop of generation and fine-tuning to enhance cross-task generalization. \\
			& EDP \cite{EDP2023NIPS} & MLP & Designed an action approximation technique that enables direct reconstruction of actions during training without executing the full sampling chain, combined with the DPM-Solver, significantly improving training efficiency. \\
			& SRDP \cite{DiffusionPolicy2024OOD} & MLP & Considered out-of-distribution (OOD) generalization issues in offline reinforcement learning caused by a mismatch between real-world states and the training distribution by using GDMs to learn generalizable state reconstruction features. \\
			& FISOR \cite{safeOfflineRL2024} & MLP & Proposed feasibility-guided safe offline reinforcement learning to balance high reward and strict safety requirements in safety-critical DRL tasks. \\
			& DiffuserLite\cite{DiffuserLite2024NIPS} & DiT & Designed a planning refinement process that significantly increased decision frequency and reduced model size. \\
			& SODP \cite{taskAgnosticDiffusionPlanner2025} & CNN, U-Net1D, MLP & Pre-trained a task-agnostic GDM policy using sub-optimal data, then performed few-shot fine-tuning on downstream tasks, enabling a versatile and reusable diffusion-based policy framework. \\
		\end{tblr}
        \vspace{-0.25cm}
		\label{tab_GDM_SG_DRL_AI}
	\end{table*}
	
	\subsection{GDMs as DRL Policy Networks}
	Many network optimization problems can be modeled as MDP problems, but they differ from traditional robotic control tasks in two significant ways. First, the reward functions (or objective functions) and constraints in network optimization are typically far more complex than those in robotic control. While robot control tasks often have clear terminal states, making it easier to learn critic modules, network optimization problems often involve non-differentiable objectives and constraints, which cannot be directly optimized and must be approximated via external relaxation. These surrogate modules are sensitive to the objective formulation and hyperparameter settings. Second, in terms of real-time performance, dynamic network optimization typically requires decision-making within milliseconds to avoid system failures. In contrast, the actions generated by GDMs in robotic control are typically handled at higher-level planning layers, with real-time control handled at lower levels, thereby relaxing real-time constraints on GDMs.
	
	GDMOPT \cite{survey4_du} is one of the earliest works to introduce GDMs as policy networks for DRL from the AI domain to network optimization. In their framework, GDMs function as actors, either using a learnable reward function derived from the original network optimization objective or leveraging an additional neural module to approximate the value function, enabling loss computation for maximizing expected rewards. D2SAC \cite{D2SAC2024} is a practical realization of this framework in the context of AI-generated content (AIGC) service provider selection. Rather than generating direct actions, the GDM in D2SAC outputs a target action distribution from which executable actions are sampled, resembling conventional DRL but differing from the typical usage of GDMs in robot control. In \cite{GDMPolicy2025WirelessComm}, various GenAI models are compared as DRL policy networks, and GDMs are shown to outperform others. SemSharing \cite{duD2SAC_ContractDesign2023jsac} applies GDMs in the metaverse-virtual reality (VR) context to generate incentive-sharing contracts for users, aiming to reduce redundant scene rendering and thereby save energy and computation. G-DDPG \cite{D2SAC2025LLMasUserFeature} targets distributed diffusion-based image generation services and uses a GDM-based deep deterministic policy gradient (DDPG) to adjust the cloud–edge denoising process based on intermediate feedback. DSAC-QMIX \cite{D2SAC2025TWC} addresses wireless resource competition in tunnel construction by modeling it as a stochastic optimization problem, using GDMs to allocate resources based on construction consistency and age of information (AoI).
	
	gMADRL-VCS \cite{GroundAirSpace2024MultiAgentGDM} focuses on post-disaster information collection in integrated ground-air-space networks, using GDMs as policy networks in an Actor-Critic framework to maximize spatial information coverage, balance fairness, and minimize energy consumption. In \cite{GDMPolicy2025VehicleEmbodied}, a contract-theoretic model is proposed for roadside unit (RSU) selection in autonomous driving, where GDMs serve as DRL actors to design contracts that incentivize RSUs to support mobile vehicle intelligence. In \cite{GDMPolicy2025MetaverseIoT}, the combination of AIGC metaverse services and internet of things (IoT) is explored in the context of parallel offloading, using GDMs as DRL policy networks to minimize rendering and fusion latency for multi-modal sensing data. Meta-DSAC \cite{TVT2025MetaverseGDMPolicy} focuses on resource management for multiple unmanned aerial vehicle (UAV)-based AIGC service providers in the metaverse.
	
	DLRS \cite{GDMPolicy2025LLMQoS} addresses the use of vector databases for request caching in edge-deployed LLMs. Observing that existing methods overlook previous search results, \cite{GDMPolicy2025LLMQoS} proposes a new MDP-based quality of service (QoS) optimization framework called DLRS, improving both similar and novel request response frequencies in a cloud-edge cooperative setting. In \cite{GDMPolicy2025DownlinkSemantic}, a GDM serves as a contract generator in a downlink semantic communication system, jointly optimizing rate-splitting multiple access (RSMA)-based resource allocation and contract design, achieving higher rewards than traditional DRL and supporting downstream semantic segmentation tasks. AC-GDM \cite{SecureBeamIRS2025GDMPolicy} targets secure beamforming in a multi-user IoT system with imperfect channels, using GDM-based DRL to jointly optimize precoding at the base station and phase shifts at an intelligent reflecting surface (IRS), outperforming traditional methods in simulations. GAI-Asynchronous \cite{GDMPolicy2025LowAltitudeEdgeInference} considers a low-altitude multi-task over-the-air edge inference system, modeling the joint optimization of task batching and beamforming, and uses GDMs as DRL policies to perform batch-wise beamforming decisions.
	
	In summary, GDMs demonstrate superior performance as DRL policy networks in MDP-formulated network-optimization problems, thanks to their powerful ability to model complex, multimodal action distributions. However, as in the broader AI community, the high latency of GDMs due to their iterative denoising process remains a limitation. Furthermore, the complexity of network optimization objectives and constraints presents substantial challenges for developing general-purpose GDM-based DRL approaches and achieving cross-problem generalization in the network domain.
	
	\subsection{GDMs for Other Network Tasks}
\begin{figure}[h]
		\centering
		\includegraphics[width=0.98\linewidth]{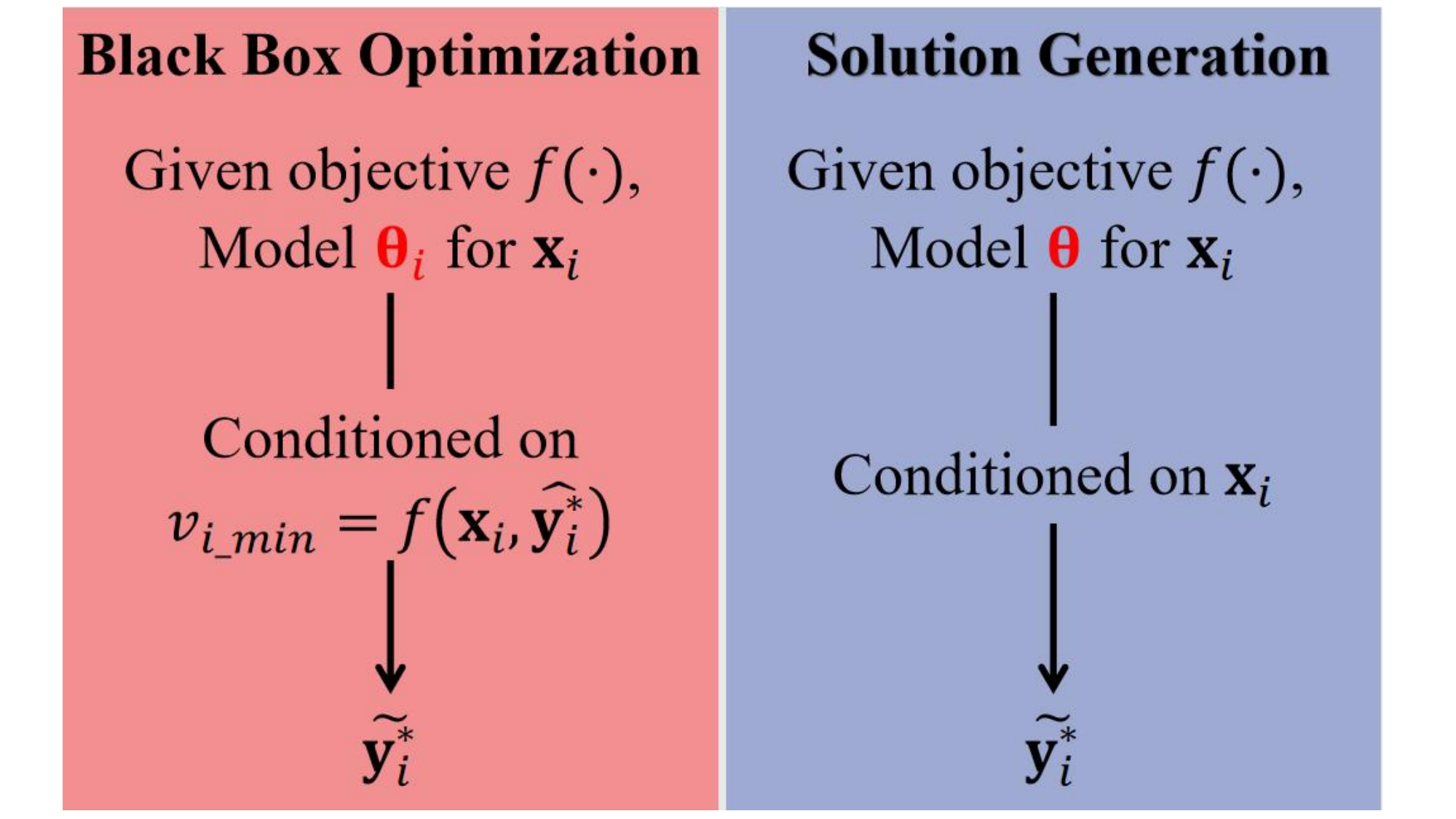}
		\caption{{Comparison between BBO and solution generation in the networking domain. Here, {$\mathbf{\hat{y_i^*}}$} denotes the hypothetical optimal solution for input {$\mathbf{x}_i$}, while {$\mathbf{\tilde{y_i^*}}$} represents the corresponding generated solution.}}
		\label{fig_BBO_SG_comp}
        \vspace{-0.3cm}
	\end{figure}

	While GDMs have shown strong capabilities as a solution generator in combinatorial optimization and network optimization, its applications are not limited to these areas. GDMs have also been extended to tackle black-box optimization (BBO) problems commonly found in AI4S scenarios \cite{GDMSurvey2024CoveredBBO}. Many BBO problems are leveraging GDMs as a solution generator in materials science, molecular design \cite{BBO2023NIPS,BBO2023PMLR,BBO2024ConstrainedSampler}, and topology optimization \cite{AAAIBeatGAN2023TopOptimization,GDM2023TopOptimization,BBO2023NIPS}. As shown in Fig. \ref{fig_BBO_SG_comp}, BBO differs significantly from network optimization in both objectives and efficiency requirements. Specifically, the objective functions in BBO are unknown and cannot be expressed through explicit mathematical modeling, thus the only available feedback is the function value associated with a candidate solution, obtained through system evaluation. Moreover, BBO problems often have extremely large and complex solution spaces, which typically rely on large-scale random sampling of solution–value pairs to train models and use empirically low values to guide the generation process. For each distinct parameter configuration, a separate GDM-based generator must be constructed, which makes both training and inference inefficient and falls short of meeting the stringent efficiency demands of network optimization.
	
	In addition, GDMs continue to demonstrate innovation in conventional data synthesis tasks, particularly in areas such as channel estimation \cite{CDDM2023globecom}, channel denoising \cite{GDMDenoiseSensing2024JSAC}, and time-frequency sequence generation \cite{RF_Diffusion,Privacy2025infocomGDM}. 
	
	\subsection{Summary}
	
	As shown in Table \ref{tab_GDM_SG_DRL_NO}, both the solution generator and the DRL policy network of GDMs in network optimization continue to evolve. However, the majority still focus on using GDMs as DRL policies. This trend primarily stems from the fact that many near-real-time resource allocation tasks in network optimization can be naturally formulated as MDP problems, making them well-suited for DRL-based approaches. Nevertheless, {one-shot} optimal optimization problems remain prevalent in the field, where using  GDMs as solution generators shows particular promise in enabling large-scale optimization and cross-problem generalization.
	
	By comparing Table \ref{tab_GDM_SG_DRL_AI} and Table \ref{tab_GDM_SG_DRL_NO}, an interesting contrast emerges regarding the diversity of neural backbones used in GDM-based solution generators and policies across domains. In the AI domain, GDMs, as solution generators, are mainly applied to classical combinatorial optimization problems, where graph convolutional networks (GCNs) dominate as the backbone architecture. In contrast, GDMs, as policy networks in this domain, must address challenges such as multi-modal inputs and cross-task generalization, prompting exploration of a wider variety of neural architectures. In the networking domain, however, the situation is reversed. GDM-based solution generators are increasingly being designed for cross-problem generalization, leading to the adoption of new neural backbones. On the other hand, works that use GDMs as DRL policies typically focus more on novel scenario design and thus tend to adopt more established and stable network architectures for policy implementation.
	
	Overall, GDMs have contributed to a range of advances in network optimization, demonstrating effectiveness both as a solution generator for one-shot problems and as a policy network for MDP-based tasks. However, realizing its potential often requires meticulous problem-specific design. Despite the growing body of work, most existing research disregards the probabilistic nature of GDMs, overlooking its inherent limitations when applied to tasks that demand precise mathematical reasoning and deterministic accuracy. Moreover, innovation in neural architectures and learning paradigms tailored for network optimization remains limited. To truly unlock GDM's capabilities in this domain, future research should go beyond direct adaptation and explore architectures and training strategies that better align with the structural and precision demands of network optimization tasks.
	
	\section{LPTMs in Network Optimization}\label{sec_LPTM}
	
	\begin{table*}[t]
		\centering
		\caption{Relevant works of GDMs as solution generators and DRL policies in network optimization.}
		\begin{tblr}{
				colspec={X[0.14,l] X[0.22,c] X[0.15,c] X[0.63,c]},
				cells = {c},
				hlines, vlines,
				rowsep = {2pt},
				row{1} = {0ex},
				cell{2}{1} = {r=8}{}, 
				cell{10}{1} = {r=14}{}, 
			}
			\textbf{Type} & \textbf{Reference} & \textbf{Neural Backbone} & \textbf{Contribution} \\
			Solution generator & GDMOPT \cite{survey4_du} & MLP & Confirmed the feasibility of the GDM solution generator. \\
			& DiffSG \cite{liang2024diffsg} & U-Net1D & Demonstrated the broad applicability and cross-domain generalization potential of GDMs. \\
			& Analysis of DiffSG \cite{liang2024iotj} & U-Net1D & Provided mechanistic insights into how GenAI models work in optimization problems, along with analysis on hyperparameter tuning in network optimization context. \\
			& DFSS \cite{SG_2024DFSS} & GAT and MLP & Proposed GDMs for generating system topologies and safeguarded CSI. \\
			& GDSG \cite{liang2024gdsg} & GCN & Demonstrated the robustness of GenAI-based solution generator against training data quality. \\
			& DEITSP \cite{DEITSP2025KDD} & GAT & Designed a graph diffusion model that achieves more efficient denoising, broader exploration, and smaller model size. \\
			& GDPlan \cite{GDPlan2025TON} & GAT & Proposed a topology generation model for network planning with generated solutions refinement. \\
			& PAD \cite{liang2025pad} & Transformer & Validated the effectiveness of mathematical information guided cross-problem generalization. \\
			DRL policy & GDMOPT \cite{survey4_du} & MLP & Proposed a generalized implementation framework for GDMs with DRL. \\
			& D2SAC \cite{D2SAC2024} & MLP & Applied GDMs with DRL to utility optimization in AIGC service systems. \\
			& Comparison of different GenAIs for DRL \cite{GDMPolicy2025WirelessComm} & MLP & Demonstrated the superiority of GDMs over other GenAI approaches as a DRL policy network in network optimization tasks. \\
			& SemSharing \cite{duD2SAC_ContractDesign2023jsac} & MLP & Applied GDMs with DRL to incentive contract generation for information sharing in metaverse-VR environments. \\
			& G-DDPG \cite{D2SAC2025LLMasUserFeature} & MLP & Combined GDMs with DDPG to optimize the denoising process in distributed diffusion model-based image generation. \\
			& DSAC-QMIX \cite{D2SAC2025TWC} & MLP & Applied GDMs with DRL to optimize resource allocation and information freshness in tunnel construction wireless networks. \\
			& gMADRL-VCS \cite{GroundAirSpace2024MultiAgentGDM} & MLP & Integrated GDMs with multi-agent DRL to optimize post-disaster information collection in integrated space-air-ground networks. \\
			& GDM-based Multi-Dimensional Contract Design \cite{GDMPolicy2025VehicleEmbodied} & MLP & Formulated a contract theory and applied GDMs with DRL to optimize incentive contracts for embodied vehicle intelligence in cooperative information sharing. \\
			& GDM-based Multi-Modal Offloading \cite{GDMPolicy2025MetaverseIoT} & Attention-based LSTM & Applied GDMs with DRL to parallel offloading and resource optimization in AIGC-metaverse-IoT systems. \\
			& Meta-DSAC \cite{TVT2025MetaverseGDMPolicy} & MLP & Applied GDMs with DRL to service provider selection optimization for AIGC-metaverse applications in UAV networks. \\
			& DLRS \cite{GDMPolicy2025LLMQoS} & MLP & Used GDMs with DRL to optimize cloud-edge collaborative deployment of LLMs. \\
			& GDM-driven Incentive Mechanism \cite{GDMPolicy2025DownlinkSemantic} & MLP & Applied GDMs with DRL to resource allocation optimization for downstream tasks in semantic communication systems. \\
			& AC-GDM \cite{SecureBeamIRS2025GDMPolicy} & MLP & Applied GDMs with DRL to IRS-assisted secure beamforming. \\
			& GAI-Asynchronous \cite{GDMPolicy2025LowAltitudeEdgeInference} & MLP & Applied GDMs with DRL to batch-wise beamforming decision generation in low-altitude over-the-air edge inference systems. \\
		\end{tblr}
        \vspace{-0.25cm}
		\label{tab_GDM_SG_DRL_NO}
	\end{table*}
	
	Due to the inherent limitations of their data processing architectures, current research trends favor adapting general-purpose LPTMs into domain-specific LPTMs tailored for networking tasks \cite{survey2_pushing6G,netgpt2023tenissues,understanding2023telecomLLM,Telecom2024MustLarge}. In this direction, researchers have constructed specialized question-answering datasets covering areas such as network optimization, goal interpretation, and mathematical modeling \cite{TeleQnA2023,TSpec_LLM2024,WirelessMathBench2025XinLi}. However, most of these network-oriented LPTMs \cite{TelecomGPT2024,Mobile_LLaMA,MobileFoundationFirmware2024,WirelessGPT2025Pengcheng} primarily focus on upstream or auxiliary tasks related to network optimization, rather than directly solving optimization problems themselves.
	
	Specifically, TelecomGPT \cite{TelecomGPT2024} addresses the lack of domain knowledge in general LLMs for telecom applications by building three datasets for pre-training, supervised fine-tuning, and style alignment. It targets tasks such as professional question-answering, document classification, code generation, code summarization, and mathematical modeling, achieving performance that surpasses that of general-purpose models, despite using models with no more than 8B parameters. Mobile LLaMA \cite{Mobile_LLaMA} focuses on network data analytics function and fine-tunes LLaMA 2 13B \cite{LLaMA} on datasets covering IP routing analysis, packet inspection, and Python code diagnostics. M4 \cite{MobileFoundationFirmware2024} proposes a firmware-like foundational model jointly managed by mobile operating systems and hardware, enabling support for dozens of multi-modal mobile AI tasks. M4 achieves accuracy comparable to baseline methods across most tasks while offering significantly reduced storage, memory, and operational complexity. WirelessGPT \cite{WirelessGPT2025Pengcheng} targets channel-related tasks in wireless communications by introducing Traciverse, a large-scale wireless channel dataset for unsupervised pre-training. It encodes multi-dimensional channel data into unified representations while capturing inter-dimensional correlations. The model delivers strong performance in channel estimation, prediction, human activity recognition, and wireless signal reconstruction, with demonstrated scalability in model size. These foundational models developed for the networking domain provide valuable design insights and a rich set of benchmarks for applying LPTMs to network optimization tasks.
	
	\subsection{LPTMs as Solution Generators}
	Although the natural language processing architecture of LLMs poses inherent limitations for directly solving network optimization problems (such as imprecision, high uncertainty, and significant computational overhead), several studies have nevertheless achieved meaningful results by directly generating solutions through external feedback and prompt engineering. Specifically, \cite{survey7_GenerativeOpimization} provides a forward-looking overview of generative optimization, suggesting that LLMs can offer hybrid improvements in three key areas: mitigating the computational cost of traditional optimization, enhancing initial solution quality through iterative exploration, and unifying inputs across different modalities. Furthermore, \cite{LLMasOptimizer2024} proposes optimization by prompting (OPRO), which leverages an external objective evaluator to generate solution-score pairs. These historical evaluations are then used as prompts to iteratively guide the generation of new solutions. Similarly, \cite{NonConvexLLMOpt2025} extends this approach to non-convex network optimization problems, demonstrating effective application beyond the simpler linear and low-dimensional cases considered in \cite{LLMasOptimizer2024}.
	
	Overall, while LPTMs have demonstrated some effectiveness in directly generating solutions for network optimization through prompt engineering and iterative refinement, the computational overhead remains prohibitively high. Instead, their extensive pre-trained knowledge and strong cross-problem generalization capabilities hold greater promise as key enablers for improving the upper performance of GenAI-enabled network optimization.
	
	\subsection{LPTMs as DRL Policy Networks}
	Some studies have also explored leveraging LPTMs to achieve general-purpose DRL implementations. Specifically, \cite{GDMPolicy2025WirelessComm} is one of the earlier works in the networking domain to investigate the use of LPTMs as DRL policy networks through exploratory validation. \cite{lambo2024} proposes a framework where LPTMs serve as DRL actors, guided by feedback from an external critic. NetLLM \cite{netllm2024sigcomm} utilizes existing pre-trained LPTMs as actors, augmented with task-specific encoders, prediction heads, and low-rank adaptation (LoRA)-based \cite{LORA} specialization layers to adapt them to various downstream network optimization tasks.
	
	Overall, while LPTMs, as DRL policy networks, demonstrate impressive performance and often significantly outperform traditional architectures in terms of reward, they exhibit a poor performance-cost ratio, as the performance gains from network optimization do not justify the substantial training and computational overhead. Applying LPTMs to tasks such as intent decomposition and understanding, as well as high-level decision-making in time-insensitive scenarios, can be a more suitable direction. 
	
	\subsection{LPTMs for Other Network Tasks}
	At the current stage, LPTMs are still primarily used for understanding user intent \cite{LLM2024NO_VQR,D2SAC2025LLMasUserFeature,LAMeTA2025,ToolAided2025FL}, designing and selecting customized network modules \cite{LLMind2024,LLMsubtask2025hotnets}, generating network algorithm code {\cite{LLMOptiRA2025}},\cite{LLMGraphOperation2023NetworkManage,NADA2024HotNets}, and processing network data \cite{Ottor2023}.
	
	For user intent understanding, early work such as \cite{LLM2024NO_VQR} proposes shifting from a system-centric to a user-centric paradigm. They build a pipeline of agents based on LLaMA-3B \cite{LLaMA} to extract user intent and construct corresponding problem instances, which are then passed to external optimization solvers. \cite{D2SAC2025LLMasUserFeature} focuses on the distributed deployment of diffusion model-based image generation services. In this setting, LPTMs are used to construct cloned agents that reflect users’ personalized features, and the simulated feedback from these agents is leveraged to schedule the denoising process of cloud-based diffusion models. LAMeTA \cite{LAMeTA2025} aims to enable intent-driven agentic networks. It distills a general-purpose LPTM into multiple edge-level LPTMs, each functioning as an agent within an agent graph. The core innovation lies in defining four theoretical metrics from the AI domain, including capability, information loss, latency, and error probability, to quantify service QoS requirements as 4-dimensional scalar vectors. These vectors are then used by an external DRL algorithm to plan the execution trajectory for each service request. T-ELLM \cite{ToolAided2025FL} addresses challenges in federated learning caused by dynamically evolving network conditions. Based on natural-language descriptions of service requirements, a fine-tuned LLM generates text responses with device selection results. External numerical solvers are then used to perform resource allocation.
	
	\begin{table*}[t]
		\centering
		\caption{Relevant works on LPTMs in network optimization.}
		\begin{tblr}{
				colspec={X[0.13,l] X[0.1,c] X[0.45,c] X[0.6,c]},
				cells = {c},
				hlines, vlines,
				rowsep = {2pt},
				row{1} = {0ex},
				cell{2}{1} = {r=3}{},
				cell{5}{1} = {r=3}{},
				cell{8}{1} = {r=4}{},
				cell{12}{1} = {r=10}{},
			}
			\textbf{Type} & \textbf{Reference} & \textbf{Characteristics} & \textbf{Contribution} \\
			Solution generator & \cite{survey7_GenerativeOpimization} & Perspective of GenAI for optimization & Proposes three paradigms for integrating GenAI into optimization. \\
			& \cite{LLMasOptimizer2024} & Feedback–solution iteration & Introduces OPRO to use LLMs for generating solutions iteratively based on historical trajectories. \\
			& \cite{NonConvexLLMOpt2025} & Feedback–solution iteration & Extends LLMs as solution generators to non-convex optimization problems. \\
			DRL policy & \cite{GDMPolicy2025WirelessComm} & Actor guided by evaluation prompts & Early proposal of LPTMs as policy networks in DRL. \\
			& \cite{lambo2024} & Actor guided by evaluation prompts & Proposes a DRL framework where LPTMs act as actors guided by an external critic. \\
			& \cite{netllm2024sigcomm} & Foundation for downstream tasks & Enhances LPTMs with minimal additional layers to support downstream network optimization tasks efficiently. \\
			Foundational model & \cite{TelecomGPT2024} & General-purpose LLM for telecom & Builds question-answering and domain document datasets to enhance LLMs with telecommunications knowledge. \\
			& \cite{Mobile_LLaMA} & LLM for network data analytics & Fine-tunes general LLMs to create models specialized for network data analysis. \\
			& \cite{MobileFoundationFirmware2024} & Firmware LPTM for mobile devices & Constructs efficient firmware-based foundation models for mobile AI tasks. \\
			& \cite{WirelessGPT2025Pengcheng} & Foundation model for wireless tasks & Develops wireless-specific foundation models using mature datasets with highly flexible model sizes. \\
			Assistance tasks & \cite{LLM2024NO_VQR} & User intent understanding & Uses feature vectors of user requests to implicitly guide LLMs in understanding user intent. \\
			& \cite{D2SAC2025LLMasUserFeature} & User personality cloning & Uses LPTMs to replicate user preferences for text-to-image generation and simulate feedback to guide generation. \\
			& \cite{LAMeTA2025} & User intent understanding & Proposes four scalar metrics to evaluate edge LPTM agent performance and describe QoS requirements of services. \\
			& \cite{ToolAided2025FL} & User intent understanding & Selects devices in federated learning networks based on natural language service descriptions. \\
			& \cite{LLMind2024} & Domain-specific AI selection & Proposes a framework for coordinating and selecting domain-relevant AI modules. \\
			& \cite{LLMsubtask2025hotnets} & Network configuration generation & Designs a method to generate network configurations based on user requirements and topologies. \\
			& \cite{LLMGraphOperation2023NetworkManage} & Network algorithm code generation & Early work using LPTMs for generating network algorithm code and establishing the NeMoEval benchmark. \\
			& \cite{NADA2024HotNets} & Network algorithm code generation & Proposes an LLM framework to generate and select neural code blocks for dynamic network environments. \\
			& \cite{LLMOptiRA2025} & Network algorithm code generation & Proposes an LLM framework to convert non-convex problems into convex problems and generate executable solver code, and establish a new dataset for testing. \\
			& \cite{Ottor2023} & Network data processing & Proposes an end-to-end LPTM pipeline compatible with commercial embedded platforms to process sensor data from edge devices. \\
		\end{tblr}
		\label{tab_LLM_SG_DRL_NO}
        \vspace{-0.25cm}
	\end{table*}
	
	For customized network module design and selection, LLMind \cite{LLMind2024} proposes an intelligent framework to coordinate domain-specific AI modules. \cite{LLMsubtask2025hotnets} introduces a framework that enhances interpretability in network synthesis by leveraging local sub-specifications corresponding to different network components. In particular, it incorporates constraint-simplification techniques that, given configuration requirements and network topology, generate local network configurations and assemble them into a complete setup.
	
	For network algorithm code generation, \cite{LLMGraphOperation2023NetworkManage} uses LPTMs to generate code for network management tasks and releases a benchmark and dataset named NeMoEval. NADA \cite{NADA2024HotNets} performs block-wise code generation based on status prompts and neural architecture instructions. It then compiles and checks candidate code blocks, enabling flexible selection and assembly to meet evolving environmental demands. The recent LLM-OptiRA \cite{LLMOptiRA2025} framework targets non-convex optimization in network scenarios by leveraging GPT-4 to transform non-convex problems into convex ones, subsequently generating executable code for mature numerical solvers to obtain precise solutions. LLM-OptiRA further incorporates code debugging and solution feasibility correction steps, achieving impressive code execution and solution success rates on the newly proposed WireOpt dataset.
	
	For network data processing, Otter \cite{Ottor2023} presents an end-to-end solution compatible with various commercial embedded platforms and software. It processes data from embedded sensor terminals to edge devices, where LPTMs are used to prompt and filter key information during preprocessing.
	
	\subsection{Summary}
	As shown in Table \ref{tab_LLM_SG_DRL_NO}, research on LPTMs in the network domain tends to focus on building foundation models for specific task categories. Only subsequently is attention given to lower-level tasks related to network optimization, such as user intent understanding. Currently, the application of LPTMs in network optimization has evolved from initial efforts in multimodal data processing and code generation to more advanced capabilities such as user intent comprehension and complex task reasoning. This progression reflects a growing trend of adapting general-purpose LPTMs to a wide range of networking tasks, with flexible model sizes enabling practical deployment.
	
	It is worth noting, however, that these models largely follow the same architectures and training strategies as popular LLMs. While they have achieved impressive task-specific results—thanks to the scale of their training data and parameters—they also inherit the known limitations of current LPTMs, particularly in learning mathematical rules and performing compositional reasoning. Given that simply increasing data and model size no longer yields significant improvements in reasoning ability, there is a pressing need to examine and reflect on the upper bounds of their performance.
	
	\section{Theoretical Analysis of GenAI for Network Optimization}\label{sec_theory}
	
	GenAI has demonstrated remarkable success in both one-shot solution generation and MDP-based decision-making for network optimization. However, there remains a lack of theoretical understanding regarding its generalization capabilities in optimization contexts. In this section, we use GDMs as representative examples and apply the analysis of generalization capability in distribution learning to the corresponding optimization problems. Specifically, we derive bounds on the gap between the expected objective value of generated solutions and the optimal value, as well as the gap between the expected returns of generated and optimal policies. {These results demonstrate the performance convergence of GDMs in both types of problem formulations and provide an intuitive illustration of the positive and negative factors influencing the convergence outcomes.}
	
	\subsection{Analysis}
	We revisit and synthesize the AI community's analyses of GDM's generalization capabilities \cite{OnGeneralization2023Diffusion,onMemorizationOfDiff,emergence2024Diffusion,understanding2024generalizability,fewShot2024DiffusionCurse}, leveraging their established conclusions to support our objectives. Specifically, the findings of \cite{OnGeneralization2023Diffusion} serve as a core foundation for our work and are largely consistent with the empirical observations reported in \cite{onMemorizationOfDiff}. Subsequently, \cite{emergence2024Diffusion} further validated the conclusions of \cite{OnGeneralization2023Diffusion,onMemorizationOfDiff} through additional empirical studies. Although the content of \cite{understanding2024generalizability} is relatively dense, its key insight—that the strong generalization ability of GDMs stems from their ability to learn shared low-dimensional structural features across disjoint datasets—is particularly valuable, and can be used to counter the claims made in \cite{fewShot2024DiffusionCurse} regarding the curse of few-shot generalization.
	
	We introduce the generalization result of GDMs from \textit{Theorem 1} in \cite{OnGeneralization2023Diffusion}, which provides a bound on the Kullback–Leibler (KL) divergence between the model's predicted distribution and the true target distribution. Denote the target distribution as $p_0$ and the predicted distribution as $p_{0|\hat{\boldsymbol{\theta}}_n(\tau)}$, according to \cite{OnGeneralization2023Diffusion}, the bound on the KL divergence is given by
	\begin{equation}\label{eq_On_theorem1}
		\begin{aligned}
			D_\mathrm{KL}(p_0||p_{0|\hat{\boldsymbol{\theta}}_n(\tau)})\lesssim& \left(\frac{\tau^4}{mn}+\frac{\tau^3}{m^2}+\frac{1}{\tau}\right) \\
			&+\left[\frac{1}{m}+\bar{\mathcal{L}}(\bar{\boldsymbol{\theta}}^*)+\mathcal{L}(\boldsymbol{\theta}^*)\right] \\
			&+D_\mathrm{KL}(p_T||\pi),
		\end{aligned}
	\end{equation}
	where $\hat{\boldsymbol{\theta}}_n$ represents the model parameters at training time $\tau$. Here, $m$ denotes the model size, and $n$ denotes the number of training samples. $\bar{\mathcal{L}}$ and $\bar{\boldsymbol{\theta}}^*$ refer to the continuous version of the score function and its optimal parameterization, respectively, while $\mathcal{L}$ and $\boldsymbol{\theta}^*$ denote their generalized counterparts. $T$ is the total number of diffusion steps. $p_T$ and $\pi$ represent the noise distribution after $T$ steps of diffusion and the target noise distribution, respectively.
	
	{Both one-shot optimization and MDP decision-making in network optimization are modeled as conditional diffusion processes, which necessitates extending Eq. (\ref{eq_On_theorem1}) to the conditional setting. Specifically, consider the conditional GDM, where generation is conditioned on some variable $\mathbf{x}$ (input parameters), and let the joint distribution over the solution and condition be denoted by $(\mathbf{y},\mathbf{x})$. Through the chain rule of KL divergence, the expected conditional KL divergence is bounded above by the joint divergence }
	\begin{equation}\label{eq_uncond2cond_final}
		\begin{aligned}
			\mathbb{E}_{\mathbf{x} \sim p(\mathbf{x})} \left[D_{\mathrm{KL}}(p_0(\mathbf{y}|\mathbf{x})||p_{0|\hat{\boldsymbol{\theta}}_n(\tau)}(\mathbf{y}|\mathbf{x})) \right]\leq& \\ 
			D_{\mathrm{KL}}(p_0(\mathbf{y},\mathbf{x})||p_{0|\hat{\boldsymbol{\theta}}_n(\tau)}&(\mathbf{y},\mathbf{x})).
		\end{aligned}
	\end{equation}
	This observation allows us to treat conditional diffusion models within the same generalization framework as unconditional ones, by simply extending the sample space to the joint variable $(\mathbf{y},\mathbf{x})$. 
    
    {Therefore, the convergence boundary of the GDM with respect to its original probabilistic distribution learning objective is established.  Based on this boundary, the following two subsections derive performance bounds for the GDM in terms of one-shot optimization solution quality and MDP decision rewards, thereby characterizing the advantages and limitations of one-shot optimization and MDP decision-making within the probabilistic distribution-learning objective.}

	\subsubsection{Bound of Solution Generation}
	Let the objective function of the one-shot optimization formulation in network optimization be denoted as $\min_{\mathbf{y}\in\mathcal{Y}}f(\mathbf{y},\mathbf{x}),\ \mathbf{x}\in\mathcal{X}$, where we omit constraints or treat them as part of the objective. For maximization objectives, an equivalent minimization formulation can be adopted.
	
	Our goal is to relate the KL divergence between the GDM-generated solution distribution and the true target solution distribution to the expected objective value of the generated solutions. Due to the asymmetry of KL divergence, we adopt the forward KL divergence, rather than the reverse KL used in Eq. (\ref{eq_uncond2cond_final}), to properly express the expected objective value under the generated distribution.
	
	Specifically, we construct a target solution distribution induced by the objective function, denoted as $p_0(\mathbf{y}|\mathbf{x})\propto e^{-\alpha f(\mathbf{y},\mathbf{x})}$. Let $p_0(\mathbf{y}|\mathbf{x})=\frac{e^{-\alpha f(\mathbf{y},\mathbf{x})}}{Z(\mathbf{x})}$, where $Z(\mathbf{x})$ serves as the normalization constant or partition function, and $\alpha>0$ is a temperature parameter that controls the preference for optimal solutions. For large $\alpha$, the distribution concentrates around the optimal solutions; as $\alpha$ approaches zero, the distribution becomes nearly uniform. Then we assume the objective function $f$ is integrable with $\mathcal{X}$ and $\mathcal{Y}$ uniformly bounded. The normalization constant $Z$ can be expressed as $Z(\mathbf{x})=\sum_{\mathbf{y}\in\mathcal{Y}}e^{-\alpha f(\mathbf{y},\mathbf{x})}$ for discrete distributions and as $Z(\mathbf{x})=\int_{\mathcal{Y}}e^{-\alpha f(\mathbf{y},\mathbf{x})}d\mathbf{y}$ for continuous distributions, respectively.
	
	The pre-assumptions are consistent with the conditions stated in \textit{Theorem 1} of \cite{OnGeneralization2023Diffusion}. Given an input $\mathbf{x}$, let the solution distribution learned by the model $\boldsymbol{\theta}$ be denoted as $p_{\boldsymbol{\theta}}:=p_{0|\hat{\boldsymbol{\theta}}_n(\tau)}$. The KL divergence between the predicted solution distribution and the target distribution is given by 
	{\begin{equation}\label{eq_KLD_boltzmann}
		\begin{aligned}
			D_{\mathrm{KL}}(&p_{\boldsymbol{\theta}}(\mathbf{y}|\mathbf{x})||p_0(\mathbf{y}|\mathbf{x}))=\int_{\mathcal{Y}} p_{\boldsymbol{\theta}}(\mathbf{y}|\mathbf{x})\log\frac{p_{\boldsymbol{\theta}}(\mathbf{y}|\mathbf{x})}{p_0(\mathbf{y}|\mathbf{x})}d\mathbf{y} \\
			&=\alpha\mathbb{E}_{p_{\boldsymbol{\theta}}}\left[f(\mathbf{y},\mathbf{x})\right]\!+\!\log Z\!+\!\int_{\mathcal{Y}} p_{\boldsymbol{\theta}}(\mathbf{y}|\mathbf{x})\log p_{\boldsymbol{\theta}}(\mathbf{y}|\mathbf{x})d\mathbf{y}.
		\end{aligned}
	\end{equation}}
	{With the entropy of $p_{\boldsymbol{\theta}}(\mathbf{y}|\mathbf{x})$ given as $\mathcal{H}(p_{\boldsymbol{\theta}}(\mathbf{y}|\mathbf{x}))=-\int_{\mathcal{Y}} p_{\boldsymbol{\theta}}(\mathbf{y}|\mathbf{x})\log p_{\boldsymbol{\theta}}(\mathbf{y}|\mathbf{x})d\mathbf{y}$, we can conclude that }
	\begin{equation}\label{eq_sg_bound_boltzmann}
		\begin{aligned}
			\mathbb{E}_{p_{\boldsymbol{\theta}}}&\left[f(\mathbf{y},\mathbf{x})\right]-f(\mathbf{y}^*,\mathbf{x})\leq \\
			&\frac{1}{\alpha}\left(D_{\mathrm{KL}}(p_{\boldsymbol{\theta}}(\mathbf{y}|\mathbf{x})||p_0(\mathbf{y}|\mathbf{x}))-\log Z+\mathcal{H}(p_{\boldsymbol{\theta}}(\mathbf{y}|\mathbf{x}))\right).
		\end{aligned}
	\end{equation}
	
	In Eq. (\ref{eq_sg_bound_boltzmann}), $\mathbf{y}^*$ denotes the optimal solution corresponding to the given $\mathbf{x}$. The normalization constant $Z$ depends only on the target solution distribution and $\mathbf{y}$, and is thus fixed under the given $\mathbf{x}$. The entropy term captures the uniformity of the model's output distribution across solutions, higher entropy indicates a more uniform distribution, greater uncertainty, and a lower expected quality (i.e., higher objective value) of the generated solution. The parameter $\alpha$ characterizes both the distance between modes and the number of modes in the high-quality region of the target distribution. Intuitively, it reflects the complexity of the optimization problem. When $\alpha$ is small, the target distribution is more complex (i.e., more modes or closely spaced modes), implying higher problem complexity and a looser bound on the right side of Eq. (\ref{eq_sg_bound_boltzmann}). The KL divergence term is always non-negative. Although Eq. (\ref{eq_sg_bound_boltzmann}) is derived using the forward KL divergence, which differs from Eq. (\ref{eq_uncond2cond_final}), the latter can still be interpreted as providing an upper bound for this term.
	
	{In summary, Eq. (\ref{eq_sg_bound_boltzmann}) quantifies the generalization boundary of the conditional GDM in one-shot optimization solution generation, and subsequently highlights several key factors that influence its generalization ability.}
	
	\subsubsection{Bound of Policy Decision Generation}
	DRL trains a policy that takes actions based on the input state to receive rewards, aiming to maximize the expected cumulative discounted return. Our objective is to leverage the KL divergence bound between the optimal policy and the learned (predicted) policy, be described by Eq. (\ref{eq_uncond2cond_final}), to quantify the gap between their corresponding value functions. 
	
	Consider an infinite horizon MDP $\mathcal{M}=\{\mathcal{S},\mathcal{A},r,P,\gamma\}$. Here, $\mathcal{S}$ and $\mathcal{A}$ denote the state space and action space, respectively; $r: (\mathbf{s}, \mathbf{a}) \rightarrow \mathbb{R}$ is the reward function; $P$ represents the state transition probability; and $\gamma$ is the reward discount factor. Specifically, the reward function is defined as $r(\mathbf{s}, \mathbf{a}) = \mathbb{E}[R_t \mid \mathbf{S}_t = \mathbf{s}, \mathbf{A}_t = \mathbf{a}]$. The policy is defined as $\pi:\mathcal{S}\rightarrow\Delta(\mathcal{A})$. {Accordingly, we denote the discounted return as $U_t$, the state–action value function (Q-function) as $Q_{\pi}(\mathbf{s}_t,\mathbf{a}_t)$, and the state value function as $V_{\pi}(\mathbf{s}_t)$.}
	
	{Next, we introduce the performance difference lemma (PDL) \cite{PDL2002}. For the optimal policy $\pi^*$ and the predicted policy $\hat{\pi}$, PDL yields the following result:}
	\begin{equation}\label{eq_DRL_V_gap_ini}
		|V_{\pi^*}(\mathbf{s}_0)-V_{\hat{\pi}}(\mathbf{s}_0)|=\frac{1}{1-\gamma}\mathbb{E}_{\mathbf{s}\sim d^{\pi^*}_{s_0}}\left[\mathbb{E}_{\mathbf{a}\sim\pi^*(\cdot|\mathbf{s})}\left[A_{\hat{\pi}}(\mathbf{s},\mathbf{a})\right]\right],
	\end{equation}
	where $A_{\hat{\pi}}(\mathbf{s},\mathbf{a})=Q_{\hat{\pi}}(\mathbf{s},\mathbf{a})-V_{\hat{\pi}}(\mathbf{s})$.
	{Expanding the innermost expectation on the right-hand side of $\mathbb{E}_{\mathbf{a}\sim\pi^*(\cdot|\mathbf{s})}\left[A_{\hat{\pi}}(\mathbf{s},\mathbf{a})\right]$, and transforming the expectation of the independent value into the form of a Q-function, we obtain}
	\begin{equation}\label{eq_A_func_final}
		\begin{aligned}
			\mathbb{E}_{\mathbf{a}\sim\pi^*(\cdot|\mathbf{s})}\left[A_{\hat{\pi}}(\mathbf{s},\mathbf{a})\right]=&\mathbb{E}_{\mathbf{a}\sim\pi^*(\cdot|\mathbf{s})}\left[Q_{\hat{\pi}}(\mathbf{s},\mathbf{a})\right] \\
			&-\mathbb{E}_{\mathbf{a}\sim\hat{\pi}(\cdot|\mathbf{s})}\left[Q_{\hat{\pi}}(\mathbf{s},\mathbf{a})\right].
		\end{aligned}
	\end{equation}
	
	According to the properties of total variation (TV) distance \cite{TV2002RelatedLipschitz}, we obtain
	\begin{equation}\label{eq_DRL_TV}
		|\mathbb{E}_{\mathbf{a}\sim\pi^*(\cdot|\mathbf{s})}\left[Q_{\hat{\pi}}(\mathbf{s},\mathbf{a})\right]-\mathbb{E}_{\mathbf{a}\sim\hat{\pi}(\cdot|\mathbf{s})}\left[Q_{\hat{\pi}}(\mathbf{s},\mathbf{a})\right]|\leq L\cdot\sqrt{\frac{\epsilon}{2}},
	\end{equation}
	where $L>0$ is the Lipschitz constant of the Q-function, $\epsilon=D_{\mathrm{KL}}(\pi^*(\mathbf{a}|\mathbf{s})||\hat{\pi}(\mathbf{a}|\mathbf{s}))$. Combine Eq. (\ref{eq_A_func_final}) and Eq. (\ref{eq_DRL_TV}), we get
	\begin{equation}\label{eq_DRL_TV_final}
		\mathbb{E}_{\mathbf{a}\sim\pi^*(\cdot|\mathbf{s})}\left[A_{\hat{\pi}}(\mathbf{s},\mathbf{a})\right]\leq L\cdot\sqrt{\frac{\epsilon}{2}}.
	\end{equation}
	
	Further, combining Eq. (\ref{eq_DRL_V_gap_ini}) and Eq. (\ref{eq_DRL_TV_final}), we can get
	\begin{equation}\label{eq_DRL_bound_res}
		\begin{aligned}
			|V_{\pi^*}(s_0)-V_{\hat{\pi}}&(s_0)|\leq\frac{1}{1-\gamma}\mathbb{E}_{\mathbf{s}\sim d_{s_0}^{\pi^*}}\left[L\cdot\sqrt{\frac{\epsilon}{2}}\right] \\
			&\leq\frac{1}{1-\gamma}\mathbb{E}_{\mathbf{s}\sim d_{s_0}^{\pi^*}}\left[L\cdot\sqrt{\frac{D_{\mathrm{KL}}(\pi^*(\mathbf{a}|\mathbf{s})||\hat{\pi}(\mathbf{a}|\mathbf{\mathbf{s}}))}{2}}\right].
		\end{aligned}
	\end{equation}
	The KL divergence term can be bounded using Eq. (\ref{eq_uncond2cond_final}). In DRL, the target value function is externally defined, and a larger discount factor $\gamma$ implies that the objective places more emphasis on long-term outcomes. As $\gamma$ approaches 1, the gap between the predicted policy and the optimal policy tends to increase over an infinite time horizon. The Lipschitz constant $L$ characterizes the smoothness or complexity of the state-value function. A larger $L$ indicates a more complex objective, which, in turn, leads to a larger gap between the predicted and optimal policies.
	
    {In summary, Eq. (\ref{eq_DRL_bound_res}) applies the existing KL divergence bound from conditional GDM distribution learning to derive a bound on the difference in state value functions between the learned and optimal policies, and further identifies several key factors that influence this bound.}

    \begin{remark}[{Generalization Bound}]
    {When a GDM serves as a one-shot optimization solution generator or an MDP policy, its generalization error is bounded. This boundary is characterized by a KL-divergence-based bound on the generalization error of the GDM’s distribution-learning objective and highlights several influencing factors, including training duration, dataset size, problem complexity, and the discount factor.}
    \end{remark}
	
	\subsection{Discussion}
	\subsubsection{Strengths and Potential}
	As demonstrated by the above results, the learning errors of GDMs in both one-shot optimization and MDP decision-making tasks admit bounded guarantees. This suggests that the distribution-learning properties of GenAI enable convergence and, when combined with existing theoretical frameworks \cite{li2023t2t,liang2024gdsg,isConditional2023,IDQL2023}, help identify the various factors that contribute positively or negatively to final learning outcomes. Consequently, the feasibility of applying GenAI to network optimization across these two paradigms is now supported by a solid theoretical foundation that also offers guidance on practical design choices.
	
	From an engineering perspective, GDMs have demonstrated training stability in both paradigms and is making progress toward task generalization across diverse network optimization problems \cite{boosting2025energyGuided,taskAgnosticDiffusionPlanner2025}. Moreover, recent work has significantly reduced GDM’s computational and memory overhead \cite{DiffuserLite2024NIPS}, making it increasingly suitable for real-world deployment in various networking scenarios. In contrast, LPTMs show promising potential for tasks such as extracting user or operator intent from natural-language problem descriptions and translating it into mathematical models for network optimization \cite{Autoformulation2024LLM,ModelingAgent2025ModelingBench}, owing to their emerging capabilities for understanding and reasoning over mathematical content.
	
	\subsubsection{Flaws and Challenges}
	At a fundamental level, GenAI still faces inherent limitations, including output uncertainty, reliance on data-driven training, and autoregressive generation. When applied to network optimization, GDM's solution and decision generation may degrade into regression \cite{CARD2022NIPS}, resulting in limited generalization. LPTMs, meanwhile, internally compress and distinguish semantics in ways that diverge significantly from human cognition and often exhibit strong biases \cite{Tokens2Thoughts2025}, as summarized in Table \ref{tab_limitations}. The dominant autoregressive generation mechanism in LPTMs is also prone to error accumulation. While recent research increasingly explores reflective behaviors in these models to mitigate error accumulation \cite{rStarMath2025}, fundamental alternatives that challenge the autoregressive paradigm remain scarce \cite{LLaDA2025}.
	
	Empirically, a substantial body of the newest research suggests that many of LPTMs’ capabilities in solving mathematical problems stem from pattern matching rather than genuine rule-based understanding \cite{MATH2021Dataset,GSM-Symbolic,DRL2025ReasoningLLM,DoPhD2025LLM,WirelessMathBench2025XinLi}. That is, for tasks involving conceptual reasoning or arithmetic, LLMs often generate answers by recalling training samples similar in form to the given question, rather than by applying step-by-step logical rules. Moreover, the Transformer architecture has been shown to exhibit theoretical limitations in tasks involving function composition, i.e., decomposing and recombining subproblems or basic semantic units of the original problem \cite{FaithAndFate2023,Representational2023TFM}. Similarly, GDMs rely on parallel sampling to generate multiple candidates and select higher-quality solutions or decisions. Like LPTMs, their performance is highly dependent on the completeness of training data, and they inherit GenAI’s broader difficulty in learning from long-tail distributions that can be particularly critical in the dynamic environments of network optimization.
	
	\begin{table*}[thb]
		\centering
		\small
		\caption{{A set of recent works that shows the limitations or potential of GenAI and provides differing attitudes on the performance of generative optimization.}}
		\begin{tblr}{
				colspec={X[0.08,l] X[0.06,c] X[0.45,c] X[0.4,c]},
				cells = {c},
				hlines, vlines,
				rowsep = {2pt},
				row{1} = {0ex},
			}
			\textbf{References} & {\textbf{Attitude}} & \textbf{Contributions} & \textbf{Implications for Network Optimization} \\
			\cite{WirelessMathBench2025XinLi} & {Negative} & Demonstrates the {\textbf{performance limitations of state-of-the-art LPTMs in mathematical modeling}} within wireless communication tasks. & Highlights the need for task-specific adaptation and hybrid symbolic-neural optimization pipelines. \\
			\cite{GSM-Symbolic} & {Negative} & Shows that current LLMs {\textbf{replicate observed reasoning patterns}} instead of performing genuine logical reasoning. & Suggests that semantic understanding and explicit constraint encoding are essential for reliable optimization. \\
			{\cite{DRL2025ReasoningLLM}} & {Negative} & {Confirms that \textbf{RLVR improves reasoning efficiency but not reasoning capability.}} & {Indicates that RLVR shows limited ability to improve GenAI's inference for network optimization.} \\
			\cite{DoPhD2025LLM} & {Negative} & Reveals that existing LLMs rely on {\textbf{pattern memorization rather than learning mathematical principles}}. & Implies the need for structure-aware optimization models that encode physical or mathematical priors. \\
			\cite{FaithAndFate2023} & {Negative} & Analyzes {\textbf{the failure of Transformers to generalize in compositional reasoning tasks}}. & Suggests that compositional generalization remains a key bottleneck in scaled network optimization. \\
			\cite{Representational2023TFM}& {Neutral} & Provides a comprehensive analysis of the {\textbf{representational strengths and weaknesses of the Transformer architecture}}. & Provides reference for the upper and lower bounds of GenAI's neural network backbone. \\
			\cite{onMemorizationOfDiff} & {Neutral} & Empirically analyzes how dataset size, dimensionality, diversity, and hyperparameters affect {\textbf{memorization in GDMs}}. & Emphasizes controlling data-model balance for stable generative optimization. \\
			\cite{OnGeneralization2023Diffusion} & {Neutral} & Provides theoretical insights into how sample size, dimensionality, and training duration influence {\textbf{GDM generalization error}}. & Suggests generalization–efficiency trade-offs for GenAI in network optimization. \\
			\cite{FormalMath2024Intrinsic} & {Positive} & Enables theorem conjecturing and proving from axioms, demonstrating {\textbf{self-improvement}} and challenging theorem generation. & Suggests self-reinforcing optimization mechanisms based on intrinsic reward signals. \\
			\cite{LLMOPT2025} & {Positive} & Proposes {\textbf{a unified 5-element framework}} (scenario, parameters, variables, objectives, constraints) for optimizing generalization. & Provides a generalizable paradigm for translating optimization problems into runnable codes. \\
			\cite{emergence2024Diffusion}& {Neutral} & Shows that GDMs {\textbf{learn distinct distributions depending on training data size}}. & Indicates the impact of sample quantity and diversity on GenAI learning results. \\
			\cite{Tokens2Thoughts2025}& {Negative} & Demonstrates that LLMs achieve {\textbf{information-theoretic compression}} at the expense of semantic detail. & Indicates that LPTM’s grasp of semantic details of encoded tokens is very limited compared to humans. \\
			{\cite{GeneralistRewardModels2025}} & {Positive}& {Theoretically proves that next-token prediction \textbf{implicitly contains a universal reward model} equivalent to one learned via inverse RL.} & {Suggests leveraging intrinsic reward modeling for unsupervised network optimization.} \\
			{\cite{ReasoningBank}} & {Positive} & {Demonstrates that combining successful and failed experiences improves LLM reasoning via \textbf{memory-aware test-time scaling}.} & {Encourages bidirectional experience replay for efficient online optimization.} \\
			{\cite{MarkovianThinker}} & {Positive} & {Proposes Markovian Thinker, enabling \textbf{stepwise RL reasoning under Markov assumptions}, achieving significant efficiency gains.} & {Suggests state-transition-aware network optimization strategies with bounded computational growth.} \\
		\end{tblr}
		\label{tab_limitations}
        \vspace{-0.25cm}
	\end{table*}
	
	\section{Future Directions}\label{sec_future}
	Despite the impressive performance of GenAI in network optimization and related AI domains, debates over its internal mechanisms remain unresolved. In recent years, both significant benchmark improvements driven by new GenAI techniques and critical analyses exposing their limitations have emerged in parallel. Even within the AI community, opinions diverge: some believe that the continuous breakthroughs of LPTMs and GDMs across various metrics indicate their potential to approach human-level intelligence; others argue that these gains are merely incremental advances fueled by data scale and engineering tricks, rather than evidence of genuine progress toward human-like reasoning. In reality, both perspectives largely rely on empirical results from specific testing, while deep theoretical understanding remains scarce, mainly because many phenomena observed in LPTMs and GDMs remain poorly understood. {In the summary of cutting-edge works presented in Table \ref{tab_limitations}, we illustrate the diverse attitudes of current researchers in this area.}
	
	In light of this, we argue that both sides of the evidence should be considered. On the one hand, we should leverage current, effective techniques to continue improving performance metrics. On the other hand, we must not let these successes overshadow the models’ limitations. It is essential to continue identifying flaws and exploring new directions. This section explores potential future directions for GenAI in network optimization, from both theoretical and implementation perspectives. {These directions not only reflect the potential application of GenAI in network optimization, but also point out some specific problems that may be helped to be solved in the current network optimization domain, such as explainability of the principles, diversity of solvable problems, and reliability of the solutions.}
	
	\subsection{Theories}
	
	\subsubsection{Disentangling the Link Between Generation and Optimization}
	Generation and optimization remain fundamentally different tasks. The former is a data-driven process aimed at approximating probability distributions, while the latter involves finding a unique optimal solution determined by objective functions and constraints. Although recent works have attempted to bridge this gap by defining high-quality solution distributions and converting solution quality into likelihood for sampling purposes \cite{sun2023difusco,liang2024gdsg}, the relationship between optimization problem types and the corresponding generative distribution types remains unclear.
	
	For example, it is not yet well understood whether common optimization problem classes (e.g., convex optimization, mixed-integer programming, combinatorial optimization) are inherently compatible with typical generative distributions (e.g., Gaussian, Poisson, or multimodal distributions). Clarifying these relationships would help define the expected target distribution when applying GenAI to network optimization problems, thereby enabling better estimates of learning difficulty and theoretical performance limits. This is especially crucial when dealing with long-tailed distributions, which are notoriously challenging for GenAI models. A deeper theoretical understanding in this direction can improve both the design and deployment of GenAI-based solvers under various problem regimes.
	
	\subsubsection{Decomposing Network Optimization Problems}
	Most existing GenAI-for-network-optimization efforts focus on decomposing the control mechanisms (e.g., agent behaviors or response strategies) rather than dissecting the structure of the optimization problems themselves. However, decomposing complex optimization problems into simpler sub-problems can greatly enhance tractability and generalization.
	
	A classical example is transforming non-convex problems into convex ones using relaxation techniques. In this vein, GenAI models could be developed as specialized solvers targeting different types of simpler sub-problems, with their outputs fused to form a high-quality overall solution. This modular approach reduces both the complexity of designing a single end-to-end GenAI optimizer and improves generalizability across tasks.
	
	While this idea reduces the modeling and data requirements, it introduces new theoretical challenges: real-world network optimization problems often exhibit mathematically intricate structures, and there is currently no unified methodology for problem decomposition with broad applicability. We argue that a practical and promising route is to start from representative problem classes and design flexible, modular sub-solution components \cite{WirelessNetworkDesign2019TCOM}. For instance, in combinatorial optimization, AI techniques can be used to replace inefficient modules in traditional algorithms \cite{LearnToCO2022infocom,FreeEnergyCO2025NatureCS}; in mixed-integer linear programming, neural networks can be employed to guide cut-plane selection \cite{LearnToCut2024TPAMI}. These targeted decompositions offer a balanced compromise between engineering feasibility and optimization theory.

    \begin{figure}[t]
		\centering
		\includegraphics[width=0.95\linewidth]{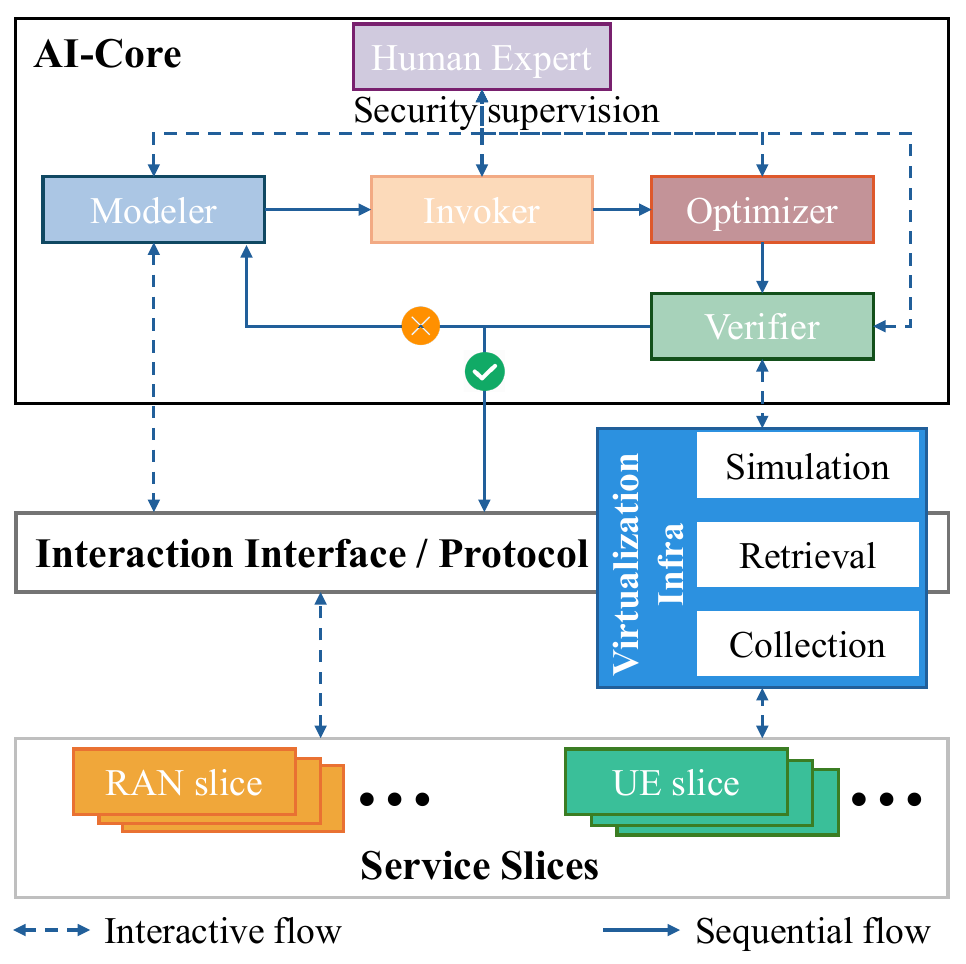}
		\caption{{Outlook on the applications of GenAI in the AI-Core and next-generation networking.}}
		\label{fig_prospect}
	\end{figure}
	
	\subsection{Implementations}
	
	\subsubsection{Rule-Guided Generation}
	Recent studies have consistently shown that current GenAI models—particularly LPTMs—do not exhibit human-like capabilities in conceptual understanding or rule capturing \cite{Tokens2Thoughts2025}. Although reinforcement learning has been used to enhance reasoning, these improvements mainly reflect behavioral preference tuning rather than genuine comprehension of underlying logic. {\cite{DRL2025ReasoningLLM} investigates whether RL with verifiable rewards (RLVR) truly enhances LLM reasoning beyond base models, finding that current RLVR methods do not elicit fundamentally new reasoning patterns but are bounded by the base model's capabilities.} Despite growing skepticism, no fundamentally new paradigm has yet emerged to replace autoregressive generation or sampling-based methods.
	
	In network optimization and related domains that demand high reliability and structure, we advocate for a practical path forward: rather than chasing elusive general intelligence, we should enhance controllability and generalization by embedding structured, rule-guided mechanisms into GenAI workflows. Traditional rule learning typically involves injecting mathematical constraints (e.g., numerical bounds) or system-level rules (e.g., protocol specifications) into loss or reward functions. However, this approach heavily relies on expert design and often lacks scalability and adaptability.
	
	To address this, we envisage two complementary strategies. First, inference-time scaling \cite{inferenceTimeScaling2025Xie}, a recently popularized method for enhancing reasoning, can be leveraged to equip GenAI models with the ability to reference and apply rules dynamically during inference, encouraging self-reflection and correction. Second, we suggest improving the rule design process itself by using GenAI models to assist or automate expert rule generation. For example, pre-trained models can be used to define energy functions or decision boundaries that guide another model’s behavior, echoing recent efforts in energy-guided generation \cite{boosting2025energyGuided}. This could substantially reduce human workload and improve the efficiency and generalizability of rule-guided generation systems.

    \begin{figure*}[t]
		\centering
		\centerline{\includegraphics[width=7.0in]{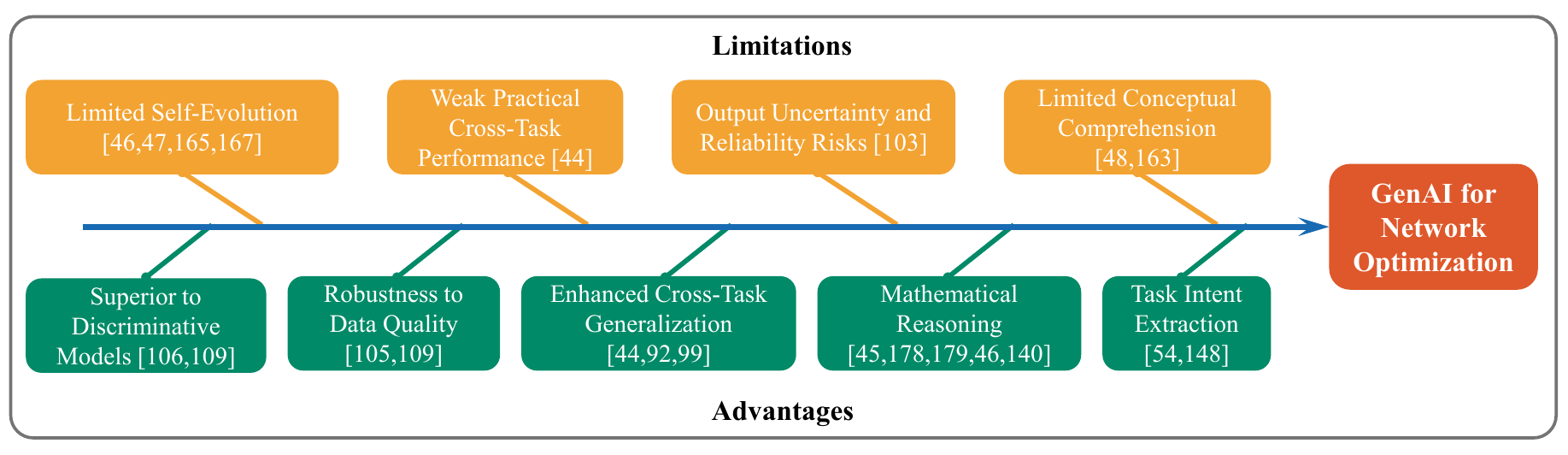}}
		\caption{{Current advantages, limitations, and the corresponding relevant works of GenAI for network optimization.}}
		\label{fig_suggestion}
	\end{figure*}
	
	\subsubsection{Collaboration Beyond Isolated Decision-Making}
	Recent advances in GenAI have demonstrated significant improvements in mathematical reasoning, formal modeling, and symbolic proof generation \cite{IntegrateFormal2023hotnets,formalMethod2022AsProblemAware,FormalMath2024Intrinsic,DeepSeekProverV2,rStarMath2025,ReasoningBeyondLimits2025Debbah}. While the underlying mechanisms behind these capabilities remain incompletely understood, engineering outcomes are increasingly competitive in specialized domains. Given GenAI’s inherent uncertainty, imprecision, and reliance on training data distributions, many of these works have adopted modular architectures that decompose the complex task of mathematical problem-solving into multiple cooperative components.
	
	In the field of network optimization, which demands structure and precision, early studies such as \cite{LLMOPT2025,LLMOptiRA2025} offer valuable insights. For instance, \cite{LLMOPT2025} decomposes an optimization problem into five components, including original context, physical parameters, optimization variables, objective functions, and constraints, using natural language prompts to formalize it. Meanwhile, \cite{LLMOptiRA2025} leverages LLMs to orchestrate the interaction with external solvers. These designs demonstrate the value of GenAI as a component within a broader, orchestrated system.
	
	Building on this perspective, we argue that true reliability and performance in network optimization may be achieved not through standalone GenAI models but through synergistic collaboration between GenAI, traditional numerical solvers, conventional neural networks, and human expertise. We envisage a five-module cooperative framework: a Modeler that interprets user intent and formulates formal optimization problems; an Optimizer that performs core numerical or data-driven solving; an Invoker that integrates external toolchains or simulation engines; a Verifier that evaluates solution validity and feasibility; and a Human Expert who intervenes when model limitations are reached, providing feedback and a corrective strategy. Such a collaborative system mitigates the intrinsic limitations of GenAI by distributing cognitive load and leveraging complementary strengths across components. Moreover, it aligns well with distributed or decentralized deployment environments common in real-world network systems \cite{DecentralizedGDM}. Importantly, this architecture enables autonomous evolution, allowing GenAI agents to learn from human interventions over time and progressively enhance their capabilities in dynamic, complex environments.
	
	\subsubsection{Virtualization-Enabled Evolution}
	Beyond the development and deployment of GenAI models, it is equally important to consider how these models evolve alongside dynamic network environments. Traditional static deployment strategies fall short in long-term performance maintenance, especially in systems that require adaptability to fluctuating traffic, topology changes, and real-time decision-making. We therefore advocate for a virtualized evolution infrastructure that enables GenAI models to be continuously refined, evaluated, and validated in a controlled, cost-efficient environment.
	
	We envision a virtual evolution platform that mirrors real-world network dynamics and enables GenAI agents to simulate their deployment scenarios before going live, akin to digital twins. This environment should be tightly synchronized with real-time data flows from the physical network, offering rich contextual information for pre-deployment testing. Recently, standardization institutes such as European Telecommunications Standards Institute (ETSI) and Next Generation Mobile Networks Alliance (NGMN) have begun integrating sandboxing, public memory, and other virtualization primitives into their Agent-based AI-Core architectures \cite{etsiGReni051,ngmn2025_6G}, recognizing the importance of controlled environments for AI development and orchestration. This approach enables GenAI models to interact with realistic network conditions, gather feedback, and adjust behavior without incurring the cost or risk of live network disruption. It also allows developers to conduct large-scale experiments and monitor model behaviors across diverse conditions, fostering robust generalization and reliability.

    \begin{table*}[t]
		\centering
		\caption{Relevant benchmarks and datasets, with the relevance to network optimization.}
		\begin{tblr}{
				colspec={X[0.15,l] X[0.15,c] X[0.08,c] X[0.6,c]},
				cells = {c},
				hlines, vlines,
				rowsep = {2pt},
				row{1} = {0ex},
				cell{2}{1} = {r=3}{},
				cell{5}{1} = {r=2}{},
				cell{7}{1} = {r=3}{},
				cell{10}{1} = {r=3}{},
			}
			\textbf{Type} & \textbf{Reference} & \textbf{Relevance} & \textbf{Content} \\
			Mathematical QA benchmarks & MATH \cite{MATH_bench} & \CIRCLE\Circle\Circle & A collection of 12.5K representative math QA problems covering challenging topics such as number theory and calculus. \\
			& GSM8K \cite{GSM8K} & \CIRCLE\Circle\Circle & Contains 8.5K elementary-level math word problems with solutions. \\
			& GSM-Symbolic \cite{GSM-Symbolic} & \CIRCLE\Circle\Circle & A variant of GSM8K where key symbols are replaced without altering the underlying problem semantics. \\
			Telecom-specific QA benchmark & TeleQnA \cite{TeleQnA2023} & \CIRCLE\CIRCLE\Circle & A set of 10K telecom-domain QA pairs constructed from standard documents and technical manuals. \\
			& TelecomGPT \cite{TelecomGPT2024} & \CIRCLE\CIRCLE\Circle & Three datasets designed for pre-training (OpenTelecom), instruction tuning (TelecomInstruct), and fine-tuning (TelecomAlign) general-purpose LLMs in the telecommunications domain. \\
			Network optimization benchmark or task data & WirelessMathBench \cite{WirelessMathBench2025XinLi} & \CIRCLE\CIRCLE\CIRCLE & Over 500 natural-language descriptions and mathematical formulations of network optimization problems extracted from papers across multiple domains. \\
			& WireOpt \cite{LLMOptiRA2025} & \CIRCLE\CIRCLE\CIRCLE & Natural-language background descriptions and variable explanations for 100 network optimization problems. \\
			& EVRQ \cite{LLM2024NO_VQR} & \CIRCLE\CIRCLE\CIRCLE & A dataset on electric vehicle routing and charging, designed for intent understanding in LLMs. \\
			Related upstream and downstream task datasets & eAIBench \cite{MobileFoundationFirmware2024} & \CIRCLE\CIRCLE\Circle & A benchmark including 38 mobile AI tasks and 50 datasets across five input modalities. \\
			& Traciverse \cite{WirelessGPT2025Pengcheng} & \CIRCLE\CIRCLE\Circle & A large-scale wireless channel dataset for pre-training general-purpose models in wireless communication. \\
			& Sensiverse \cite{Sensiverse} & \CIRCLE\CIRCLE\Circle & Multi-frequency wireless sensing data from several major cities. \\
		\end{tblr}
		\label{tab_benchmarks}
	\end{table*}
	
	While existing digital twin frameworks have been applied in networking contexts, most current efforts focus on optimizing algorithmic performance within the virtual layer, rather than designing the virtualization infrastructure needed for GenAI-driven evolution. We believe that achieving a next-generation, evolution-capable AI network requires significant advancements in three areas, including efficient collection and representation of high-dimensional, heterogeneous environment data; memory and retrieval mechanisms for distilled knowledge, enabling cross-task transfer and generalization; and fine-grained temporal and spatial simulation of network dynamics to improve model adaptability and resilience. {Fig. \ref{fig_prospect} illustrates our vision for the application scenarios and functional roles of GenAI within the ETSI AI-Core architecture \cite{etsiGReni051}.}

    \subsection{{Empirical Suggestions}}
    {In the context of the evolving landscape of GenAI for network optimization and related fields, we summarize the key milestones and latest advancements. As illustrated in Fig. \ref{fig_suggestion}, we present a synthesized view of the technology's inherent performance advantages and critical limitations. Specifically, positive experiences, such as superiority to discriminative models, robustness to data quality, enhanced cross-task generalization, mathematical reasoning, and proficient task intent extraction, are often underpinned and validated by pioneering works \cite{liang2024diffsg,liang2024gdsg,boosting2025energyGuided,WirelessMathBench2025XinLi,LLMOptiRA2025}. Conversely, several critical limitations have emerged, including limited self-evolution/correction, weak practical cross-task performance compounded by insufficient theoretical basis, output uncertainty and reliability risks related to solution validity, and limited conceptual comprehension. These weaknesses have been identified by critical studies \cite{liang2025pad,DoPhD2025LLM,safeOfflineRL2024,GeneralistRewardModels2025} which often point toward viable future research directions. By distilling these positive and negative empirical experiences, we aim to provide readers and practitioners with direct guidance for both effective deployment and focused future investigation.}

    {Based on our past research experience and the conclusions drawn from related work, we propose three valuable recommendations for effective implementation and the avoidance of common pitfalls.} First, data diversity is key to mitigating long-tail bias. The training dataset must be sufficiently diverse. Otherwise, the model will suffer from long-tail distribution bias, causing the generator to disproportionately produce the most frequent solution category during inference. Second, structural alignment in model selection. Model architectures should align with the problem type. Specifically, structured or sequential optimization problems may find GCNs and Transformer architectures more suitable. Third, handling GDM loss convergence anomalies. In GDMs for network optimization, an occasional observation is a rapid decrease in loss followed by a sharp rise and a prolonged plateau. This behavior often stems from the task's complexity, creating large high-loss planes in the parameter space, and the issue is typically resolved by simply restarting the training process.
	
	\section{Existing Benchmarks and Datasets}\label{sec_bench}
	
	To complement the theoretical and methodological discussions above, we compile and categorize a diverse set of recent benchmarks and datasets relevant to network optimization in Table \ref{tab_benchmarks}. These resources are organized into four categories: (1) {mathematical question-answering (QA) benchmarks that assess symbolic reasoning and mathematical problem-solving ability by accuracy;} (2) {telecom-specific QA datasets that assess LLM-based understanding of domain-specific language and protocols by selection accuracy and contextual accuracy;} (3) {network optimization benchmarks or task datasets that evaluate the model’s ability to interpret and solve optimization-related problems by accuracy, optimality gap and feasibility rate;} and (4) {upstream and downstream task datasets that measure modeling, perception, and interaction capabilities in network environments through multiple metrics, including accuracy, efficiency, and quality.}
	
	To guide the interpretation, we also annotate the relevance to network optimization using a three-level scale. This relevance reflects how directly the dataset supports optimization tasks, ranging from abstract mathematical skill development (e.g., MATH \cite{MATH_bench}, GSM8K \cite{GSM8K}, GSM-Symbolic \cite{GSM-Symbolic}) to datasets constructed explicitly for modeling or solving network optimization problems (e.g., WirelessMathBench \cite{WirelessMathBench2025XinLi}, WireOpt \cite{LLMOptiRA2025}). For instance, EVRQ \cite{LLM2024NO_VQR} bridges intent understanding and energy scheduling, making it highly relevant to optimization, while datasets like eAIBench \cite{MobileFoundationFirmware2024}, Traciverse \cite{WirelessGPT2025Pengcheng} and Sensiverse \cite{Sensiverse} are primarily useful for evaluation or pre-training tasks such as environment modeling or channel estimation.
	
	Collectively, these datasets underscore the growing ecosystem supporting GenAI-driven network optimization. They enable more rigorous benchmarking, facilitate model transferability, and reveal application gaps. However, challenges remain. General-purpose benchmarks often overlook optimization-specific criteria such as feasibility, constraint satisfaction, and convergence behavior. Looking ahead, there is clear demand for more comprehensive, standardized benchmarks to evaluate GenAI models across problem understanding and high-precision decision-making. Future datasets should integrate real-world conditions, include structured constraint annotations, and support multi-stage inference pipelines.
	
	\section{Conclusion}\label{sec_conclusion}
	Generative models hold transformative potential for network optimization, offering more efficient, higher-performing approaches to predicting and optimizing communication systems. {This survey has explored the theories, limitations and visions of GenAI in the field of network optimization. We began by examining relevant developments in the AI community that have inspired GenAI applications in networking, and then we categorized existing GenAI efforts within the network domain. Building on this foundation, we comprehensively reviewed the latest theoretical and implementation advancements of the two most representative GenAI models—GDMs and LPTMs—and further analyzed their roles and capabilities.} 
	
	We have revealed that the foundational theories and initial implementations of GenAI in network optimization are already in place. Both the AI and networking communities are actively pursuing all-in-one, pre-trained models for general-purpose applications. {However, we identified unresolved challenges, including limitations in LPTM's mathematical reasoning and conceptual understanding, limitations of rule learning by GDMs,  and difficulties meeting real-time requirements for network optimization.}
	
	Beyond reviewing existing work, we further have integrated and discuss recent critical perspectives from both the AI and networking communities. These contributions highlight fundamental limitations that are often overshadowed by the impressive surface-level performance of GenAI. Notably, we have presented generalization error bounds between generated and optimal solutions (in one-shot settings) and between generated and optimal policies (in MDP settings), providing theoretical guidance for GenAI model development. We then discussed GenAI’s strengths and weaknesses in network optimization and proposed future research directions from both theoretical and implementation perspectives, emphasizing the importance of bridging the generation-optimization gap and incorporating rule-guided generation ideas. We also provide valuable positive and negative research experiences, hoping to guide other researchers.
	
	In summary, this survey has not only outlined the historical evolution and current landscape of GenAI in network optimization but it also has identified promising future directions by examining the underlying principles and limitations of generative models. Thus, it is hoped that this tutorial will provide readers with a clear understanding of GenAI's core capabilities in this domain and offer guidance for advancing the field based on emerging critical insights.
	
	\vspace{-2mm}
	

	

	\bibliographystyle{IEEEtran}  

\begin{thebibliography}{100}
\providecommand{\url}[1]{#1}
\csname url@samestyle\endcsname
\providecommand{\newblock}{\relax}
\providecommand{\bibinfo}[2]{#2}
\providecommand{\BIBentrySTDinterwordspacing}{\spaceskip=0pt\relax}
\providecommand{\BIBentryALTinterwordstretchfactor}{4}
\providecommand{\BIBentryALTinterwordspacing}{\spaceskip=\fontdimen2\font plus
\BIBentryALTinterwordstretchfactor\fontdimen3\font minus \fontdimen4\font\relax}
\providecommand{\BIBforeignlanguage}[2]{{%
\expandafter\ifx\csname l@#1\endcsname\relax
\typeout{** WARNING: IEEEtran.bst: No hyphenation pattern has been}%
\typeout{** loaded for the language `#1'. Using the pattern for}%
\typeout{** the default language instead.}%
\else
\language=\csname l@#1\endcsname
\fi
#2}}
\providecommand{\BIBdecl}{\relax}
\BIBdecl

\bibitem{roadmap6G}
K.~B. Letaief, W.~Chen, Y.~Shi, J.~Zhang, and Y.-J.~A. Zhang, ``{The Roadmap to 6G: AI Empowered Wireless Networks},'' \emph{IEEE Communications Magazine}, vol.~57, no.~8, pp. 84--90, 2019.

\bibitem{shi2023optimizationIn6G}
Y.~Shi, L.~Lian, Y.~Shi, Z.~Wang, Y.~Zhou, L.~Fu, L.~Bai, J.~Zhang, and W.~Zhang, ``{Machine Learning for Large-Scale Optimization in 6G Wireless Networks},'' \emph{IEEE Communications Surveys \& Tutorials}, vol.~25, no.~4, pp. 2088--2132, 2023.

\bibitem{AI2024MultipleAccessSurvey}
X.~Cao, B.~Yang, K.~Wang, X.~Li, Z.~Yu, C.~Yuen, Y.~Zhang, and Z.~Han, ``{AI-Empowered Multiple Access for 6G: A Survey of Spectrum Sensing, Protocol Designs, and Optimizations},'' \emph{Proceedings of the IEEE}, vol. 112, no.~9, pp. 1264--1302, 2024.

\bibitem{oRAN2024Evolution6G}
Z.~Li, Q.~Wang, Y.~Wang, and T.~Chen, ``{The Architecture of AI and Communication Integration towards 6G: An O-RAN Evolution},'' in \emph{Proceedings of the 30th Annual International Conference on Mobile Computing and Networking}.\hskip 1em plus 0.5em minus 0.4em\relax Association for Computing Machinery, 2024, p. 2329–2334.

\bibitem{mec2017mao}
Y.~Mao, C.~You, J.~Zhang, K.~Huang, and K.~B. Letaief, ``{A Survey on Mobile Edge Computing: The Communication Perspective},'' \emph{IEEE Communications Surveys \& Tutorials}, vol.~19, no.~4, pp. 2322--2358, 2017.

\bibitem{NetProvision2022NSDI}
H.~Sharma, P.~Thakkar, S.~Bharadwaj, R.~Bhagwan, V.~N. Padmanabhan, Y.~Bansal, V.~Kumar, and K.~Voelbel, ``{Optimizing Network Provisioning through Cooperation},'' in \emph{Proceedings of 19th USENIX Symposium on Networked Systems Design and Implementation}.\hskip 1em plus 0.5em minus 0.4em\relax USENIX Association, Apr. 2022, pp. 1179--1194.

\bibitem{Taking5GRAN}
X.~Foukas, B.~Radunovic, M.~Balkwill, and Z.~Lai, ``{Taking 5G RAN Analytics and Control to a New Level},'' in \emph{Proceedings of the 29th Annual International Conference on Mobile Computing and Networking}, 2023.

\bibitem{yb-tmc}
B.~Yang, X.~Cao, J.~Bassey, X.~Li, and L.~Qian, ``{Computation Offloading in Multi-Access Edge Computing: A Multi-Task Learning Approach},'' \emph{IEEE Transactions on Mobile Computing}, vol.~20, no.~9, pp. 2745--2762, 2021.

\bibitem{MADRL2024xURLLCris}
A.~Paul, R.~Allu, K.~Singh, C.-P. Li, and T.~Q. Duong, ``{Hybridized MA-DRL for Serving xURLLC With Cognizable RIS and UAV Integration},'' \emph{IEEE Transactions on Wireless Communications}, vol.~23, no.~10, pp. 15\,507--15\,524, 2024.

\bibitem{DyRANSlicing2021}
W.~Wu, N.~Chen, C.~Zhou, M.~Li, X.~Shen, W.~Zhuang, and X.~Li, ``{Dynamic RAN Slicing for Service-Oriented Vehicular Networks via Constrained Learning},'' \emph{IEEE Journal on Selected Areas in Communications}, vol.~39, no.~7, pp. 2076--2089, 2021.

\bibitem{TanGo2023infocom}
L.~Luo, G.~Zhao, H.~Xu, Z.~Yu, and L.~Xie, ``{TanGo: A Cost Optimization Framework for Tenant Task Placement in Geo-distributed Clouds},'' in \emph{Proceedings of IEEE Conference on Computer Communications}, 2023, pp. 1--10.

\bibitem{Transmitting2023Mobicom}
B.~Tao, M.~Masood, I.~Gupta, and D.~Vasisht, ``{Transmitting, Fast and Slow: Scheduling Satellite Traffic through Space and Time},'' in \emph{Proceedings of the 29th Annual International Conference on Mobile Computing and Networking}, 2023.

\bibitem{Edge2023LEO}
X.~Cao, B.~Yang, Y.~Shen, C.~Yuen, Y.~Zhang, Z.~Han, H.~V. Poor, and L.~Hanzo, ``{Edge-Assisted Multi-Layer Offloading Optimization of LEO Satellite-Terrestrial Integrated Networks},'' \emph{IEEE Journal on Selected Areas in Communications}, vol.~41, no.~2, pp. 381--398, 2023.

\bibitem{VAE2020NIPS}
A.~Vahdat and J.~Kautz, ``{NVAE: a deep hierarchical variational autoencoder},'' in \emph{Proceedings of the 34th International Conference on Neural Information Processing Systems}, 2020.

\bibitem{GAN2024NIPS}
Y.~Huang, A.~Gokaslan, V.~Kuleshov, and J.~Tompkin, ``{The GAN is dead; long live the GAN! A Modern GAN Baseline},'' in \emph{Proceedings of the 38th International Conference on Neural Information Processing Systems}, vol.~37, 2024, pp. 44\,177--44\,215.

\bibitem{NormalizingFlow2021}
G.~Papamakarios, E.~Nalisnick, D.~J. Rezende, S.~Mohamed, and B.~Lakshminarayanan, ``{Normalizing Flows for Probabilistic Modeling and Inference},'' \emph{Journal of Machine Learning Research}, vol.~22, no.~57, pp. 1--64, 2021.

\bibitem{GFlowNets2021Bengio}
E.~Bengio, M.~Jain, M.~Korablyov, D.~Precup, and Y.~Bengio, ``{Flow network based generative models for non-iterative diverse candidate generation},'' in \emph{Proceedings of the 35th International Conference on Neural Information Processing Systems}, 2021.

\bibitem{VAE2022channelEst}
V.~Lauinger, F.~Buchali, and L.~Schmalen, ``{Blind Equalization and Channel Estimation in Coherent Optical Communications Using Variational Autoencoders},'' \emph{IEEE Journal on Selected Areas in Communications}, vol.~40, no.~9, pp. 2529--2539, 2022.

\bibitem{GAN2022IPheader}
Y.~Yin, Z.~Lin, M.~Jin, G.~Fanti, and V.~Sekar, ``{Practical GAN-based synthetic IP header trace generation using NetShare},'' in \emph{Proceedings of the ACM SIGCOMM 2022 Conference}, 2022, p. 458–472.

\bibitem{NF2019NoiseModeling}
A.~Abdelhamed, M.~A. Brubaker, and M.~S. Brown, ``{Noise Flow: Noise Modeling With Conditional Normalizing Flows},'' in \emph{Proceedings of the IEEE/CVF International Conference on Computer Vision (ICCV)}, October 2019.

\bibitem{gflownet2023selection}
S.~Evmorfos, Z.~Xu, and A.~Petropulu, ``{Gflownets for Sensor Selection},'' in \emph{Proceedings of 2023 IEEE 33rd International Workshop on Machine Learning for Signal Processing (MLSP)}, 2023, pp. 1--6.

\bibitem{DDPM}
J.~Ho, A.~Jain, and P.~Abbeel, ``{Denoising Diffusion Probabilistic Models},'' in \emph{Proceedings of the 34th International Conference on Neural Information Processing Systems}, 2020.

\bibitem{GPT3}
T.~Brown, B.~Mann, N.~Ryder, M.~Subbiah, J.~D. Kaplan, P.~Dhariwal, A.~Neelakantan, P.~Shyam, G.~Sastry, A.~Askell, S.~Agarwal, A.~Herbert-Voss, G.~Krueger, T.~Henighan, R.~Child, A.~Ramesh, D.~Ziegler, J.~Wu, C.~Winter, C.~Hesse, M.~Chen, E.~Sigler, M.~Litwin, S.~Gray, B.~Chess, J.~Clark, C.~Berner, S.~McCandlish, A.~Radford, I.~Sutskever, and D.~Amodei, ``{Language Models are Few-Shot Learners},'' in \emph{Proceedings of Advances in Neural Information Processing Systems}, vol.~33, 2020, pp. 1877--1901.

\bibitem{LMM-PM2024}
\BIBentryALTinterwordspacing
X.~Wang, G.~Chen, G.~Qian, P.~Gao, X.-Y. Wei, Y.~Wang, Y.~Tian, and W.~Gao, ``{Large-scale Multi-Modal Pre-trained Models: A Comprehensive Survey},'' 2024. [Online]. Available: \url{https://arxiv.org/abs/2302.10035}
\BIBentrySTDinterwordspacing

\bibitem{survey2_pushing6G}
\BIBentryALTinterwordspacing
Z.~Lin, G.~Qu, Q.~Chen, X.~Chen, Z.~Chen, and K.~Huang, ``{Pushing Large Language Models to the 6G Edge: Vision, Challenges, and Opportunities},'' 2024. [Online]. Available: \url{https://arxiv.org/abs/2309.16739}
\BIBentrySTDinterwordspacing

\bibitem{survey3_shi6G}
Y.~Shi, L.~Lian, Y.~Shi, Z.~Wang, Y.~Zhou, L.~Fu, L.~Bai, J.~Zhang, and W.~Zhang, ``{Machine Learning for Large-Scale Optimization in 6G Wireless Networks},'' \emph{IEEE Communications Surveys \& Tutorials}, vol.~25, no.~4, pp. 2088--2132, 2023.

\bibitem{survey5_iotj}
J.~Guo, M.~Wang, H.~Yin, B.~Song, Y.~Chi, F.~R. Yu, and C.~Yuen, ``{Large Language Models and Artificial Intelligence Generated Content Technologies Meet Communication Networks},'' \emph{IEEE Internet of Things Journal}, vol.~12, no.~2, pp. 1529--1553, 2025.

\bibitem{survey6_LLMTelecom}
H.~Zhou, C.~Hu, Y.~Yuan, Y.~Cui, Y.~Jin, C.~Chen, H.~Wu, D.~Yuan, L.~Jiang, D.~Wu, X.~Liu, C.~Zhang, X.~Wang, and J.~Liu, ``{Large Language Model (LLM) for Telecommunications: A Comprehensive Survey on Principles, Key Techniques, and Opportunities},'' \emph{IEEE Communications Surveys \& Tutorials}, pp. 1--1, 2024.

\bibitem{survey7_GenerativeOpimization}
\BIBentryALTinterwordspacing
C.~Picard, L.~Regenwetter, A.~H. Nobari, A.~Srivastava, and F.~Ahmed, ``{Generative Optimization: A Perspective on AI-Enhanced Problem Solving in Engineering},'' 2024. [Online]. Available: \url{https://arxiv.org/abs/2412.13281}
\BIBentrySTDinterwordspacing

\bibitem{survey12_FromLLM2AutoAIAgent}
\BIBentryALTinterwordspacing
M.~A. Ferrag, N.~Tihanyi, and M.~Debbah, ``{From LLM Reasoning to Autonomous AI Agents: A Comprehensive Review},'' 2025. [Online]. Available: \url{https://arxiv.org/abs/2504.19678}
\BIBentrySTDinterwordspacing

\bibitem{survey13_KnowledgeDrivenDL}
R.~Sun, N.~Cheng, C.~Li, W.~Quan, H.~Zhou, Y.~Wang, W.~Zhang, and X.~Shen, ``{A Comprehensive Survey of Knowledge-Driven Deep Learning for Intelligent Wireless Network Optimization in 6G},'' \emph{IEEE Communications Surveys \& Tutorials}, pp. 1--1, 2025.

\bibitem{survey15_GAIforOptXG}
F.~Khoramnejad and E.~Hossain, ``{Generative AI for the Optimization of Next-Generation Wireless Networks: Basics, State-of-the-Art, and Open Challenges},'' \emph{IEEE Communications Surveys \& Tutorials}, pp. 1--1, 2025.

\bibitem{magazine3_LLMMultiAgent6G}
F.~Jiang, Y.~Peng, L.~Dong, K.~Wang, K.~Yang, C.~Pan, D.~Niyato, and O.~A. Dobre, ``{Large Language Model Enhanced Multi-Agent Systems for 6G Communications},'' \emph{IEEE Wireless Communications}, vol.~31, no.~6, pp. 48--55, 2024.

\bibitem{magazine4_LLMAgent6GPerception}
M.~Xu, D.~Niyato, J.~Kang, Z.~Xiong, S.~Mao, Z.~Han, D.~I. Kim, and K.~B. Letaief, ``{When Large Language Model Agents Meet 6G Networks: Perception, Grounding, and Alignment},'' \emph{IEEE Wireless Communications}, vol.~31, no.~6, pp. 63--71, 2024.

\bibitem{survey10_InternetOfAgent}
\BIBentryALTinterwordspacing
Y.~Wang, S.~Guo, Y.~Pan, Z.~Su, F.~Chen, T.~H. Luan, P.~Li, J.~Kang, and D.~Niyato, ``{Internet of Agents: Fundamentals, Applications, and Challenges},'' 2025. [Online]. Available: \url{https://arxiv.org/abs/2505.07176}
\BIBentrySTDinterwordspacing

\bibitem{survey11_AGINative}
W.~Saad, O.~Hashash, C.~K. Thomas, C.~Chaccour, M.~Debbah, N.~Mandayam, and Z.~Han, ``{Artificial General Intelligence (AGI)-Native Wireless Systems: A Journey Beyond 6G},'' \emph{Proceedings of the IEEE}, pp. 1--39, 2025.

\bibitem{magazine1_AgentStandard}
S.~T. Arzo, D.~Scotece, R.~Bassoli, F.~Granelli, L.~Foschini, and F.~H. Fitzek, ``{A New Agent-Based Intelligent Network Architecture},'' \emph{IEEE Communications Standards Magazine}, vol.~6, no.~4, pp. 74--79, 2022.

\bibitem{magazine2_MobileAIAgent6G}
Z.~Chen, Q.~Sun, N.~Li, X.~Li, Y.~Wang, and C.-L. I, ``{Enabling Mobile AI Agent in 6G Era: Architecture and Key Technologies},'' \emph{IEEE Network}, vol.~38, no.~5, pp. 66--75, 2024.

\bibitem{survey1_foundation}
F.~Le, M.~Srivatsa, R.~Ganti, and V.~Sekar, ``{Rethinking data-driven networking with foundation models: challenges and opportunities},'' in \emph{Proceedings of the 21st ACM Workshop on Hot Topics in Networks}.\hskip 1em plus 0.5em minus 0.4em\relax Association for Computing Machinery, 2022, p. 188–197.

\bibitem{survey4_du}
H.~Du, R.~Zhang, Y.~Liu, J.~Wang, Y.~Lin, Z.~Li, D.~Niyato, J.~Kang, Z.~Xiong, S.~Cui, B.~Ai, H.~Zhou, and D.~I. Kim, ``{Enhancing Deep Reinforcement Learning: A Tutorial on Generative Diffusion Models in Network Optimization},'' \emph{IEEE Communications Surveys \& Tutorials}, vol.~26, no.~4, pp. 2611--2646, 2024.

\bibitem{survey8_jiang}
\BIBentryALTinterwordspacing
F.~Jiang, C.~Pan, L.~Dong, K.~Wang, M.~Debbah, D.~Niyato, and Z.~Han, ``{A Comprehensive Survey of Large AI Models for Future Communications: Foundations, Applications and Challenges},'' 2025. [Online]. Available: \url{https://arxiv.org/abs/2505.03556}
\BIBentrySTDinterwordspacing

\bibitem{survey9_wirelessLargeAI}
\BIBentryALTinterwordspacing
F.~Zhu, X.~Wang, X.~Li, M.~Zhang, Y.~Chen, C.~Huang, Z.~Yang, X.~Chen, Z.~Zhang, R.~Jin, Y.~Huang, W.~Feng, T.~Yang, B.~Bai, F.~Gao, K.~Yang, Y.~Liu, S.~Muhaidat, C.~Yuen, K.~Huang, K.-K. Wong, D.~Niyato, and M.~Debbah, ``{Wireless Large AI Model: Shaping the AI-Native Future of 6G and Beyond},'' 2025. [Online]. Available: \url{https://arxiv.org/abs/2504.14653}
\BIBentrySTDinterwordspacing

\bibitem{survey14_FromLAM2Agentic}
\BIBentryALTinterwordspacing
F.~Jiang, C.~Pan, L.~Dong, K.~Wang, O.~A. Dobre, and M.~Debbah, ``{From Large AI Models to Agentic AI: A Tutorial on Future Intelligent Communications},'' 2025. [Online]. Available: \url{https://arxiv.org/abs/2505.22311}
\BIBentrySTDinterwordspacing

\bibitem{liang2025pad}
\BIBentryALTinterwordspacing
R.~Liang, B.~Yang, P.~Chen, X.~Cao, Z.~Yu, H.~V. Poor, and C.~Yuen, ``{Cross-Problem Solving for Network Optimization: Is Problem-Aware Learning the Key?}'' 2025. [Online]. Available: \url{https://arxiv.org/abs/2505.05067}
\BIBentrySTDinterwordspacing

\bibitem{WirelessMathBench2025XinLi}
\BIBentryALTinterwordspacing
X.~Li, M.~Liu, L.~Wei, J.~An, M.~Debbah, and C.~Yuen, ``{WirelessMathBench: A Mathematical Modeling Benchmark for LLMs in Wireless Communications},'' 2025. [Online]. Available: \url{https://arxiv.org/abs/2505.14354}
\BIBentrySTDinterwordspacing

\bibitem{GSM-Symbolic}
I.~Mirzadeh, K.~Alizadeh, H.~Shahrokhi, O.~Tuzel, S.~Bengio, and M.~Farajtabar, ``{GSM-Symbolic: Understanding the Limitations of Mathematical Reasoning in Large Language Models},'' in \emph{Proceedings of International Conference on Learning Representations}, 2025.

\bibitem{DRL2025ReasoningLLM}
\BIBentryALTinterwordspacing
Y.~Yue, Z.~Chen, R.~Lu, A.~Zhao, Z.~Wang, Y.~Yue, S.~Song, and G.~Huang, ``{Does Reinforcement Learning Really Incentivize Reasoning Capacity in LLMs Beyond the Base Model?}'' 2025. [Online]. Available: \url{https://arxiv.org/abs/2504.13837}
\BIBentrySTDinterwordspacing

\bibitem{DoPhD2025LLM}
\BIBentryALTinterwordspacing
Y.~Yan, Y.~Lu, R.~Xu, and Z.~Lan, ``{Do PhD-level LLMs Truly Grasp Elementary Addition? Probing Rule Learning vs. Memorization in Large Language Models},'' 2025. [Online]. Available: \url{https://arxiv.org/abs/2504.05262}
\BIBentrySTDinterwordspacing

\bibitem{ReasoningBeyondLimits2025Debbah}
\BIBentryALTinterwordspacing
M.~A. Ferrag, N.~Tihanyi, and M.~Debbah, ``{Reasoning Beyond Limits: Advances and Open Problems for LLMs},'' 2025. [Online]. Available: \url{https://arxiv.org/abs/2503.22732}
\BIBentrySTDinterwordspacing

\bibitem{FaithAndFate2023}
N.~Dziri, X.~Lu, M.~Sclar, X.~L. Li, L.~Jiang, B.~Y. Lin, P.~West, C.~Bhagavatula, R.~Le~Bras, J.~D. Hwang, S.~Sanyal, S.~Welleck, X.~Ren, A.~Ettinger, Z.~Harchaoui, and Y.~Choi, ``{Faith and fate: limits of transformers on compositionality},'' in \emph{Proceedings of the 37th International Conference on Neural Information Processing Systems}, 2023.

\bibitem{Representational2023TFM}
C.~Sanford, D.~Hsu, and M.~Telgarsky, ``{Representational strengths and limitations of transformers},'' in \emph{Proceedings of the 37th International Conference on Neural Information Processing Systems}, 2023.

\bibitem{onMemorizationOfDiff}
\BIBentryALTinterwordspacing
X.~Gu, C.~Du, T.~Pang, C.~Li, M.~Lin, and Y.~Wang, ``{On Memorization in Diffusion Models},'' 2025. [Online]. Available: \url{https://arxiv.org/abs/2310.02664}
\BIBentrySTDinterwordspacing

\bibitem{OnGeneralization2023Diffusion}
P.~Li, Z.~Li, H.~Zhang, and J.~Bian, ``{On the Generalization Properties of Diffusion Models},'' in \emph{Proceedings of Advances in Neural Information Processing Systems}, vol.~36, 2023, pp. 2097--2127.

\bibitem{LLMOptiRA2025}
\BIBentryALTinterwordspacing
X.~Peng, Y.~Liu, Y.~Cang, C.~Cao, and M.~Chen, ``{LLM-OptiRA: LLM-Driven Optimization of Resource Allocation for Non-Convex Problems in Wireless Communications},'' 2025. [Online]. Available: \url{https://arxiv.org/abs/2505.02091}
\BIBentrySTDinterwordspacing

\bibitem{BLO2023Introduction}
Y.~Zhang, P.~Khanduri, I.~Tsaknakis, Y.~Yao, M.~Hong, and S.~Liu, ``{An Introduction to Bilevel Optimization: Foundations and applications in signal processing and machine learning},'' \emph{IEEE Signal Processing Magazine}, vol.~41, no.~1, pp. 38--59, 2024.

\bibitem{StochasticOptimization}
J.~Zhu and S.~Wang, ``{QoS-Guaranteed Resource Allocation in Mobile Communications: A Stochastic Network Calculus Approach},'' \emph{IEEE/ACM Transactions on Networking}, vol.~31, no.~2, pp. 1234--1245, 2023.

\bibitem{Unet2015}
O.~Ronneberger, P.~Fischer, and T.~Brox, ``{U-Net: Convolutional Networks for Biomedical Image Segmentation},'' in \emph{Proceedings of Medical Image Computing and Computer-Assisted Intervention -- MICCAI 2015}.\hskip 1em plus 0.5em minus 0.4em\relax Springer International Publishing, 2015, pp. 234--241.

\bibitem{ClassifierGuidance}
P.~Dhariwal and A.~Nichol, ``{Diffusion models beat GANs on image synthesis},'' in \emph{Proceedings of the 35th International Conference on Neural Information Processing Systems}, 2021.

\bibitem{ClassifierFree}
\BIBentryALTinterwordspacing
J.~Ho and T.~Salimans, ``{Classifier-Free Diffusion Guidance},'' 2022. [Online]. Available: \url{https://arxiv.org/abs/2207.12598}
\BIBentrySTDinterwordspacing

\bibitem{DDIM2022song}
\BIBentryALTinterwordspacing
J.~Song, C.~Meng, and S.~Ermon, ``{Denoising Diffusion Implicit Models},'' 2022. [Online]. Available: \url{https://arxiv.org/abs/2010.02502}
\BIBentrySTDinterwordspacing

\bibitem{DPM_solver}
C.~Lu, Y.~Zhou, F.~Bao, J.~Chen, C.~Li, and J.~Zhu, ``{DPM-solver: a fast ODE solver for diffusion probabilistic model sampling in around 10 steps},'' in \emph{Proceedings of the 36th International Conference on Neural Information Processing Systems}, 2022.

\bibitem{GaussianSolvers2023NIPS}
H.~Guo, C.~Lu, F.~Bao, T.~Pang, S.~Yan, C.~Du, and C.~Li, ``{Gaussian mixture solvers for diffusion models},'' in \emph{Proceedings of the 37th International Conference on Neural Information Processing Systems}, 2023.

\bibitem{ConsistencyModel2023}
Y.~Song, P.~Dhariwal, M.~Chen, and I.~Sutskever, ``{Consistency models},'' in \emph{Proceedings of the 40th International Conference on Machine Learning}.\hskip 1em plus 0.5em minus 0.4em\relax JMLR.org, 2023.

\bibitem{ParallelGDM2023NIPS}
A.~Shih, S.~Belkhale, S.~Ermon, D.~Sadigh, and N.~Anari, ``{Parallel sampling of diffusion models},'' in \emph{Proceedings of the 37th International Conference on Neural Information Processing Systems}, 2023.

\bibitem{StableLatentDiffusion}
R.~Rombach, A.~Blattmann, D.~Lorenz, P.~Esser, and B.~Ommer, ``{High-Resolution Image Synthesis with Latent Diffusion Models},'' in \emph{Proceedings of 2022 IEEE/CVF Conference on Computer Vision and Pattern Recognition (CVPR)}, 2022, pp. 10\,674--10\,685.

\bibitem{ControlNet}
L.~Zhang, A.~Rao, and M.~Agrawala, ``{Adding Conditional Control to Text-to-Image Diffusion Models},'' in \emph{Proceedings of 2023 IEEE/CVF International Conference on Computer Vision (ICCV)}, 2023, pp. 3813--3824.

\bibitem{DiscreteDiffusion}
J.~Austin, D.~D. Johnson, J.~Ho, D.~Tarlow, and R.~van~den Berg, ``{Structured denoising diffusion models in discrete state-spaces},'' in \emph{Proceedings of the 35th International Conference on Neural Information Processing Systems}, 2021.

\bibitem{GraphTransformer2019}
S.~Yun, M.~Jeong, R.~Kim, J.~Kang, and H.~J. Kim, ``{Graph Transformer Networks},'' in \emph{Proceedings of Advances in Neural Information Processing Systems}, vol.~32, 2019.

\bibitem{DiT_Xie}
W.~Peebles and S.~Xie, ``{Scalable Diffusion Models with Transformers},'' in \emph{Proceedings of the IEEE/CVF International Conference on Computer Vision (ICCV)}, October 2023, pp. 4195--4205.

\bibitem{LLaMA}
\BIBentryALTinterwordspacing
H.~Touvron, T.~Lavril, G.~Izacard, X.~Martinet, M.-A. Lachaux, T.~Lacroix, B.~Rozière, N.~Goyal, E.~Hambro, F.~Azhar, A.~Rodriguez, A.~Joulin, E.~Grave, and G.~Lample, ``{LLaMA: Open and Efficient Foundation Language Models},'' 2023. [Online]. Available: \url{https://arxiv.org/abs/2302.13971}
\BIBentrySTDinterwordspacing

\bibitem{Gemini}
\BIBentryALTinterwordspacing
Gemini-Team, R.~Anil, and S.~B. et~al, ``{Gemini: A Family of Highly Capable Multimodal Models},'' 2025. [Online]. Available: \url{https://arxiv.org/abs/2312.11805}
\BIBentrySTDinterwordspacing

\bibitem{DeepSeekV3}
\BIBentryALTinterwordspacing
C.~Zhao, C.~Deng, C.~Ruan, D.~Dai, H.~Gao, J.~Li, L.~Zhang, P.~Huang, S.~Zhou, S.~Ma, W.~Liang, Y.~He, Y.~Wang, Y.~Liu, and Y.~X. Wei, ``{Insights into DeepSeek-V3: Scaling Challenges and Reflections on Hardware for AI Architectures},'' 2025. [Online]. Available: \url{https://arxiv.org/abs/2505.09343}
\BIBentrySTDinterwordspacing

\bibitem{DeepSeek-R1}
\BIBentryALTinterwordspacing
DeepSeek-AI, D.~Guo, and D.~Y. et~al, ``{DeepSeek-R1: Incentivizing Reasoning Capability in LLMs via Reinforcement Learning},'' 2025. [Online]. Available: \url{https://arxiv.org/abs/2501.12948}
\BIBentrySTDinterwordspacing

\bibitem{rStarMath2025}
\BIBentryALTinterwordspacing
X.~Guan, L.~L. Zhang, Y.~Liu, N.~Shang, Y.~Sun, Y.~Zhu, F.~Yang, and M.~Yang, ``{rStar-Math: Small LLMs Can Master Math Reasoning with Self-Evolved Deep Thinking},'' 2025. [Online]. Available: \url{https://arxiv.org/abs/2501.04519}
\BIBentrySTDinterwordspacing

\bibitem{DeepSeekProverV2}
\BIBentryALTinterwordspacing
Z.~Z. Ren, Z.~Shao, J.~Song, H.~Xin, H.~Wang, W.~Zhao, L.~Zhang, Z.~Fu, Q.~Zhu, D.~Yang, Z.~F. Wu, Z.~Gou, S.~Ma, H.~Tang, Y.~Liu, W.~Gao, D.~Guo, and C.~Ruan, ``{DeepSeek-Prover-V2: Advancing Formal Mathematical Reasoning via Reinforcement Learning for Subgoal Decomposition},'' 2025. [Online]. Available: \url{https://arxiv.org/abs/2504.21801}
\BIBentrySTDinterwordspacing

\bibitem{FormalMath2024Intrinsic}
G.~Poesia, D.~Broman, N.~Haber, and N.~D. Goodman, ``{Learning formal mathematics from intrinsic motivation},'' in \emph{Proceedings of the 38th International Conference on Neural Information Processing Systems}, 2025.

\bibitem{LLMOPT2025}
\BIBentryALTinterwordspacing
C.~Jiang, X.~Shu, H.~Qian, X.~Lu, J.~Zhou, A.~Zhou, and Y.~Yu, ``{LLMOPT: Learning to Define and Solve General Optimization Problems from Scratch},'' 2025. [Online]. Available: \url{https://arxiv.org/abs/2410.13213}
\BIBentrySTDinterwordspacing

\bibitem{LLaDA2025}
S.~Nie, F.~Zhu, Z.~You, X.~Zhang, J.~Ou, J.~Hu, J.~Zhou, Y.~Lin, J.-R. Wen, and C.~Li, ``{Large Language Diffusion Models},'' in \emph{Proceedings of International Conference on Learning Representations}, 2025.

\bibitem{MMaDA}
\BIBentryALTinterwordspacing
L.~Yang, Y.~Tian, B.~Li, X.~Zhang, K.~Shen, Y.~Tong, and M.~Wang, ``{MMaDA: Multimodal Large Diffusion Language Models},'' 2025. [Online]. Available: \url{https://arxiv.org/abs/2505.15809}
\BIBentrySTDinterwordspacing

\bibitem{V-JEPA-2}
\BIBentryALTinterwordspacing
M.~Assran, A.~Bardes, and D.~F. et~al, ``{V-JEPA 2: Self-Supervised Video Models Enable Understanding, Prediction and Planning},'' 2025. [Online]. Available: \url{https://arxiv.org/abs/2506.09985}
\BIBentrySTDinterwordspacing

\bibitem{ModelingAgent2025ModelingBench}
\BIBentryALTinterwordspacing
C.~Qian, H.~Du, H.~Wang, X.~Chen, Y.~Zhang, A.~Sil, C.~Zhai, K.~McKeown, and H.~Ji, ``{ModelingAgent: Bridging LLMs and Mathematical Modeling for Real-World Challenges},'' 2025. [Online]. Available: \url{https://arxiv.org/abs/2505.15068}
\BIBentrySTDinterwordspacing

\bibitem{LLMasOptimizer2024}
\BIBentryALTinterwordspacing
C.~Yang, X.~Wang, Y.~Lu, H.~Liu, Q.~V. Le, D.~Zhou, and X.~Chen, ``{Large Language Models as Optimizers},'' 2024. [Online]. Available: \url{https://arxiv.org/abs/2309.03409}
\BIBentrySTDinterwordspacing

\bibitem{CLIP}
A.~Radford, J.~W. Kim, C.~Hallacy, A.~Ramesh, G.~Goh, S.~Agarwal, G.~Sastry, A.~Askell, P.~Mishkin, J.~Clark, G.~Krueger, and I.~Sutskever, ``{Learning Transferable Visual Models From Natural Language Supervision},'' in \emph{Proceedings of the 38th International Conference on Machine Learning}, vol. 139.\hskip 1em plus 0.5em minus 0.4em\relax PMLR, 18--24 Jul 2021, pp. 8748--8763.

\bibitem{SegmentAnything}
A.~Kirillov, E.~Mintun, N.~Ravi, H.~Mao, C.~Rolland, L.~Gustafson, T.~Xiao, S.~Whitehead, A.~C. Berg, W.-Y. Lo, P.~Dollár, and R.~Girshick, ``{Segment Anything},'' in \emph{Proceedings of 2023 IEEE/CVF International Conference on Computer Vision (ICCV)}, 2023, pp. 3992--4003.

\bibitem{CARTE}
M.~J. Kim, L.~Grinsztajn, and G.~Varoquaux, ``{CARTE: pretraining and transfer for tabular learning},'' in \emph{Proceedings of the 41st International Conference on Machine Learning}, 2024.

\bibitem{GraphEdit}
\BIBentryALTinterwordspacing
Z.~Guo, L.~Xia, Y.~Yu, Y.~Wang, K.~Lu, Z.~Huang, and C.~Huang, ``{GraphEdit: Large Language Models for Graph Structure Learning},'' 2025. [Online]. Available: \url{https://arxiv.org/abs/2402.15183}
\BIBentrySTDinterwordspacing

\bibitem{any2any}
Z.~Tang, Z.~Yang, C.~Zhu, M.~Zeng, and M.~Bansal, ``{Any-to-Any Generation via Composable Diffusion},'' in \emph{Proceedings of the 37th International Conference on Neural Information Processing Systems}, vol.~36, 2023, pp. 16\,083--16\,099.

\bibitem{sun2023difusco}
Z.~Sun and Y.~Yang, ``{DIFUSCO: Graph-based Diffusion Solvers for Combinatorial Optimization},'' in \emph{Proceedings of Neural Information Processing Systems}, vol.~36, 2023, pp. 3706--3731.

\bibitem{accelerating2023afterDIFUSCO}
\BIBentryALTinterwordspacing
J.~Huang, Z.~Sun, and Y.~Yang, ``{Accelerating Diffusion-based Combinatorial Optimization Solvers by Progressive Distillation},'' 2023. [Online]. Available: \url{https://arxiv.org/abs/2308.06644}
\BIBentrySTDinterwordspacing

\bibitem{li2023t2t}
Y.~Li, J.~Guo, R.~Wang, and J.~Yan, ``{T2T: From Distribution Learning in Training to Gradient Search in Testing for Combinatorial Optimization},'' in \emph{Proceedings of Neural Information Processing Systems}, vol.~36, 2023, pp. 50\,020--50\,040.

\bibitem{unsupervised2024CO}
S.~Sanokowski, S.~Hochreiter, and S.~Lehner, ``{A diffusion model framework for unsupervised neural combinatorial optimization},'' in \emph{Proceedings of the 41st International Conference on Machine Learning}, 2024.

\bibitem{boosting2025energyGuided}
\BIBentryALTinterwordspacing
H.~Lei, K.~Zhou, Y.~Li, Z.~Chen, and F.~Farnia, ``{Boosting Generalization in Diffusion-Based Neural Combinatorial Solver via Energy-guided Sampling},'' 2025. [Online]. Available: \url{https://arxiv.org/abs/2502.12188}
\BIBentrySTDinterwordspacing

\bibitem{DIMES2022}
R.~Qiu, Z.~Sun, and Y.~Yang, ``{DIMES: a differentiable meta solver for combinatorial optimization problems},'' in \emph{Proceedings of the 36th International Conference on Neural Information Processing Systems}, 2022.

\bibitem{Diffusion_robot_Survey}
M.~Song, X.~Deng, Z.~Zhou, J.~Wei, W.~Guan, and L.~Nie, ``{A survey on diffusion policy for robotic manipulation: Taxonomy, analysis, and future directions},'' \emph{Authorea Preprints}, 2025.

\bibitem{GDM_DRL_survey}
\BIBentryALTinterwordspacing
Z.~Zhu, H.~Zhao, H.~He, Y.~Zhong, S.~Zhang, H.~Guo, T.~Chen, and W.~Zhang, ``{Diffusion Models for Reinforcement Learning: A Survey},'' 2024. [Online]. Available: \url{https://arxiv.org/abs/2311.01223}
\BIBentrySTDinterwordspacing

\bibitem{isConditional2023}
A.~Ajay, Y.~Du, A.~Gupta, J.~Tenenbaum, T.~Jaakkola, and P.~Agrawal, ``{Is Conditional Generative Modeling all you need for Decision-Making?}'' in \emph{Proceedings of International Conference on Learning Representations}, 2023.

\bibitem{IDQL2023}
\BIBentryALTinterwordspacing
P.~Hansen-Estruch, I.~Kostrikov, M.~Janner, J.~G. Kuba, and S.~Levine, ``{IDQL: Implicit Q-Learning as an Actor-Critic Method with Diffusion Policies},'' 2023. [Online]. Available: \url{https://arxiv.org/abs/2304.10573}
\BIBentrySTDinterwordspacing

\bibitem{DiffusionQL2023}
Z.~Wang, J.~J. Hunt, and M.~Zhou, ``{Diffusion Policies as an Expressive Policy Class for Offline Reinforcement Learning},'' in \emph{Proceedings of International Conference on Learning Representations}, 2023.

\bibitem{he2023MTDIFF}
H.~He, C.~Bai, K.~Xu, Z.~Yang, W.~Zhang, D.~Wang, B.~Zhao, and X.~Li, ``{Diffusion Model is an Effective Planner and Data Synthesizer for Multi-Task Reinforcement Learning},'' in \emph{Proceedings of Advances in Neural Information Processing Systems}, vol.~36, 2023, pp. 64\,896--64\,917.

\bibitem{AdaptDiffuser2023}
\BIBentryALTinterwordspacing
Z.~Liang, Y.~Mu, M.~Ding, F.~Ni, M.~Tomizuka, and P.~Luo, ``{AdaptDiffuser: Diffusion Models as Adaptive Self-evolving Planners},'' 2023. [Online]. Available: \url{https://arxiv.org/abs/2302.01877}
\BIBentrySTDinterwordspacing

\bibitem{EDP2023NIPS}
B.~Kang, X.~Ma, C.~Du, T.~Pang, and S.~Yan, ``{Efficient Diffusion Policies For Offline Reinforcement Learning},'' in \emph{Proceedings of Advances in Neural Information Processing Systems}, vol.~36, 2023, pp. 67\,195--67\,212.

\bibitem{DiffusionPolicy2024OOD}
S.~E. Ada, E.~Oztop, and E.~Ugur, ``{Diffusion Policies for Out-of-Distribution Generalization in Offline Reinforcement Learning},'' \emph{IEEE Robotics and Automation Letters}, vol.~9, no.~4, pp. 3116--3123, 2024.

\bibitem{safeOfflineRL2024}
Y.~Zheng, J.~Li, D.~Yu, Y.~Yang, S.~E. Li, X.~Zhan, and J.~Liu, ``{Safe Offline Reinforcement Learning with Feasibility-Guided Diffusion Model},'' in \emph{Proceedings of International Conference on Learning Representations}, 2024.

\bibitem{DiffuserLite2024NIPS}
Z.~Dong, J.~Hao, Y.~Yuan, F.~Ni, Y.~Wang, P.~Li, and Y.~Zheng, ``{DiffuserLite: Towards Real-time Diffusion Planning},'' in \emph{Proceedings of Advances in Neural Information Processing Systems}, vol.~37, 2024, pp. 122\,556--122\,583.

\bibitem{taskAgnosticDiffusionPlanner2025}
\BIBentryALTinterwordspacing
C.~Fan, C.~Bai, Z.~Shan, H.~He, Y.~Zhang, and Z.~Wang, ``{Task-agnostic Pre-training and Task-guided Fine-tuning for Versatile Diffusion Planner},'' 2025. [Online]. Available: \url{https://arxiv.org/abs/2409.19949}
\BIBentrySTDinterwordspacing

\bibitem{liang2024diffsg}
R.~Liang, B.~Yang, Z.~Yu, B.~Guo, X.~Cao, M.~Debbah, H.~V. Poor, and C.~Yuen, ``{DiffSG: A Generative Solver for Network Optimization with Diffusion Model},'' \emph{IEEE Communications Magazine}, vol.~63, no.~6, pp. 16--24, 2025.

\bibitem{liang2024iotj}
R.~Liang, B.~Yang, P.~Chen, X.~Li, Y.~Xue, Z.~Yu, X.~Cao, Y.~Zhang, M.~Debbah, H.~V.~Poor, and C.~Yuen, ``{Diffusion Models as Network Optimizers: Explorations and Analysis},'' \emph{IEEE Internet of Things Journal}, vol.~12, no.~10, pp. 13\,183--13\,193, 2025.

\bibitem{SG_2024DFSS}
J.~Wang, H.~Du, Y.~Liu, G.~Sun, D.~Niyato, S.~Mao, D.~In~Kim, and X.~Shen, ``{Generative AI Based Secure Wireless Sensing for ISAC Networks},'' \emph{IEEE Transactions on Information Forensics and Security}, vol.~20, pp. 5195--5210, 2025.

\bibitem{liang2024gdsg}
R.~Liang, B.~Yang, P.~Chen, X.~Cao, Z.~Yu, M.~Debbah, D.~Niyato, H.~V. Poor, and C.~Yuen, ``{GDSG: Graph Diffusion-based Solution Generator for Optimization Problems in MEC Networks},'' \emph{IEEE Transactions on Mobile Computing}, pp. 1--14, 2025.

\bibitem{DEITSP2025KDD}
M.~Wang, Y.~Zhou, Z.~Cao, Y.~Xiao, X.~Wu, W.~Pang, Y.~Jiang, H.~Yang, P.~Zhao, and Y.~Li, ``{An Efficient Diffusion-based Non-Autoregressive Solver for Traveling Salesman Problem},'' in \emph{Proceedings of the 31st ACM SIGKDD Conference on Knowledge Discovery and Data Mining V.1}.\hskip 1em plus 0.5em minus 0.4em\relax Association for Computing Machinery, 2025, p. 1469–1480.

\bibitem{GDPlan2025TON}
N.~Kan, S.~Yan, J.~Zou, W.~Dai, X.~Gao, C.~Li, and H.~Xiong, ``{GDPlan: Generative Network Planning via Graph Diffusion Model},'' \emph{IEEE Transactions on Networking}, pp. 1--16, 2025.

\bibitem{D2SAC2024}
H.~Du, Z.~Li, D.~Niyato, J.~Kang, Z.~Xiong, H.~Huang, and S.~Mao, ``{Diffusion-Based Reinforcement Learning for Edge-Enabled AI-Generated Content Services},'' \emph{IEEE Transactions on Mobile Computing}, vol.~23, no.~9, pp. 8902--8918, 2024.

\bibitem{GDMPolicy2025WirelessComm}
G.~Sun, W.~Xie, D.~Niyato, F.~Mei, J.~Kang, H.~Du, and S.~Mao, ``{Generative AI for Deep Reinforcement Learning: Framework, Analysis, and Use Cases},'' \emph{IEEE Wireless Communications}, pp. 1--10, 2025.

\bibitem{duD2SAC_ContractDesign2023jsac}
H.~Du, J.~Wang, D.~Niyato, J.~Kang, Z.~Xiong, and D.~I. Kim, ``{AI-Generated Incentive Mechanism and Full-Duplex Semantic Communications for Information Sharing},'' \emph{IEEE Journal on Selected Areas in Communications}, vol.~41, no.~9, pp. 2981--2997, 2023.

\bibitem{D2SAC2025LLMasUserFeature}
H.~Du, R.~Zhang, D.~Niyato, J.~Kang, Z.~Xiong, and D.~I. Kim, ``{Reinforcement Learning with Large Language Models (LLMs) Interaction for Network Services},'' in \emph{2024 International Conference on Computing, Networking and Communications (ICNC)}, 2024, pp. 799--803.

\bibitem{D2SAC2025TWC}
P.~Ning, H.~Wang, T.~Tang, J.~Zhang, H.~Du, D.~Niyato, and F.~Richard~Yu, ``{Diffusion-Based Deep Reinforcement Learning for Resource Management in Connected Construction Equipment Networks: A Hierarchical Framework},'' \emph{IEEE Transactions on Wireless Communications}, vol.~24, no.~4, pp. 2847--2861, 2025.

\bibitem{GroundAirSpace2024MultiAgentGDM}
Y.~Zhao, C.~H. Liu, T.~Yi, G.~Li, and D.~Wu, ``{Energy-Efficient Ground-Air-Space Vehicular Crowdsensing by Hierarchical Multi-Agent Deep Reinforcement Learning With Diffusion Models},'' \emph{IEEE Journal on Selected Areas in Communications}, vol.~42, no.~12, pp. 3566--3580, 2024.

\bibitem{GDMPolicy2025VehicleEmbodied}
Y.~Zhong, J.~Kang, J.~Wen, D.~Ye, J.~Nie, D.~Niyato, X.~Gao, and S.~Xie, ``{Generative Diffusion-Based Contract Design for Efficient AI Twin Migration in Vehicular Embodied AI Networks},'' \emph{IEEE Transactions on Mobile Computing}, vol.~24, no.~5, pp. 4573--4588, 2025.

\bibitem{GDMPolicy2025MetaverseIoT}
W.~Zeng, J.~Zheng, L.~Gao, J.~Niu, J.~Ren, H.~Wang, R.~Cao, and S.~Ji, ``{Generative AI-Aided Multimodal Parallel Offloading for AIGC Metaverse Service in IoT Networks},'' \emph{IEEE Internet of Things Journal}, vol.~12, no.~10, pp. 13\,273--13\,285, 2025.

\bibitem{TVT2025MetaverseGDMPolicy}
Z.~Zhang, J.~Wang, J.~Chen, H.~Fu, Z.~Tong, and C.~Jiang, ``{Diffusion-Based Reinforcement Learning for Cooperative Offloading and Resource Allocation in Multi-UAV Assisted Edge-Enabled Metaverse},'' \emph{IEEE Transactions on Vehicular Technology}, pp. 1--13, 2025.

\bibitem{GDMPolicy2025LLMQoS}
Z.~Yao, Z.~Tang, W.~Yang, and W.~Jia, ``{Enhancing LLM QoS through Cloud-Edge Collaboration: A Diffusion-based Multi-Agent Reinforcement Learning Approach},'' \emph{IEEE Transactions on Services Computing}, pp. 1--17, 2025.

\bibitem{GDMPolicy2025DownlinkSemantic}
D.~Liu, Y.~Liu, L.~Zhang, A.~Hafid, L.~Khoukhi, M.~Li, Y.~Liu, and Y.~Dong, ``{Generative AI-Driven Incentive Mechanism for Semantic Communications in RSMA Networks},'' \emph{IEEE Transactions on Cognitive Communications and Networking}, pp. 1--1, 2025.

\bibitem{SecureBeamIRS2025GDMPolicy}
J.~Zhang, Z.~Liu, X.~Feng, H.~Yang, and S.~Liang, ``{Enhanced Secure Beamforming for IRS-Assisted IoT Communication Using a Generative-Diffusion-Model-Enabled Optimization Approach},'' \emph{IEEE Internet of Things Journal}, vol.~12, no.~10, pp. 13\,398--13\,414, 2025.

\bibitem{GDMPolicy2025LowAltitudeEdgeInference}
Y.~Fu, P.~Qin, Y.~Wang, L.~Chen, M.~Li, and X.~Zhao, ``{Over-the-Air Edge Inference for Low-Altitude Airspace: Generative AI-Aided Multi-Task Batching and Beamforming Design},'' \emph{IEEE Transactions on Communications}, pp. 1--1, 2025.

\bibitem{GDMSurvey2024CoveredBBO}
\BIBentryALTinterwordspacing
M.~Chen, S.~Mei, J.~Fan, and M.~Wang, ``{An Overview of Diffusion Models: Applications, Guided Generation, Statistical Rates and Optimization},'' 2024. [Online]. Available: \url{https://arxiv.org/abs/2404.07771}
\BIBentrySTDinterwordspacing

\bibitem{BBO2023NIPS}
H.~Yuan, K.~Huang, C.~Ni, M.~Chen, and M.~Wang, ``{Reward-Directed Conditional Diffusion: Provable Distribution Estimation and Reward Improvement},'' in \emph{Proceedings of Advances in Neural Information Processing Systems}, vol.~36, 2023, pp. 60\,599--60\,635.

\bibitem{BBO2023PMLR}
S.~Krishnamoorthy, S.~M. Mashkaria, and A.~Grover, ``{Diffusion Models for Black-Box Optimization},'' in \emph{Proceedings of the 40th International Conference on Machine Learning}, vol. 202.\hskip 1em plus 0.5em minus 0.4em\relax PMLR, 23--29 Jul 2023, pp. 17\,842--17\,857.

\bibitem{BBO2024ConstrainedSampler}
\BIBentryALTinterwordspacing
L.~Kong, Y.~Du, W.~Mu, K.~Neklyudov, V.~D. Bortoli, D.~Wu, H.~Wang, A.~Ferber, Y.-A. Ma, C.~P. Gomes, and C.~Zhang, ``{Diffusion Models as Constrained Samplers for Optimization with Unknown Constraints},'' 2024. [Online]. Available: \url{https://arxiv.org/abs/2402.18012}
\BIBentrySTDinterwordspacing

\bibitem{AAAIBeatGAN2023TopOptimization}
F.~Mazé and F.~Ahmed, ``{Diffusion Models Beat GANs on Topology Optimization},'' in \emph{Proceedings of the AAAI Conference on Artificial Intelligence}, vol.~37, no.~8, Jun. 2023, pp. 9108--9116.

\bibitem{GDM2023TopOptimization}
G.~Giannone, A.~Srivastava, O.~Winther, and F.~Ahmed, ``{Aligning Optimization Trajectories with Diffusion Models for Constrained Design Generation},'' in \emph{Proceedings of Advances in Neural Information Processing Systems}, vol.~36, 2023, pp. 51\,830--51\,861.

\bibitem{CDDM2023globecom}
T.~Wu, Z.~Chen, D.~He, L.~Qian, Y.~Xu, M.~Tao, and W.~Zhang, ``{CDDM: Channel Denoising Diffusion Models for Wireless Communications},'' in \emph{Proceedings of 2023 IEEE Global Communications Conference}, 2023, pp. 7429--7434.

\bibitem{GDMDenoiseSensing2024JSAC}
J.~Wang, H.~Du, D.~Niyato, Z.~Xiong, J.~Kang, B.~Ai, Z.~Han, and D.~In~Kim, ``{Generative Artificial Intelligence Assisted Wireless Sensing: Human Flow Detection in Practical Communication Environments},'' \emph{IEEE Journal on Selected Areas in Communications}, vol.~42, no.~10, pp. 2737--2753, 2024.

\bibitem{RF_Diffusion}
G.~Chi, Z.~Yang, C.~Wu, J.~Xu, Y.~Gao, Y.~Liu, and T.~X. Han, ``{RF-Diffusion: Radio Signal Generation via Time-Frequency Diffusion},'' in \emph{Proceedings of the 30th Annual International Conference on Mobile Computing and Networking}.\hskip 1em plus 0.5em minus 0.4em\relax Association for Computing Machinery, 2024, p. 77–92.

\bibitem{Privacy2025infocomGDM}
W.~Ningning, Z.~Tianya, M.~Shiwen, and W.~Xuyu, ``{Privacy-Preserving Wi-Fi Data Generation via Differential Privacy in Diffusion Models},'' in \emph{Proceedings of IEEE International Conference on Computer Communications}, 2025, pp. 1--10.

\bibitem{netgpt2023tenissues}
\BIBentryALTinterwordspacing
W.~Tong, C.~Peng, T.~Yang, F.~Wang, J.~Deng, R.~Li, L.~Yang, H.~Zhang, D.~Wang, M.~Ai, L.~Yang, G.~Liu, Y.~Yang, Y.~Xiao, L.~Yue, W.~Sun, Z.~Li, and W.~Sun, ``{Ten issues of NetGPT},'' 2023. [Online]. Available: \url{https://arxiv.org/abs/2311.13106}
\BIBentrySTDinterwordspacing

\bibitem{understanding2023telecomLLM}
L.~Bariah, H.~Zou, Q.~Zhao, B.~Mouhouche, F.~Bader, and M.~Debbah, ``{Understanding Telecom Language Through Large Language Models},'' in \emph{Proceedings of 2023 IEEE Global Communications Conference}, 2023, pp. 6542--6547.

\bibitem{Telecom2024MustLarge}
N.~Piovesan, A.~De~Domenicoo, and F.~Ayed, ``{Telecom Language Models: Must They Be Large?}'' in \emph{Proceedings of 2024 IEEE 35th International Symposium on Personal, Indoor and Mobile Radio Communications (PIMRC)}, 2024, pp. 1--6.

\bibitem{TeleQnA2023}
\BIBentryALTinterwordspacing
A.~Maatouk, F.~Ayed, N.~Piovesan, A.~D. Domenico, M.~Debbah, and Z.-Q. Luo, ``{TeleQnA: A Benchmark Dataset to Assess Large Language Models Telecommunications Knowledge},'' 2023. [Online]. Available: \url{https://arxiv.org/abs/2310.15051}
\BIBentrySTDinterwordspacing

\bibitem{TSpec_LLM2024}
\BIBentryALTinterwordspacing
R.~Nikbakht, M.~Benzaghta, and G.~Geraci, ``{TSpec-LLM: An Open-source Dataset for LLM Understanding of 3GPP Specifications},'' 2024. [Online]. Available: \url{https://arxiv.org/abs/2406.01768}
\BIBentrySTDinterwordspacing

\bibitem{TelecomGPT2024}
\BIBentryALTinterwordspacing
H.~Zou, Q.~Zhao, Y.~Tian, L.~Bariah, F.~Bader, T.~Lestable, and M.~Debbah, ``{TelecomGPT: A Framework to Build Telecom-Specfic Large Language Models},'' 2024. [Online]. Available: \url{https://arxiv.org/abs/2407.09424}
\BIBentrySTDinterwordspacing

\bibitem{Mobile_LLaMA}
K.~B. Kan, H.~Mun, G.~Cao, and Y.~Lee, ``{Mobile-LLaMA: Instruction Fine-Tuning Open-Source LLM for Network Analysis in 5G Networks},'' \emph{IEEE Network}, vol.~38, no.~5, pp. 76--83, 2024.

\bibitem{MobileFoundationFirmware2024}
J.~Yuan, C.~Yang, D.~Cai, S.~Wang, X.~Yuan, Z.~Zhang, X.~Li, D.~Zhang, H.~Mei, X.~Jia, S.~Wang, and M.~Xu, ``{Mobile Foundation Model as Firmware},'' in \emph{Proceedings of the 30th Annual International Conference on Mobile Computing and Networking}.\hskip 1em plus 0.5em minus 0.4em\relax Association for Computing Machinery, 2024, p. 279–295.

\bibitem{WirelessGPT2025Pengcheng}
T.~Yang, P.~Zhang, M.~Zheng, Y.~Shi, L.~Jing, J.~Huang, and N.~Li, ``{WirelessGPT: A Generative Pre-trained Multi-task Learning Framework for Wireless Communication},'' \emph{IEEE Network}, pp. 1--1, 2025.

\bibitem{NonConvexLLMOpt2025}
H.~Li, M.~Xiao, K.~Wang, D.~I. Kim, and M.~Debbah, ``{Large Language Model Based Multi-Objective Optimization for Integrated Sensing and Communications in UAV Networks},'' \emph{IEEE Wireless Communications Letters}, vol.~14, no.~4, pp. 979--983, 2025.

\bibitem{lambo2024}
L.~Dong, F.~Jiang, Y.~Peng, K.~Wang, K.~Yang, C.~Pan, and R.~Schober, ``{LAMBO: Large AI Model Empowered Edge Intelligence},'' \emph{IEEE Communications Magazine}, vol.~63, no.~4, pp. 88--94, 2025.

\bibitem{netllm2024sigcomm}
D.~Wu, X.~Wang, Y.~Qiao, Z.~Wang, J.~Jiang, S.~Cui, and F.~Wang, ``{NetLLM: Adapting Large Language Models for Networking},'' in \emph{Proceedings of the ACM SIGCOMM 2024 Conference}.\hskip 1em plus 0.5em minus 0.4em\relax Association for Computing Machinery, 2024, p. 661–678.

\bibitem{LORA}
E.~J. Hu, Y.~Shen, P.~Wallis, Z.~Allen-Zhu, Y.~Li, S.~Wang, L.~Wang, W.~Chen \emph{et~al.}, ``{LoRA: Low-rank adaptation of large language models.}'' in \emph{Proceedings of International Conference on Learning Representations}, vol.~1, no.~2, 2022, p.~3.

\bibitem{LLM2024NO_VQR}
T.~Mongaillard, S.~Lasaulce, O.~Hicheur, C.~Zhang, L.~Bariah, V.~S. Varma, H.~Zou, Q.~Zhao, and M.~Debbah, ``{Large Language Models for Power Scheduling: A User-Centric Approach},'' in \emph{Proceedings of 2024 22nd International Symposium on Modeling and Optimization in Mobile, Ad Hoc, and Wireless Networks (WiOpt)}, 2024, pp. 321--328.

\bibitem{LAMeTA2025}
\BIBentryALTinterwordspacing
Y.~Liu, G.~Liu, J.~Wang, R.~Zhang, D.~Niyato, G.~Sun, Z.~Xiong, and Z.~Han, ``{LAMeTA: Intent-Aware Agentic Network Optimization via a Large AI Model-Empowered Two-Stage Approach},'' 2025. [Online]. Available: \url{https://arxiv.org/abs/2505.12247}
\BIBentrySTDinterwordspacing

\bibitem{ToolAided2025FL}
\BIBentryALTinterwordspacing
C.~Tan, R.~Wen, R.~Li, Z.~Zhao, E.~Hossain, and H.~Zhang, ``{Tool-Aided Evolutionary LLM for Generative Policy Toward Efficient Resource Management in Wireless Federated Learning},'' 2025. [Online]. Available: \url{https://arxiv.org/abs/2505.11570}
\BIBentrySTDinterwordspacing

\bibitem{LLMind2024}
H.~Cui, Y.~Du, Q.~Yang, Y.~Shao, and S.~C. Liew, ``{LLMind: Orchestrating AI and IoT with LLM for Complex Task Execution},'' \emph{IEEE Communications Magazine}, vol.~63, no.~4, pp. 214--220, 2025.

\bibitem{LLMsubtask2025hotnets}
A.~Nazari, Y.~Zhang, M.~Raghothaman, and H.~Chen, ``{Localized Explanations for Automatically Synthesized Network Configurations},'' in \emph{Proceedings of the 23rd ACM Workshop on Hot Topics in Networks}.\hskip 1em plus 0.5em minus 0.4em\relax Association for Computing Machinery, 2024, p. 52–59.

\bibitem{LLMGraphOperation2023NetworkManage}
S.~K. Mani, Y.~Zhou, K.~Hsieh, S.~Segarra, T.~Eberl, E.~Azulai, I.~Frizler, R.~Chandra, and S.~Kandula, ``{Enhancing Network Management Using Code Generated by Large Language Models},'' in \emph{Proceedings of the 22nd ACM Workshop on Hot Topics in Networks}.\hskip 1em plus 0.5em minus 0.4em\relax Association for Computing Machinery, 2023, p. 196–204.

\bibitem{NADA2024HotNets}
Z.~He, A.~Gottipati, L.~Qiu, X.~Luo, K.~Xu, Y.~Yang, and F.~Y. Yan, ``{Designing Network Algorithms via Large Language Models},'' in \emph{Proceedings of the 23rd ACM Workshop on Hot Topics in Networks}.\hskip 1em plus 0.5em minus 0.4em\relax Association for Computing Machinery, 2024, p. 205–212.

\bibitem{Ottor2023}
S.~Waskito, K.~J. Leow, P.~Medaranga, T.~Gupta, S.~Chakrabarty, M.~Gulati, and A.~Varshney, ``{Otter: Simplifying Embedded Sensor Data Collection and Analysis using Large Language Models},'' in \emph{Proceedings of the 29th Annual International Conference on Mobile Computing and Networking}.\hskip 1em plus 0.5em minus 0.4em\relax Association for Computing Machinery, 2023.

\bibitem{emergence2024Diffusion}
H.~Zhang, J.~Zhou, Y.~Lu, M.~Guo, P.~Wang, L.~Shen, and Q.~Qu, ``{The Emergence of Reproducibility and Consistency in Diffusion Models},'' in \emph{Proceedings of Forty-first International Conference on Machine Learning}, 2024.

\bibitem{understanding2024generalizability}
X.~Li, Y.~Dai, and Q.~Qu, ``{Understanding Generalizability of Diffusion Models Requires Rethinking the Hidden Gaussian Structure},'' in \emph{Proceedings of Advances in Neural Information Processing Systems}, vol.~37, 2024, pp. 57\,499--57\,538.

\bibitem{fewShot2024DiffusionCurse}
R.~Yang, B.~Jiang, C.~Chen, R.~Jin, B.~Wang, and S.~Li, ``{Few-Shot Diffusion Models Escape the Curse of Dimensionality},'' in \emph{Proceedings of Advances in Neural Information Processing Systems}, vol.~37, 2024, pp. 68\,528--68\,558.

\bibitem{PDL2002}
S.~Kakade and J.~Langford, ``{Approximately Optimal Approximate Reinforcement Learning},'' in \emph{Proceedings of the Nineteenth International Conference on Machine Learning}.\hskip 1em plus 0.5em minus 0.4em\relax Morgan Kaufmann Publishers Inc., 2002, p. 267–274.

\bibitem{TV2002RelatedLipschitz}
\BIBentryALTinterwordspacing
A.~L. Gibbs and F.~E. Su, ``{On choosing and bounding probability metrics},'' 2002. [Online]. Available: \url{https://arxiv.org/abs/math/0209021}
\BIBentrySTDinterwordspacing

\bibitem{Autoformulation2024LLM}
\BIBentryALTinterwordspacing
N.~Astorga, T.~Liu, Y.~Xiao, and M.~van~der Schaar, ``{Autoformulation of Mathematical Optimization Models Using LLMs},'' 2024. [Online]. Available: \url{https://arxiv.org/abs/2411.01679}
\BIBentrySTDinterwordspacing

\bibitem{CARD2022NIPS}
X.~Han, H.~Zheng, and M.~Zhou, ``{CARD: Classification and Regression Diffusion Models},'' in \emph{Proceedings of Advances in Neural Information Processing Systems}, vol.~35, 2022, pp. 18\,100--18\,115.

\bibitem{Tokens2Thoughts2025}
\BIBentryALTinterwordspacing
C.~Shani, D.~Jurafsky, Y.~LeCun, and R.~Shwartz-Ziv, ``{From Tokens to Thoughts: How LLMs and Humans Trade Compression for Meaning},'' 2025. [Online]. Available: \url{https://arxiv.org/abs/2505.17117}
\BIBentrySTDinterwordspacing

\bibitem{MATH2021Dataset}
\BIBentryALTinterwordspacing
D.~Hendrycks, C.~Burns, S.~Kadavath, A.~Arora, S.~Basart, E.~Tang, D.~Song, and J.~Steinhardt, ``{Measuring Mathematical Problem Solving With the MATH Dataset},'' 2021. [Online]. Available: \url{https://arxiv.org/abs/2103.03874}
\BIBentrySTDinterwordspacing

\bibitem{GeneralistRewardModels2025}
\BIBentryALTinterwordspacing
Y.-C. Li, T.~Xu, Y.~Yu, X.~Zhang, X.-H. Chen, Z.~Ling, N.~Chao, L.~Yuan, and Z.-H. Zhou, ``{Generalist Reward Models: Found Inside Large Language Models},'' 2025. [Online]. Available: \url{https://arxiv.org/abs/2506.23235}
\BIBentrySTDinterwordspacing

\bibitem{ReasoningBank}
\BIBentryALTinterwordspacing
S.~Ouyang, J.~Yan, I.-H. Hsu, Y.~Chen, K.~Jiang, Z.~Wang, R.~Han, L.~T. Le, S.~Daruki, X.~Tang, V.~Tirumalashetty, G.~Lee, M.~Rofouei, H.~Lin, J.~Han, C.-Y. Lee, and T.~Pfister, ``{ReasoningBank: Scaling Agent Self-Evolving with Reasoning Memory},'' 2025. [Online]. Available: \url{https://arxiv.org/abs/2509.25140}
\BIBentrySTDinterwordspacing

\bibitem{MarkovianThinker}
\BIBentryALTinterwordspacing
M.~Aghajohari, K.~Chitsaz, A.~Kazemnejad, S.~Chandar, A.~Sordoni, A.~Courville, and S.~Reddy, ``{The Markovian Thinker},'' 2025. [Online]. Available: \url{https://arxiv.org/abs/2510.06557}
\BIBentrySTDinterwordspacing

\bibitem{WirelessNetworkDesign2019TCOM}
A.~Zappone, M.~Di~Renzo, and M.~Debbah, ``{Wireless Networks Design in the Era of Deep Learning: Model-Based, AI-Based, or Both?}'' \emph{IEEE Transactions on Communications}, vol.~67, no.~10, pp. 7331--7376, 2019.

\bibitem{LearnToCO2022infocom}
Z.~Shao, J.~Yang, C.~Shen, and S.~Ren, ``{Learning for Robust Combinatorial Optimization: Algorithm and Application},'' in \emph{Proceedings of IEEE Conference on Computer Communications}, 2022, pp. 930--939.

\bibitem{FreeEnergyCO2025NatureCS}
Z.-S. Shen, F.~Pan, Y.~Wang, Y.-D. Men, W.-B. Xu, M.-H. Yung, and P.~Zhang, ``{Free-energy machine for combinatorial optimization},'' \emph{Nature Computational Science}, pp. 1--11, 2025.

\bibitem{LearnToCut2024TPAMI}
J.~Wang, Z.~Wang, X.~Li, Y.~Kuang, Z.~Shi, F.~Zhu, M.~Yuan, J.~Zeng, Y.~Zhang, and F.~Wu, ``{Learning to Cut via Hierarchical Sequence/Set Model for Efficient Mixed-Integer Programming},'' \emph{IEEE Transactions on Pattern Analysis and Machine Intelligence}, vol.~46, no.~12, pp. 9697--9713, 2024.

\bibitem{inferenceTimeScaling2025Xie}
\BIBentryALTinterwordspacing
N.~Ma, S.~Tong, H.~Jia, H.~Hu, Y.-C. Su, M.~Zhang, X.~Yang, Y.~Li, T.~Jaakkola, X.~Jia, and S.~Xie, ``{Inference-Time Scaling for Diffusion Models beyond Scaling Denoising Steps},'' 2025. [Online]. Available: \url{https://arxiv.org/abs/2501.09732}
\BIBentrySTDinterwordspacing

\bibitem{IntegrateFormal2023hotnets}
F.~Gong, D.~Raghunathan, A.~Gupta, and M.~Apostolaki, ``{Towards Integrating Formal Methods into ML-Based Systems for Networking},'' in \emph{Proceedings of the 22nd ACM Workshop on Hot Topics in Networks}.\hskip 1em plus 0.5em minus 0.4em\relax Association for Computing Machinery, 2023, p. 48–55.

\bibitem{formalMethod2022AsProblemAware}
M.~Krichen, A.~Mihoub, M.~Y. Alzahrani, W.~Y.~H. Adoni, and T.~Nahhal, ``{Are Formal Methods Applicable To Machine Learning And Artificial Intelligence?}'' in \emph{Proceedings of 2nd International Conference of Smart Systems and Emerging Technologies (SMARTTECH)}, 2022, pp. 48--53.

\bibitem{DecentralizedGDM}
\BIBentryALTinterwordspacing
D.~McAllister, M.~Tancik, J.~Song, and A.~Kanazawa, ``{Decentralized Diffusion Models},'' 2025. [Online]. Available: \url{https://arxiv.org/abs/2501.05450}
\BIBentrySTDinterwordspacing

\bibitem{etsiGReni051}
``{Experiential Networked Intelligence (ENI); Study on AI Agents based Next-generation Network Slicing},'' European Telecommunications Standards Institute (ETSI), ETSI Group Report ENI 051 V4.1.1, Feb. 2025, \url{https://www.etsi.org/deliver/etsi_gr/ENI/001_099/051/04.01.01_60/gr_ENI051v040101p.pdf}.

\bibitem{ngmn2025_6G}
``{NETWORK ARCHITECTURE EVOLUTION TOWARDS 6G},'' Next Generation Mobile Networks Alliance (NGMN), NGMN Network Architecture Evolution towards 6G 1.0, Feb. 2025, \url{https://www.ngmn.org/wp-content/uploads/250218_Network_Architecture_Evolution_towards_6G_V1.0.pdf}.

\bibitem{MATH_bench}
D.~Hendrycks, C.~Burns, S.~Kadavath, A.~Arora, S.~Basart, E.~Tang, D.~Song, and J.~Steinhardt, ``{Measuring Mathematical Problem Solving With the MATH Dataset},'' in \emph{Proceedings of Advances in Neural Information Processing Systems}, 2021.

\bibitem{GSM8K}
\BIBentryALTinterwordspacing
K.~Cobbe, V.~Kosaraju, M.~Bavarian, M.~Chen, H.~Jun, L.~Kaiser, M.~Plappert, J.~Tworek, J.~Hilton, R.~Nakano, C.~Hesse, and J.~Schulman, ``{Training Verifiers to Solve Math Word Problems},'' 2021. [Online]. Available: \url{https://arxiv.org/abs/2110.14168}
\BIBentrySTDinterwordspacing

\bibitem{Sensiverse}
\BIBentryALTinterwordspacing
J.~Luo, B.~Zhou, Y.~Yu, P.~Zhang, X.~Peng, J.~Ma, P.~Zhu, J.~Lu, and W.~Tong, ``{Sensiverse: A dataset for ISAC study},'' 2023. [Online]. Available: \url{https://arxiv.org/abs/2308.13789}
\BIBentrySTDinterwordspacing

\end{thebibliography}

	
	
	
	
	

	\vfill
	
\end{document}